\documentclass[12pt]{article}
\pdfoutput=1 
\usepackage{amssymb,amsmath,bm,bbm}
\usepackage{epsf}
\usepackage{epsfig}
\usepackage[normalem]{ulem}
\usepackage{afterpage}
\usepackage{longtable}
\usepackage{latexsym, mathrsfs}
\usepackage{graphics}
\usepackage{subfigure}
\usepackage{appendix}
\usepackage{color}
 \usepackage[bookmarks=true]{hyperref}

\def\beq{\begin{equation}}
\def\eeq{\end{equation}}
\def\hhref#1{\href{http://arxiv.org/abs/#1}{#1}} 

\newcommand{\be}{\begin{equation}}
\newcommand{\ee}{\end{equation}}
\newcommand{\bea}{\begin{eqnarray}}
\newcommand{\eea}{\end{eqnarray}}

\newcommand{\bi}{\begin{itemize}}
\newcommand{\ei}{\end{itemize}}

\newcommand{\gev}{\text{ GeV}}

\makeatletter
\def\art{\@ifnextchar[{\eart}{\oart}}
\def\eart[#1]#2#3#4#5#6{{\rm #2}, {#3 #4} {\rm (#6) #5} [{\em arXiv:\hhref{#1}}]}
\def\hepart[#1]#2{{{\rm #2}, \em arXiv:\hhref{#1}}}
\newcommand{\oart}[5]{{\rm #1}, {#2 #3} {\rm (#5) #4}}

\def\linkart[#1]{{\em arXiv:\hhref{#1}}}

\begin{document}
\begin{titlepage}
\vspace*{-0.5truecm}

\begin{flushright}
{MPP-2010-41}\\
{TUM-HEP-755/10}
\end{flushright}

\vfill

\begin{center}
\boldmath

{\Large\textbf{Solving the $\mu$ problem with a heavy Higgs boson}}\\
\unboldmath
\end{center}

\vspace{0.3truecm}
		
\begin{center}
{\bf Roberto Franceschini$^{a}$ and Stefania Gori$^{b,c}$}
\vspace{0.4truecm}

{\footnotesize
 $^a${\sl Institute de Theorie des Phenomenes Physiques,
Ecole Polytechnique Federale de Lausanne,\\
CH-1015 Lausanne, Switzerland}\vspace{0.2cm}

 $^b${\sl Physik Department, Technische Universit\"at M\"unchen,
D-85748 Garching, Germany}\vspace{0.2cm}
}

$^c${\sl Max-Planck-Institut f{\"u}r Physik (Werner-Heisenberg-Institut), \\
D-80805 M{\"u}nchen, Germany}\vspace{0.2cm}

\end{center}
\vspace{0.6cm}

\begin{abstract}
We discuss the generation of the $\mu$-term in a class of supersymmetric models characterized by a low energy effective superpotential containing a term $\lambda S H_1 H_2$ with a large coupling $\lambda\sim 2$. These models generically predict a lightest Higgs boson well above the LEP limit of 114 GeV and have been shown to be compatible with the unification of gauge couplings. Here we discuss a specific example where the superpotential has no dimensionful parameters and we point out the relation between the generated $\mu$-term and the mass of the lightest Higgs boson.
We discuss the fine-tuning of the model and we find that the generation of a phenomenologically viable $\mu$-term fits very well with a heavy lightest Higgs boson and a low degree of fine-tuning.

We discuss experimental constraints from collider direct searches, precision data, thermal relic dark matter abundance, and WIMP searches finding that the most natural region of the parameter space is still allowed by current experiments. 
We analyse bounds on the masses of the superpartners coming from Naturalness arguments and discuss the main signatures of the model for the LHC and future WIMP searches.
\end{abstract}
\end{titlepage}

\setcounter{page}{1}
\pagenumbering{arabic}

\section{Introduction}

At the dawn of the LEP era hopes were on the discovery of electroweak (EW) scale supersymmetry,
which provided a symmetry principle as solution of the hierarchy problem
of the Standard Model (SM) and a serious chance for the Higgs boson
to be in the reach of that machine.  The measurements of the gauge couplings performed at LEP turned out to be in very good agreement with the idea of a unification at scales $M_{GUT}\sim 10^{16} \gev$, as foreseen in the Minimal Supersymmetric Standard Model (MSSM). Unfortunately, the striking picture emerging from the study of the gauge couplings did not find a counterpart in the observation of either a light Higgs boson or of any of the numerous superpartners.

However one has to acknowledge that the expectation for SUSY, and in particular for a light Higgs boson, to be discovered at LEP was mostly based on strict requirements of minimality in model building.
The MSSM has in fact a scalar potential entirely fixed by the gauge symmetry and this yields the famous result that the lightest Higgs boson of the MSSM has to be light. Such lightness of the MSSM Higgs boson can be understood noting that at the tree-level the lightest Higgs boson mass can be upper-bounded by 

\begin{equation}
m_{Z}\cos2\beta\,,\label{boundMSSM}
\end{equation}

\noindent and therefore can even be equal to 0 if $\tan\beta=1$.

In the MSSM further contributions to the lightest Higgs mass arise at the one loop level and can lift the Higgs mass above the experimental lower-bound \cite{LEPhiggs} 

\beq
m_{h_\textrm{SM}}>114 \gev\,.\label{mhLEP}
\eeq

 \noindent However such large radiative corrections require soft masses much 
 larger than the ElectroWeak Symmetry Breaking (EWSB) scale, which
 loosens 
 the Naturalness argument to motivate supersymmetry at the TeV scale.

The necessity of large soft masses in the MSSM motivates the study of supersymmetric models that extend the MSSM and that can possibly alleviate the tension with LEP direct searches. Indeed a study of extensions of the MSSM with effective operators \cite{EFTBMSSM} shows that, departing from the minimal model, one can have substantial corrections to the bound on the lightest Higgs mass in eq.(\ref{boundMSSM}). A similar conclusion is drawn in Refs. \cite{Tobe:2002zj,Espinosa:1998re} where the issues of the Higgs mass and of the unification of gauge couplings have been considered in specific renormalizable extensions of the MSSM.

Refs. \cite{newvectors,dterms,SU2nondecDterms} are concrete examples of such models in which the tension with the LEP bound is alleviated introducing either extra vectorial fermions or extra gauge structure. In these models particular care is taken to preserve the unification of the gauge couplings in a manner that is as close as possible to that of the MSSM.  However the result is still not completely satisfactory because these models either are fine-tuned or mildly affect the maximal value of the mass of the Higgs \cite{Barbieri:2010pd}.

\bigskip

As a matter of fact, relaxing the requirement of strictly perturbative unification in general leads to  non-minimal models with significant changes for the phenomenology \cite{Lodone:2010kt}. Indeed, an attempt along this line has been made in \cite{fathiggs}, where the self-coupling of the Higgs sector has a strong coupling phase at some intermediate scale between the EWSB and the GUT scale. Handling such a strong coupling phase for the Higgs sector puts one in position to substantially increase the mass of the Higgs at the tree-level. The concrete examples in \cite{fathiggs} show that this can be done compatibly with the cherished unification of couplings. The idea has been further elaborated in \cite{lUV}  where it has been discussed a unification-compatible UV completion for the  so-called Next to Minimal Supersymmetric Standard Model (NMSSM), {\em i.e.} for a model described by a superpotential of the form
\beq
W=\lambda S H_1 \cdot H_2 + f(S)\, , \label{generalNMSSM}
\eeq
where the superfields $H_1$ and $H_2$ are Higgs doublets and $S$ is a SM singlet.

With respect to the commonly studied case with perturbative unification, the low energy NMSSM of \cite{fathiggs,lUV} can have larger couplings that become non-perturbative much below the gauge coupling unification scale. The extended range of acceptable couplings allows the mass of the lightest Higgs boson to reach 200-300 GeV, which leads to dramatic differences in the phenomenology of the Higgs sector with respect to the MSSM.

Indeed, if one considers a model with a superpotential including the term $ \lambda S H_1\cdot H_2$, as in (\ref{generalNMSSM}), the upper-bound on the lightest Higgs mass is given by

\begin{equation}
\sqrt{m_{Z}^2\cos^22\beta+\lambda^2 v^2\sin^22\beta}\,,\label{boundNMSSM}
\end{equation}

 \noindent which does not vanish for any value of $\tan\beta$ and becomes larger as one takes larger values for $\lambda$.

However the maximal value of $\lambda$ at the EW scale that does not lead to a Landau pole below the GUT scale is $\sim 0.7$  \cite{BarbieriPQ,lambdamax}, which only modestly affects the upper-bound on the Higgs mass, if compared to the MSSM. This motivates the choice to study the regime of the NMSSM where the coupling $\lambda$ can be $\mathcal{O}(1)$, namely the regime of the model in which the Higgs becomes heavy. Furthermore, the large Higgs mass attainable for $\lambda\sim 1$ automatically reduces the sensitivity of the mass of the Higgs to UV effects, and consequently the model turns out to be less fine tuned. This feature adds further motivation to pursue this regime.

\bigskip

Motivated by this exciting possibility, Ref. \cite{lsusy} analysed the framework where (\ref{generalNMSSM}) is taken as a low energy effective superpotential below a scale of $\mathcal{O}(10 \textrm{ TeV})$. Given the great importance covered by the coupling $\lambda$, this framework goes under the name $\lambda$SUSY.

In Ref. \cite{lsusy} a very detailed analysis of the Electroweak Precision Tests (EWPT) has been performed with the result that $\lambda$SUSY can be in agreement with LEP data.
Ref. \cite{lsusy} also studied the issue of the Naturalness of the model, finding that this kind of models typically do not need to be tuned.  The absence of tuning is somehow expected because the major source of fine-tuning in the MSSM is due to the need to generate large loop corrections to raise the Higgs mass above $m_Z$. 
Including the tree-level contribution coming from the large coupling $\lambda$, this need is no longer a concern and one is not forced to push the model to an unnatural region of its parameter space because of the LEP bound on the Higgs mass.
In this sense $\lambda$SUSY is a remarkable candidate for a natural supersymmetric theory of  Electroweak Symmetry Breaking.

Furthermore it has been shown that in $\lambda$SUSY a singlino-like lightest supersymmetric particle (LSP) can be a weakly interacting massive particle (WIMP) dark matter candidate with the correct thermal relic abundance. Additionally,   $\lambda$SUSY has a strikingly different Higgs sector with respect to the one of the MSSM and the one of the perturbative NMSSM, which leads to testable distinctive signals for the LHC \cite{lsusylhc}.

\bigskip

An additional virtue of the NMSSM, if compared to the MSSM, is the possibility of generating the Higgsino mass term $\mu$ dynamically. Indeed if the singlet $S$ takes a VEV, the interaction in eq. (\ref{generalNMSSM}) generates an effective mass for the Higgsinos $\mu=\lambda \langle S \rangle$.
\noindent In the NMSSM the VEV of the singlet depends on the same soft masses that trigger the VEV of the doublets. Therefore the $\mu$ term is generated by the same dynamics that breaks the EW symmetry, rendering evident why $\mu$ is of the same order of the EW scale instead of being  zero or of the order of some other energy scale that characterizes the (unspecified) UV theory.

In this sense the solution of the $\mu$ problem in the NMSSM is very economical and constitutes a testable alternative to other mechanisms that generate the $\mu$ term through the same mechanism that generates the soft masses~\cite{DGP,musol,csakietal, GiuRa}\cite{GiuMa}
\footnote{As a matter of fact explicit models of supersymmetry breaking generically have difficulties \cite{DGP} to generate correctly the $\mu$ term and special solutions for the generation of $\mu$ are needed. In this sense the NMSSM appears more suitable for an economic dynamical generation of $\mu$. Indeed it has already been considered  the possibility to generate $\mu$ from an NMSSM superpotential  with soft masses generated by gauge mediated supersymmetry braking \cite{DGS}.}.

Previous works on $\lambda$SUSY \cite{lsusy} did not try to study the dynamical generation of $\mu$ and simply put the $\mu$ term by hand in the superpotential. In this work we shall address the issue of the dynamical generation of the $\mu$ term in $\lambda$SUSY, which constitutes an important piece of information to complete the current picture of  supersymmetry without a light Higgs boson.

In our work we will discuss analytically the relation between the mass term $\mu$ generated thanks to the interaction $\lambda S H_1 \cdot H_2$ and the other mass scales of the model. In particular we will discuss to what extent the dynamical generation of the $\mu$ term through the VEV of the singlet scalar requires the doublets $H_{1}$ and $H_{2}$ to be mixed with the singlet. Because of the necessity of non-negligible mixing between doublets and singlet, we will generalize the analysis of \cite{lsusy} including the effects of the mixing where appropriate, as for instance in the EWPT analysis, in the consideration of limits from the direct searches of dark matter, and in the phenomenology at the LHC.

A recent study of the NMSSM in the large $\lambda$ regime has been  performed in Ref. \cite{CY}, where the importance of a dynamical generation of a $\mu$ term is not highlighted and a numeric scan of the parameter space allowed by the many experimental constraints is performed. Contrary to this work, we will pursue an analytic approach as much as we can.  Moreover the set of constraints that we will consider will be slightly different with respect to \cite{CY}. Notably, we will not require the NMSSM to provide an explanation for the current discrepancy between the experimental value and  the SM prediction of the $g-2$ of the muon. Besides, we do not impose the thermal production of lightest neutralinos to account for the observed relic dark matter abundance, as we content ourself to not overclose the Universe with the lightest supersymmetric particle. At variance with \cite{CY} we will take into account limits coming from direct dark matter searches through the elastic scattering of a weakly-interacting massive particle on a nucleus. 
 
\bigskip 

Our paper is organized as follows. In the first part, we present the model and its theoretical predictions concerning the spectrum, the generation of the $\mu$ term and the level of fine-tuning required. In particular in Section \ref{Sec:model} we describe the model and fix our notation.  In Sections \ref{global} and \ref{sec:EWSB} we discuss the conditions for the global stability of the potential, for the preservation of the CP invariance and for a realistic breaking of the electroweak symmetry. In Section \ref{sec:constraintsvacuum} we summarize the constraints on the scalar potential for a correct electroweak symmetry breaking and for the preservation of the CP invariance, showing which is the allowed subspace of parameter space.
In Section \ref{Sec:masses} we discuss the spectrum of the model. In Section \ref{solution} we analyse the generation of an effective $\mu$ term. In Section \ref{FT} we discuss the fine-tuning of the model and the predictions on the sparticle masses coming from Naturalness arguments.

 In the second part of the paper we discuss the experimental bounds that the theory has to satisfy. In particular, in Section \ref{sec:LEP} we  discuss  the limits from LEP direct searches. Then, in Section \ref{EWPT} we study the indirect constraints coming from Electroweak Precision Tests. In Sections \ref{relicDM} and \ref{directdetection} we discuss the relic abundance of the LSP of the model and its detection in current experiments. 
 
  Finally in {the last part of the paper (Section \ref{LHC}) we discuss the phenomenology of the model and its possible signatures at the LHC and in Section \ref{conclusions} we give our conclusions.
\section{The model}\label{Sec:model}
A recent review on the NMSSM can be found in \cite{review}, but for sake of completeness and to fix our notation, we give here some details on the model. 
We start considering the most general theory with $SU(3)_C\times SU(2)_L\times U(1)_Y$ gauge invariance that contains, in addition to the Higgs fields of the MSSM, a singlet chiral field $S$. The most general superpotential that we can assign to the theory is
%
 $W_{gen}(\Phi_i)=\mu H_{1}\cdot H_{2}+\frac{M}{2}S^{2}+\lambda SH_{1}\cdot H_{2}+\frac{k}{3}S^{3}\,,$
%
where $H_1$ and $H_2$ are the Higgs superfields with $-1/2$ and $1/2$ hypercharge respectively. This superpotential generalizes the superpotential of the MSSM and still contains dimensionful parameters such as $\mu$ and $M$. Although these parameters are  protected by holomorphy, their presence is typically seen as a problem, the ``$\mu$ problem'', because their size has to be fixed by hand, and in particular the $\mu$ parameter should be fixed to a value closed to the $Z$ boson mass. To address this issue we impose a continuous R-symmetry such to avoid  all the terms with a dimensionful constant. Consequently the theory is determined by  

\begin{equation}\label{superpotential}
 W_{\textrm{NMSSM}}(\Phi_i)=\lambda SH_{1}\cdot H_{2}+\frac{k}{3}S^{3}\,.
\end{equation}

 This superpotential plus the corresponding soft supersymmetry breaking potential

\begin{equation}\label{eq:softterms}
 V_\text{soft}=m_1^2\left|H_1\right|^2+m_2^2\left|H_2\right|^2+\mu_S^2\left|S\right|^2-(A\lambda SH_1H_2+G\frac{k}{3}S^3+h.c.)\,,
\end{equation}

\noindent and the D-terms $V_D$ coming from gauge interactions fully define our model. 

The D-terms of our model are exactly the same of the MSSM:

\begin{eqnarray}
V_{Y}&=&\frac{1}{8}g_{1}^{2}(\left|H_{2}\right|^{2}-\left|H_{1}\right|^{2})^{2}\,,\\
V_{2}&=&\frac{1}{8}g_{2}^{2}\left(H_{1}^{\dagger}T^{i}H_{1}+H_{2}^{\dagger}T^{i}H_{2}\right)^{2}\,,
\end{eqnarray}

\noindent where $T^i=\frac{\sigma^i}{2}$. Using the well known identity for canonical generators of $SU(2)$ $\sum_{i}T_{ab}^{i}T_{cd}^{i}=2\delta_{ad}\delta_{bc}-\delta_{ab}\delta_{cd}$, we can write again the total gauge potential as

\begin{equation}
V_{D}\equiv V_{Y}+V_{2}=\frac{1}{8}\left(g_{2}^{2}+g_{1}^{2}\right)(\left|H_{2}\right|^{2}-\left|H_{1}\right|^{2})^{2}+\frac{1}{2}g_{2}^{2}|H_{1}^{\dagger}H_{2}|^{2}\,.\label{eq:VD}
\end{equation}

The total scalar potential of the theory 
will also include several terms involving the squark and slepton fields, but for our purposes it will not be necessary to deal with them.
In the following  we will assume for simplicity that all the parameters of the Higgs-Higgsino potential are real. 

 In addition, we conventionally choose the  couplings $\lambda$ and $k$  to be positive.

\section{Stability of the scalar potential}\label{global}
The first requirement we have to impose to the scalar potential is that it is bounded from below. At large values of the fields the quartic part of the potential dominates and, since its coupling is not negative, the only way to destabilize the potential is to make the quartic part vanishing along some direction. However, the quartic potential of the NMSSM is the sum of D-terms and F-terms which are each positive definite and, as we will recall in the following, they cannot vanish at the same time, if not for the trivial field configuration. As such, the global stability of the potential is guaranteed by the supersymmetric structure of the theory and no constraint for the soft terms emerges.

More in detail, the global stability of the potential can be shown as follows. The quartic part of the potential gets contributions from the F-terms and D-terms, $V^{(4)}=V_F+V_D$. Since  $V_F=\sum_i\left|\frac{\partial W}{\partial\Phi_i}\right|^2\equiv\sum_i \left| F_i \right|^2$, the F-term part of the potential vanishes only if all the $F_i$ vanish. This condition is satisfied, in general, along the direction
\begin{equation}\label{eq:Fflatk}
 S=0\,,\,\,\,\,\,H_1\cdot H_2=0\,,
\end{equation}
and for $k=0$ also along the direction of generic $S$ and 

\begin{equation}\label{eq:Fflat}
 H_1=0\,,\,\,\,\,\, H_2=0\,.
\end{equation}     

 $V_D$  is explicitly given in (\ref{eq:VD}) and is a sum of positive terms. Requiring each term to vanish, one finds that $V_D$ vanishes only along the non-trivial direction
\begin{eqnarray}\label{Dflat}
& H_{1}^{\dagger}H_{2}=0\,, \\
& \left| H_1 \right| - \left| H_2 \right| = 0 . \nonumber
\end{eqnarray}

 Consequently, for $k\neq 0$ the function $V^{(4)}$ vanishes only if both the conditions (\ref{eq:Fflatk}) and (\ref{Dflat}) hold, and this is only possible for the trivial configuration
\begin{equation}
H_1=0\,,\,\,\,\,\,H_2=0 \,,\,\,\,\,\, S=0\,.
\end{equation}
Namely, in the $k\neq 0$ case there is no non-trivial direction in field space along which the whole quartic potential vanishes. Consequently the potential is always positive at large values of the fields, and there are no constraints on the parameter space resulting from the condition of stability of the potential.

 In the $k=0$ case, instead, there is the additional F-flat  direction of (\ref{eq:Fflat}). Along this direction the condition for vanishing $V_D$  (eq. (\ref{Dflat})) is always valid and therefore the whole quartic potential vanishes. As such, the large field behaviour of the potential along this direction is dictated by the soft terms. Requiring the potential to be positive for large field values yields the condition $\mu_S^2>0$.

\section{The minimum of the potential }\label{sec:EWSB}
The minimum of the potential has to be a stationary point, therefore the extremal point conditions with respect to $H_i=(H_1^0,H_2^0,S)$ must hold in a non-trivial point $(v_1,v_2,s)$ 
\begin{equation}\label{eq:minimumcondition}
\frac{\partial }{\partial H_i} \left. V \right|_{H_1^0=v_1,\,H_2^0=v_2,\,S=s}=0\, ,
\end{equation}

\noindent which are equivalent to
\begin{eqnarray}
\label{eq:sin2beta}
\cos^2\beta&=&\frac{m_2^2 + m_Z^2/2 + \lambda^2 s^2}{m_1^2 + m_2^2 + m_Z^2 + 2 \lambda^2 s^2}\,,\\\label{eq:v} 
\lambda^2v^2&=& \frac{2s \lambda (A-k s)}{\sin 2\beta}+  \frac{m_1^2-m_2^2}{\cos 2 \beta }+m_Z^2\,,
\end{eqnarray}
and 
\begin{equation}
\label{eq:s}
4 k^2 s^3-2 G k s^2+2 s \left(v^2 \lambda  (k \sin 2 \beta +\lambda )+\mu _S^2\right)- A v^2 \lambda  \sin 2\beta =0\, .
\end{equation}

The three conditions (\ref{eq:sin2beta})-(\ref{eq:s}) will be used as relations to trade $v$, $\tan\beta$ and $s$ for the soft parameters $m_{1},m_{2}$ and $\mu_S$.

For $k=0$ we can write explicitly the solution of the last equation for the VEV $s$

\begin{equation}\label{eq:sk0}
s=\frac{A\lambda v^2 \sin2\beta}{2\left(\mu_{S}^{2}+v^{2}\lambda^{2} \right)}\,.
\end{equation} 

Instead in the general case of $k\neq 0$ there are three different solutions of (\ref{eq:s}). As in the following we will impose the CP invariance of the vacuum, we will take the only real solution for $s$. This, however, will not be reported in the text, because of its quite complicated and not transparent expression.

This stationary point has to be a minimum and therefore we will require the Hessian to be positive definite. In addition we want it to be a global minimum, hence we require it to be deeper than the origin
\begin{equation}\label{eq:minabs}
V(v_{1},v_{2},s)<V(0,0,0)=0\,.
\end{equation}
This latter condition reads 
\beq\label{eq:minabsformula}
\mu_S^2s^2+k^2s^4-2G\frac{k}{3}s^3-\frac{\lambda^2}{4}v^4 \sin^2 2\beta-\frac{m_Z^2v^2}{4}\cos^2 2\beta<0\,,
\eeq

\noindent while the condition on the Hessian happens to have a rather complicated expression when $k \neq 0 $ and therefore we do not give it explicitly here.

\bigskip

For the potential to break the electroweak symmetry down to electromagnetism, we have to ensure that the neutral Higgs boson potential has a global minimum in a point with non vanishing VEVs $v_1$ and $v_2$. Additionally, to ensure electromagnetism is not broken we have to impose a vanishing  VEV for the electrically charged Higgs boson. 
Furthermore we will require the absence of spontaneous breaking of the CP symmetry, which is a necessary condition for the theory to not have tachyons at the extremal point \cite{romao}. This requirement amounts to impose that the VEVs of the scalar fields at the minimum of the potential have zero imaginary parts and will allow to treat the real and imaginary part of the scalar fields as non-mixing fields such that their mass matrices will be respectively a 3-by-3 and a 2-by-2 matrices, as opposed to the general case where a 5-by-5 mass matrix is needed.

  The requirement of positive VEVs $v_1$ and $v_2$ is given by

\bea\label{eq:SU2positive}
\lambda s (A-ks)> \sqrt{(A-ks)\lambda s\tan\beta-\lambda^2v^2\sin^2	\beta-\frac{m_Z^2}{2}\cos 2\beta}\\\nonumber
\times \sqrt{\lambda s \frac{A-ks}{\tan\beta}-\lambda^2v^2\cos^2	\beta+\frac{m_Z^2}{2}\cos 2\beta}\,,
\eea

\noindent where $s$ is expressed by the real solution of (\ref{eq:s}).

The discussion of the conservation of the $U(1)_{em}$ symmetry is simplified in the basis where only one Higgs doublet gets a VEV, $v$. In addition we exploit gauge invariance to set to zero the charged component of this same Higgs doublet such that we are left with a single charged scalar field that we call $\phi^\pm$. The absence of a VEV for $\phi^\pm$ is then expressed by the condition

\begin{equation}
\left. \frac{\partial^2 V}{\partial \phi^\pm\partial\phi^{\pm\dagger}}\right|_{\phi^\pm=0}>0\, ,
\end{equation}
 which yields the condition on the soft breaking parameters

\beq
A>\frac{\lambda v^2}{2s}\sin 2\beta +ks-\frac{m_W^2}{2\lambda s}\sin 2\beta \,.\label{eq:conditionemm2}
\eeq

For the discussion of spontaneous CP breaking we take for simplicity all the parameter of the Higgs potential to be real, we assume  
the condition (\ref{eq:conditionemm2}) for the conservation of the electromagnetism to hold, so that we can write the potential at the minimum as
\begin{eqnarray}\nonumber 
 V_\text{neutral}&=&\lambda^2|S|^2(|H_1^0|^2+|H_2^0|^2)+\lambda^2|H_1^0 H_2^0|^2+m_1^2|H_1^0|^2+m_2^2|H_2^0|^2+\mu_S^2|S|^2+k^2|S|^4+\\\label{eq:vtotmin}
&-&(A\lambda SH_1^0H_2^0+G\frac{k}{3} S^3-\lambda kS^2H_1^{0\dagger}H_2^{0\dagger}+h.c.)+\frac{g_1^2+g_2^2}{8}(|H_1^0|^2-|H_2^0|^2)^2\,.
\end{eqnarray} 
 
With a suitable $U(1)$ rotation of the scalar fields we can cancel the imaginary part of $H_1^0$, so that we can write the three Higgs fields at the minimum as 
 \begin{eqnarray}\label{complexVEVs}
H_1 = ( v_1, 0 ) \, ,\\ 
H_2 = (0, v_2 e^{\imath \phi}) \, , \nonumber \\
S=s e^{\imath \theta} \, , \nonumber
\end{eqnarray}
where $v_1$, $v_2$ and $s$ are taken real and positive. 

Replacing these fields in the scalar potential (\ref{eq:vtotmin}), and requiring a minimum in correspondence of the point $\phi=\theta=0$, we find that for $k\neq 0$ the conditions 
\begin{eqnarray}\label{eq:CP}
3Gks^{2}-5k\lambda v_{1}v_{2}s+2A\lambda v_{1}v_{2}&>&0 \,, \\\label{eq:CP1}
\lambda k\left( Gs(A-ks)-3A\lambda v_{1}v_{2}\right)&>&0\, 
\end{eqnarray} 
 must hold. In the particular case of $k=0$, these conditions get simplified to the unique requirement of $A>0$.

\section{Parameter space}\label{sec:constraintsvacuum}
In this section we give explicit bounds on the parameter space arising from the the various constraints discussed in Section \ref{sec:EWSB}.
Each point of the parameter space is fixed by coordinates  $\lambda,k, \tan\beta,s,A$ and $G$ that are in principle unconstrained, however we will focus on particular regions described and motivated in the following.

As motivated in the Introduction, we are particularly interested in the large $\lambda$ regime that helps in pushing the Higgs sector beyond LEP direct searches. Ref.~\cite{lsusy} shows that this can be done for $\lambda>1$ and a small $\tan\beta$. Remarkably this is precisely the regime where  contributions to the precision observables from the Higgs sector are better fitting LEP data. In the same work, all the contributions to the $T$ parameter were worked out in a particular model coming from a superpotential of the type in eq. (\ref{generalNMSSM}) that contains the mass term for the singlet $M S^{2}$. 
This superpotential in the limit $M\to 0$ gains a $U(1)$ symmetry under which the two doublets transforms with the same charge and the singlet with twice the charge of the doublet. This symmetry is customarily called Peccei-Quinn $U(1)_{PQ}$ symmetry and is also a symmetry of our superpotential in the limit $k\to 0$.  In the case of Ref.~\cite{lsusy}  the $U(1)_{PQ}$-symmetric limit $M\to 0$ resulted in too large contributions to the EW oblique parameter $T$ from the Higgsino sector.  In close analogy, we will explore the breaking of the $U(1)_{PQ}$ symmetry through the coupling $k$, with particular attention to the case of large $k$, where the symmetry is dramatically broken~\footnote{We remark that the large breaking of the $U(1)_{PQ}$ guarantees that there are no nearly massless pseudo-scalar to worry about in cosmology.}.

A representative point of the interesting region of the parameter space is then
\beq
\lambda=2,~~\tan\beta=1.5,\,~~ k=1.2\,, \label{point}
\eeq
which is singled out by the request of a moderate $\tan\beta$ and the maximal values of $k$ and $\lambda$, such that the theory will stay perturbative at least up to $O(10\,\, \text{TeV})$~\cite{RGE}.
Having fixed a region of the parameters  space in the surroundings of  the point (\ref{point}), the remaining parameter space is spanned by the VEV $s$ and the soft breaking parameters $A$ and $G$. Since the soft breaking terms are naturally expected to be of the same order of magnitude, we decide to make the simplifying assumption \beq G=A \label{assumptionGA}\;. \eeq
In fact, a large hierarchy between $A$ and $G$ is not possible since the requirement of absolute minimum given in (\ref{eq:minabsformula})  does not allow $G\gg A$ and the constraint from CP invariance of (\ref{eq:CP1}) does not allow $G\ll A$. Variations on the assumption (\ref{assumptionGA}) are conceivable but we checked that they do not lead to any important change in our analysis. 

Finally  we decide to  parametrize the two dimensional parameter space $s$, $A$ in terms of the
parameters  
$\mu$ and $m_{H^+}$
 that are introduced through the relations
\beq
\label{eq:newparameters}
\mu=\lambda s \,,~~~
 m_{H^+}^2=m_W^2-\lambda^2 v^2+\frac{2\mu(A-\frac{k}{\lambda}\mu)}{\sin 2\beta}\, .
 \eeq
For later convenience we also introduce the combination of parameters
\beq \tilde{m}_A^2=m_{H^+}^2-m_W^2+\lambda^2 v^2 \,.\label{mAtilde}\eeq 
We will see in Section \ref{Sec:masses} that $\mu$ and $m_{H^+}$ are parameters with direct relevance for the spectrum of the model and indeed they also allow a straightforward representation of the bounds we have to impose. This is evident in Figure \ref{Fig:constraintvacuumk} where  we show the physical parameter space for the particular choice of parameters in (\ref{point}) and (\ref{assumptionGA}). The lower-bound  on $m_{H^+}$ is given by the condition of absence of spontaneous CP violation given in (\ref{eq:CP1}), that in the new variables, after imposing the condition $G=A$, reads simply 
\beq
 m_{H^+}^2 > 2 \lambda^2 v^2 +m_W^2  \,.\label{mAtildemin}
\eeq
The other  bound mainly shaping the triangle is given by the condition of absolute minimum of (\ref{eq:minabsformula}) that in the new variables has the approximate simple form

\beq
m_{H^+}^2 < m_W^2 + 4 \frac{\mu^2 }{\sin^2 2\beta}+\frac{m_Z^2}{ \tan ^2 2\beta}+ O\left(\frac{1}{\lambda}\right) \,\label{absoluteminimum}.
   \eeq

The union of the requirements (\ref{mAtildemin}) and (\ref{absoluteminimum}) gives a minimal allowed value for $\mu$ on which we comment later in Section \ref{solution}.

\begin{figure}
\center
\includegraphics[width=.4\textwidth]{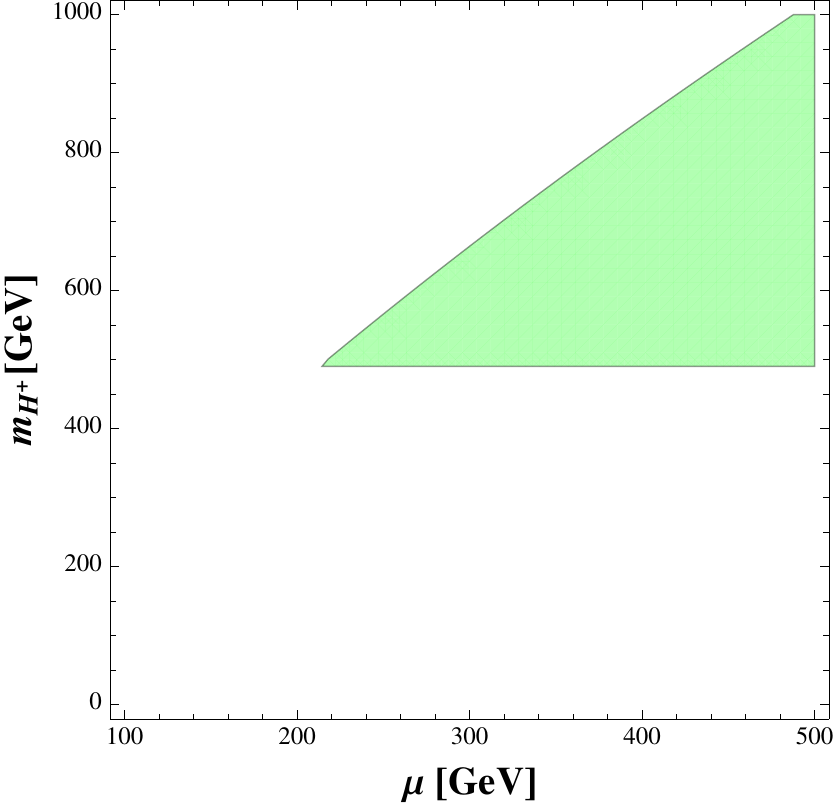}

\caption{%
Physical region of the $\mu,m_{H^+}$ parameter space for parameters fixed according to eqs. (\ref{point}) and (\ref{assumptionGA}).}\label{Fig:constraintvacuumk}
\end{figure}

\section{Spectrum }\label{Sec:masses}

The Higgs sector of the theory contains seven bosonic degrees of freedom. In particular, in the mass eigenstates basis we expect one charged Higgs, three neutral scalar fields and two neutral pseudo-scalars, that do not mix with the scalars, since we imposed CP invariance. In order to investigate the spectrum of these six particles, it is convenient to express the scalar potential with the Higgs fields expressed by

\begin{equation}
S=s+{S_1+iS_2 \over \sqrt{2} } \,,\hspace{0.5cm}
H_1=e^{-\frac{i}{\sqrt{2}}\frac{\bar{\sigma}\bar{\pi}_1}{v_1}}
\left(\begin{array}{cc}
v_1+{h_1 \over \sqrt{2}}\\
0
\end{array}\right)\,,\hspace{0.5cm}
H_2=e^{-\frac{i}{\sqrt{2}}\frac{\bar{\sigma}\bar{\pi}_2}{v_2}}
\left(\begin{array}{cc}
0\\
v_2+{ h_2 \over \sqrt{2}}
\end{array}\right)\,.
\end{equation}

After some algebra, we obtain the mass matrices given in the following, where we have used $s_\beta\equiv\sin\beta$, $c_\beta\equiv\cos\beta$, $t_\beta\equiv\tan\beta$, $s_{2\beta}\equiv \sin{2\beta}$ and $\mu$ and $\tilde{m}_A$ as defined in (\ref{eq:newparameters}) and (\ref{mAtilde}) respectively. 
For  the scalar Higgs bosons, in the basis $(h_1,h_2,S_1)$ we get 

\beq
\label{Mscalar}
M_{S}^2=\left( \begin{array}{ccc}
 c_{\beta }^2 m_Z^2+s_{\beta }^2 \tilde{m}_A^2 & ~~c_{\beta } s_{\beta } \left(2 v^2
   \lambda ^2-m_Z^2-\tilde{m}_A^2\right) & \mu v ( 2  \lambda c_{\beta }+ s_{\beta } k )
   -s_{\beta }^2  c_{ \beta } v \lambda \frac{ \tilde{m}_A^2}{\mu } \\
 . &    m_Z^2 s_{\beta }^2+ c_{\beta }^2 \tilde{m}_A^2 & -\frac{v \lambda  c_{\beta }^2 s_{\beta }
   \tilde{m}_A^2}{\mu }+v \mu( 2 \lambda s_{\beta }+k c_{\beta }) \\
 . & . & 
4 \frac{k^2}{\lambda^2} \mu^2 -G \frac{k}{\lambda} \mu + \lambda ^2 v^2 s_{2 \beta }^2 \frac{ \tilde{m}_A^2 }{4 \mu ^2}+ \frac{\lambda k}{2} v^2 s_{2 \beta }
\end{array}
\right)\,.
\eeq

For the pseudo-scalar  we use the basis $(\pi^{\left(3\right)},S_2)$, where $\pi^{\left(3\right)}$ is defined by $\pi^{\left(3\right)}=\sin\beta\pi_1^{\left(3\right)}-\cos\beta\pi_2^{\left(3\right)}$ and we get

\begin{equation}
M_{PS}^2=
\left(
\begin{array}{cc}
 \tilde{m}_A^2 & -\frac{v \lambda  c_{\beta } s_{\beta } \tilde{m}_A^2}{\mu }-3 k v \mu  \\
 .  & ~~~ \lambda ^2 v^2 c_{\beta
   }^2 s_{\beta }^2\frac{ \tilde{m}_A^2 }{\mu ^2}-3 k \lambda  c_{\beta } s_{\beta } v^2+ 3 G \mu \frac{
 k }{\lambda }
\end{array}
\right)\,.
\end{equation}

\begin{figure}
\center
\includegraphics[width=.32\textwidth]{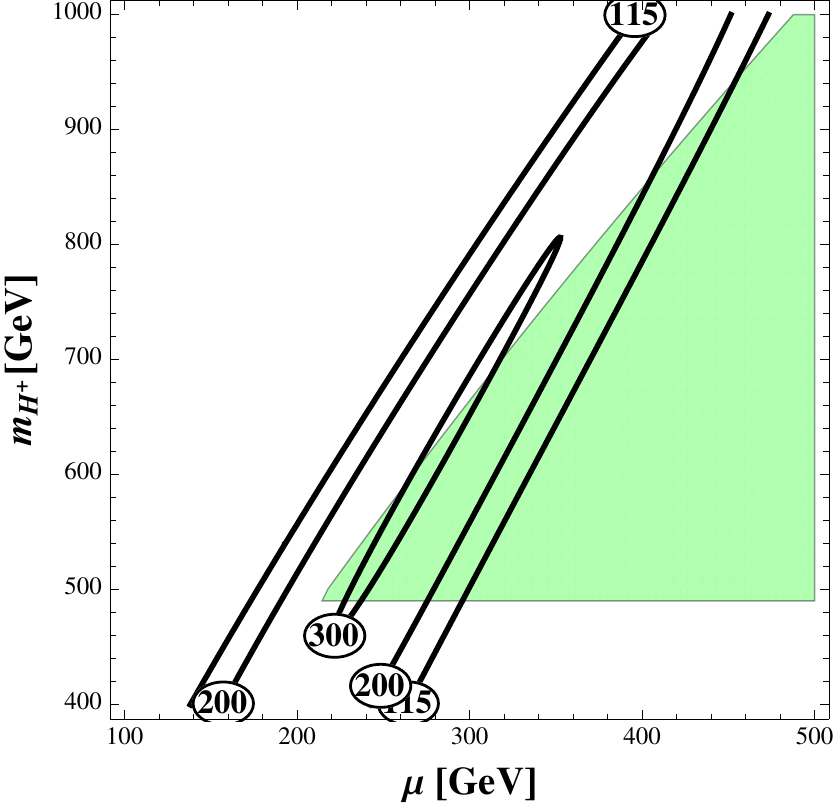}
\includegraphics[width=.32\textwidth]{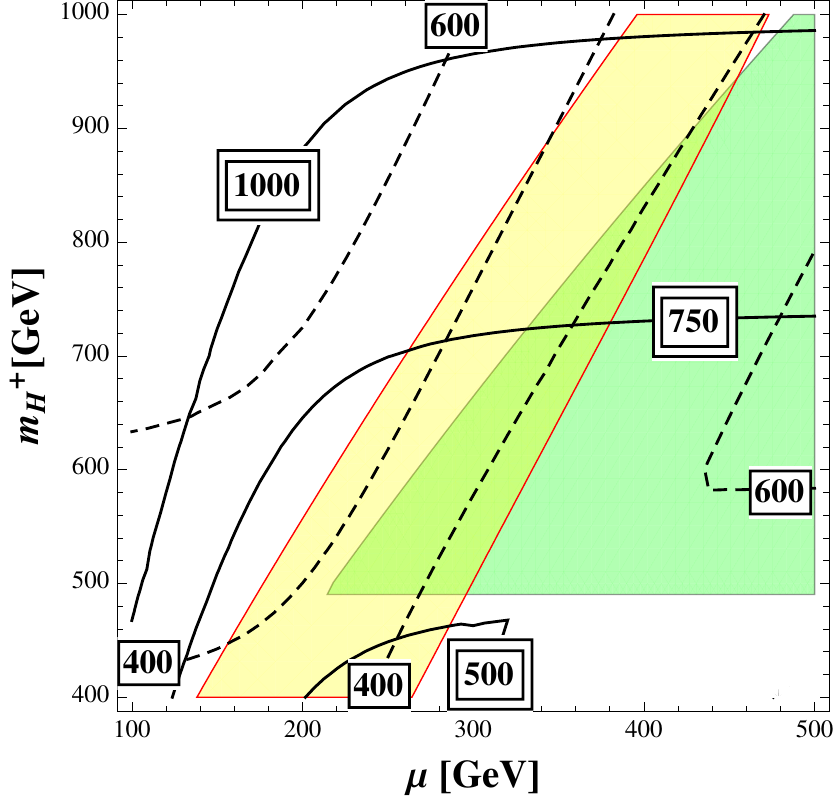}
\includegraphics[width=.32\textwidth]{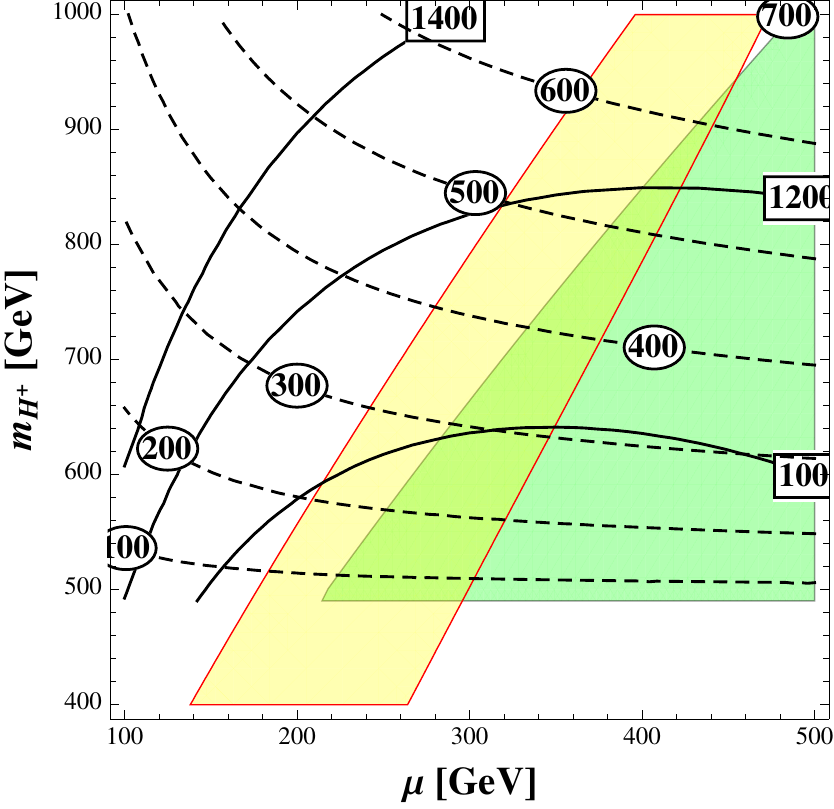}
\caption{Masses (in GeV) of the neutral scalars in the plane $\mu,m_{H^+}$ for  parameters fixed as in (\ref{point}): (from the left to the right) the mass of the lightest CP-even ($m_{s_1}$), the masses of the two heavy CP-even ($m_{s_{2,3}}$), and the masses of the two CP-odd scalars ($m_{a_{1,2}}$).
 In all the panels the overlaid yellow area corresponds to $m_{s_1}>114$ GeV and the overlaid green area corresponds to the parameter space where the $SU(2)$ breaking vacuum is stable and CP is conserved.
}\label{higgsmasses}
\end{figure}

\begin{figure}
\center

\includegraphics[width=.35\textwidth]{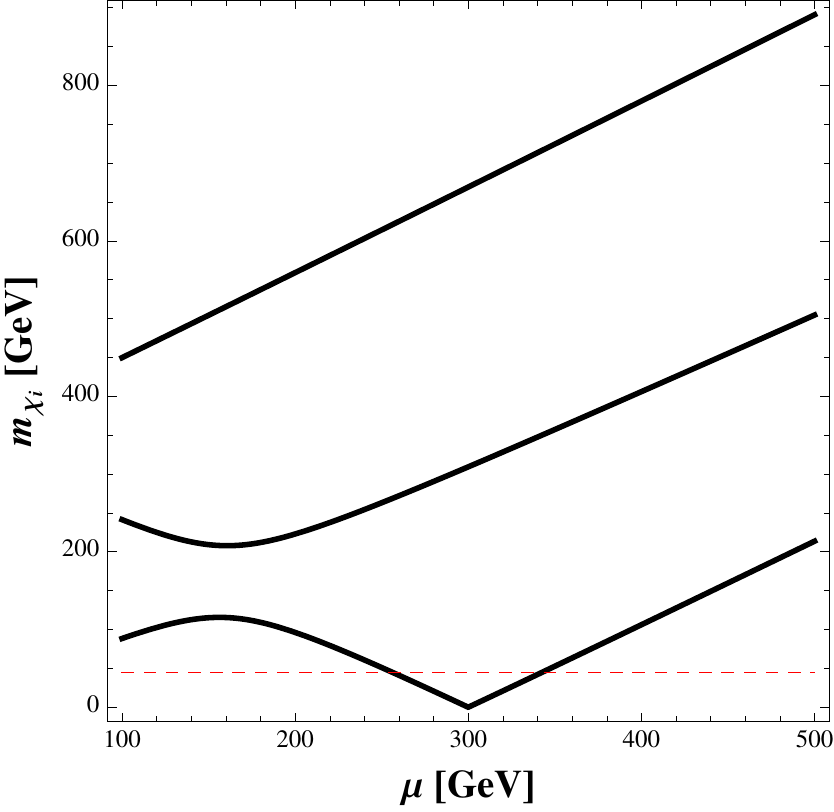}
\caption{%
Higgsino masses (in GeV) as function of the chargino mass $\mu$ for fixed  parameters as in (\ref{point}). The dashed line corresponds to $m_Z/2$ which is taken as limit from LEP (eq. (\ref{mZmezzi})).
}\label{higgsinomasses}
\end{figure}

The charged Higgs boson  mass is just equal to the parameter $m_{H^+}$
\beq 
m_{H^\pm}^2=m_{H^+}^2=m_W^2-\lambda^2 v^2+\frac{2\mu(A-\frac{k}{\lambda}\mu)}{\sin 2\beta}\,.
\eeq

For the neutralino and chargino sector we assume that the gaugino mass parameters $M_1,M_2$ are large. In this case the only light chargino is a pure Higgsino and has exactly the mass 
\beq
 m_{\chi^+}=\mu \, .
 \eeq 
For the neutralinos we choose the basis defined by
\beq
N_1=\frac{1}{\sqrt 2}\left(\tilde H_1-\tilde H_2\right)\,,\hspace{0.5cm}
N_2=\frac{1}{\sqrt 2}\left(\tilde H_1+\tilde H_2\right)\,,\hspace{0.5cm}
N_3=\tilde S\,, \label{N123}
\eeq
and the mass matrix reads

\begin{equation}
M_{N}=\left(
\begin{array}{ccc}

\mu &\,\, 0 &\,\, \frac{v}{\sqrt{2}}\lambda(c_\beta-s_\beta) \\
0 &\,\, -\mu  &\,\, -\frac{v}{\sqrt{2}} \lambda  (c_\beta +s_\beta ) \\
\frac{v}{\sqrt{2}} \lambda (c_\beta -s_\beta) &\,\, -\frac{v}{\sqrt{2}} \lambda  (c_\beta +s_\beta ) &\,\, -2\frac{k}{\lambda}\mu

\end{array} 
\right) \, .  \label{eq:higgsinomasses}
\end{equation}
From this mass matrix one can see that there is a massless Higgsino state if
\beq 
\mu^2=\frac{\lambda}{k}\frac{v^2 \lambda ^2 \sin 2 \beta }{2 }\, ,\label{zeroLSPmass}
\eeq
and that 
\beq
\sum_{i=1,2,3}m^2_{\chi_i}=2 \left[\mu ^2 \left(\frac{2 k^2}{\lambda ^2}+1\right)+v^2 \lambda ^2\right]\, \label{tracehiggsinos}.
\eeq

\bigskip

All the masses given in this section are plotted in  Figures \ref{higgsmasses} and \ref{higgsinomasses} in the plane $\mu,m_{H^+}$ for the choice of parameters in (\ref{point}).  Few comments are in order.
Firstly, we note that, due to the large value of $\lambda$, the spectrum consists of relatively heavy Higgs bosons with a lightest CP-even mass of roughly (200-300) GeV  and, due to the large value of $k$, there is no light state in the  CP-odd sector, in fact the lightest CP-odd has mass of few hundreds GeV. This shows how this realization of the NMSSM is rather at odds with the widely studied case of small $\lambda$.

 Another interesting point is the fact that the requirement of a mass for the lightest Higgs in accordance with the LEP bound tends to clash with the requirement of absolute minimum of (\ref{eq:minabsformula}). This implies that the model cannot have an arbitrarily large value of $\mu$ because this would yield a negative mass squared for the lightest Higgs boson. The existence of a maximal allowed $\mu$ can be understood taking the the CP-even mass matrix (\ref{Mscalar}) in the large $\mu$ and large $\lambda$ limit and observing that all the diagonal submatrices have negative eigenvalues independently of the other parameters. This fact is very welcome in view of the need to generate a $\mu$ term of the order of $m_Z$ and we will study the consequences of this fact in Section \ref{solution}.
 
\bigskip
For later convenience, we conclude this section fixing some notation. 
We call the scalar and pseudo-scalar mass eigenstates (from the lightest to the heaviest)  $s_1,s_2,s_3$ and $a_1,a_2$ respectively and we will denote the Higgsino mass eigenstates as $\chi_1,\chi_2,\chi_3$ (still from the lightest to the heaviest).  

We introduce the combination of doublet fields that take VEV, $h=h_1 \cos\beta+h_2 \sin\beta$, and its orthogonal one $H=h_1 \sin\beta-h_2 \cos\beta$. The rotation $U$ from the mass basis to the basis $(H,h,S_1)$ is defined by
 \beq \left(\begin{array}{c}s_1 \\s_2 \\s_3\end{array}\right) = U \left(\begin{array}{c}H \\h \\S_1\end{array}\right)\,. \label{Umatrix}\eeq

Analogously for the pseudo-scalars the rotation matrix $P$ is defined such that

\beq\left(\begin{array}{c}a_1 \\a_2 \end{array}\right) = P \left(\begin{array}{c} \pi^{\left(3\right)}\\S_2\end{array}\right)\,. \label{Pmatrix}\eeq

For the Higgsinos we call $V$ the rotation matrix from the basis $(N_1,N_2,N_3)$ to the mass eigenstates $\chi_m$ such that

\beq 
\chi_m = V_{im} N_i\,.\label{Vmatrix}
\eeq

Finally, we introduce the rotation matrix $R^{(x)}$ for a generic angle $x$

\beq
R^{(x)}=\left(  
\begin{array}{lll}
\cos x & -\sin x & 0\\
\sin x & \cos x  &0  \\
0      &  0      & 1
  \end{array}\right)\,,
  \eeq
 such that the states $N_n$ can be written as 

\begin{equation}
N_n=R^{(\pi/4)}_{n3} \tilde S+\sum_{i=1,2}R^{(\pi/4)}_{ni} \tilde{H}_i\,.\hspace{0.9 cm}
\end{equation}

\section{Generation of the $\mu$ term}\label{solution}

One of the original motivations of the NMSSM is the possibility to dynamically generate an effective $\mu$ term in the Lagrangian at the Fermi scale. This possibility is very interesting because in the MSSM such a term has to be fixed by hand  to a value close to $m_Z$ or the model would be badly unrealistic. Indeed such a term gives mass to the chargino and is crucial to have a correct EWSB.
On the other hand in the NMSSM an effective $\mu$-term is dynamically generated by EWSB through the VEV of the Higgs singlet resulting in

\begin{equation}\label{mu_eff}
\mu =  \lambda s\,.
\end{equation}
This dynamical generation of $\mu$ in connection to EWSB is particularly welcome also in view of fine-tuning problems. In facts the minimization equation (\ref{eq:sin2beta}) can be rewritten as 
\beq
\mu^2+\frac{1}{2}m_Z^2-\frac{m_1^2-m_2^2\tan^2\beta}{\tan^2\beta-1}=0\, ,
\eeq
which shows that in a natural theory $\mu\sim m_Z$.
Generating the $\mu$ term via EWSB relates the size of $\mu$ to that of the soft terms and therefore solves at the same time the issue of the presence of the term and the issue of its size.

In the traditional approach to the $\mu$-problem within the NMSSM one requires that the coupling $\lambda$  has to stay perturbative up the GUT scale, which means $\lambda\lesssim 0.7$ at the EW scale~\cite{BarbieriPQ,lambdamax}. As such, to obtain a value of $\mu$ compatible with current searches and with the requirement of EWSB, one has to go in the regime  $s\gtrsim v$. 
In particular, to have $ \lambda s\sim m_Z $ when $\lambda\to 0$ one expects $s\gg v$, so that the minimization equation (\ref{eq:s}) has an approximate stable solution~\cite{largeS} 
\beq s\simeq\frac{1}{4k}(G+\sqrt{G^2-8\mu_S^2})\, . \label{ssmalllambda}\eeq

This approximate solution generates larger values of $\mu$ as one takes larger $\lambda$ and therefore seems to signal that the theory will be fine-tuned when a too large $\lambda$ is considered. However this is not worrisome because the approximate solution (\ref{ssmalllambda}) cannot be valid for any large value of $\lambda$ as the approximation $s\gg v$ breaks down.

Indeed one can consider the minimization equation (\ref{eq:s}) in the large $\lambda$ regime for generic $s$ and find  an approximate solution 
\beq 
s\simeq \frac{1}{2} \frac{A \sin 2 \beta }{k \sin 2 \beta+\lambda } \, ,
\eeq

\noindent that in the limit $\lambda \to \infty$ gives 

$$\mu_\infty \equiv \lambda \lim_{\lambda\to\infty} s  =\frac{1}{2}A\sin2\beta\,.$$

The existence of such finite limit is not surprising as we have already noticed in Section \ref{Sec:masses} that $\mu$ cannot be taken arbitrarily large because of the incompatibility of the requirement of absolute minimum (\ref{eq:minabs}) and the requirement of positive masses squared of the CP-even scalars. Indeed one can be more quantitative and show that the presence of such maximal value of $\mu$ is a generic feature of the model due to the largeness of $\lambda$ and that the value of the maximal $\mu$ is linked to  the  maximal mass of lightest Higgs boson.

The actual maximal allowed value of $\mu$ can be estimated observing that in the region close to the boundary defined by the condition  of absolute minimum given in (\ref{eq:minabsformula}) and for $k < \lambda $ the condition $m^2_{s_1}>0$ can be approximated by
\beq \label{mh1positive}
\sqrt{ m_{H^+}^2-m_W^2+\lambda^2v^2} >\mu \frac{3- k/\lambda }{\sin2\beta}+\frac{3}{2} v \lambda (k/\lambda -1)\,. \eeq
This is a lower-bound on $m_{H^+}$ that, for values of $\mu$ large enough, gets incompatible with the condition (\ref{eq:minabsformula}) for the minimum of the potential. The value of $\mu$ where the two conditions get incompatible can be estimated  taking  (\ref{eq:minabsformula}) in the approximate form given by the upper-bound on $m_{H^+}$ of (\ref{absoluteminimum}), yielding a relatively simple condition in terms of $\rho\equiv k/\lambda$ 

\beq
\mu<\frac{v \lambda \sin2\beta}{2}  \frac{3 (\rho -4) \rho + \sqrt{8(\rho -1) (5 \rho -7)}+9
  }{(\rho -5)(\rho -1)} \simeq \frac{3}{2}v \lambda \sin2\beta+O\left(\frac{k}{\lambda}\right)\,,\label{mumax}
\eeq

\noindent from which we observe that for generic $\tan\beta$ and generic $k<\lambda$ the model has a maximal allowed value of $\mu$ of the order of $\lambda v$.

Furthermore, from the requirements of absolute minimum and of the absence of spontaneous CP breaking  given in (\ref{eq:minabsformula}) and (\ref{eq:CP1}), we can see that there is a minimum allowed value for $\mu$ that can be estimated taking the approximate condition in (\ref{absoluteminimum}) in place of (\ref{eq:minabsformula}) giving

\beq
\mu^2> \frac{\lambda^2v^2}{2}\sin ^22\beta- \frac{m_Z^2}{4}\cos ^22\beta\,.\label{mumin}
\eeq

Taking together this result and the condition (\ref{mumax}) for the absolute minimum, we find that the model is consistent only for values of $\mu$ within an interval that, neglecting the term $m_Z\cos2\beta/2$, reads
\beq
\frac{\lambda v}{\sqrt{2}}\sin2\beta \lesssim \mu \lesssim \frac{3\lambda v}{2}\sin2\beta\, .\label{murange}
\eeq

In the large $\lambda$ regime this relation automatically ensures that  the chargino with mass $\mu$ is  above the LEP bound (see later eq. (\ref{eq:chargino})) and, at the same time, gives an upper-bound for $\mu$ linked to the mass of the lightest Higgs boson. In this sense eq. (\ref{murange}) shows that, specializing the generic $\lambda$SUSY superpotential in (\ref{generalNMSSM}) to a superpotential without dimensionful parameter gives a model where the $\mu$ term is phenomenologically acceptable and is necessarily close to $m_Z$, thus solving the so-called ``$\mu$-problem''. It is important to stress that eq. (\ref{murange}) holds only for $k<\lambda$ and that this is generically  the case in the large $\lambda$ regime of $\lambda$SUSY.

\vspace{0.1cm}
Furthermore, we can use (\ref{murange}) to establish a relation between the mass of the chargino and the mass of the lightest Higgs
\beq
m_{\chi^+}\sim m_{s_1}\,,
\eeq
which is of course of phenomenological interest for collider searches.
\bigskip

\section{Naturalness}\label{FT}

As mentioned in the Introduction, one of the motivations to consider this model is the attempt to address the Naturalness problem of the MSSM in the Higgs sector. Therefore, of particular importance is the study of the level of fine-tuning required, in order to satisfy the various constraints on the parameters of the model. 

But what do we mean with {\it Naturalness constraints}? Speaking in broad generality, the Fermi scale is a function of the several dimensionful parameters $a_j$ of the Lagrangian: $v^2=v^2(a_j)$. We require, for a small variation of the parameters, also the variation of $v^2$ not to be large as well. In particular, for a fixed maximum amount of fine-tuning $\frac{1}{\Delta}$, we impose~\cite{Barbieri:1987fn}

\beq\label{eq:finetuning}
\Delta_{a_j}\equiv\left|\frac{a_j^2}{v^2}\frac{d\,v^2(a_i)}{d\,a_j^2}\right|<\Delta\,.
\eeq

For the particular theory analysed in this paper, the set of dimensional parameters is given by $a_j=(\mu_S,m_1,m_2,G,A)$. To compute the several logarithmic derivatives in (\ref{eq:finetuning}), we have to consider again the conditions of minimization of the potential (\ref{eq:sin2beta}) and (\ref{eq:s}), which show the dependence of the Fermi scale (\ref{eq:v}) on the several dimensional parameters through $\tan\beta$ and $s$, respectively. Therefore, we take (\ref{eq:sin2beta}) for the angle $\beta$ and we replace then the value of $\beta$ in eqs. (\ref{eq:v}) and (\ref{eq:s}), to eliminate their dependence on $\tan\beta$. Subsequently, we eliminate the dependence on the VEV $v$ in the equation for $s$ (\ref{eq:s}), using (\ref{eq:v}). From the such obtained eq. (\ref{eq:s}), it is then possible to compute the several derivatives of $s$ with respect to the dimensional parameters of the Lagrangian $\partial s/\partial a_j^2$. Finally, for the level of fine-tuning $\Delta_{a_j}$ we obtain

\beq\label{eq:derivatives}
\Delta_{a_j}=\left|\frac{a_j^2}{v^2}\frac{dv^2(a_i,s)}{da_j^2}\right|=\left|\frac{a_j^2}{v^2}\left(\frac{\partial v^2(a_i,s)}{\partial a_j^2}+\frac{\partial v^2(a_i,s)}{\partial s}\frac{\partial s}{\partial a_j^2}\right)\right|\,.
\eeq

From our analysis, it turns out that, considering just the dimensional parameters $m_1,m_2$, $G$ and $\mu_S$ the level of fine-tuning required is in general very small in all the parameter space allowed by the conditions on the scalar potential analysed in Section \ref{sec:EWSB} and by the LEP bound on the mass of the lightest Higgs given in (\ref{mhLEP}). 

The Fermi scale is more sensitive to variations of the soft parameter $A$, but also the fine-tuning required on this parameter is still under control. For example, for the reference point in (\ref{point}) it is always smaller than $\sim 25$ for a lightest Higgs in agreement with the LEP bound. Our result for the fine-tuning on $A$ is shown in Figure \ref{Naturalness}, as a function of the mass of the lightest Higgs, for the representative point of the parameter space presented in (\ref{point}), once that the mass of the chargino is fixed to two reference values: $\mu=230\,\text{GeV}$ on the left panel and $\mu=400\,\text{GeV}$ on the right panel. Indeed, fixing these two values for $\mu$, we do not loose of generality, since in the region allowed by LEP
the fine-tuning depends only mildly on the value of $\mu$. Interestingly enough, we notice that, increasing the mass of the lightest Higgs, the fine-tuning decreases considerably. Consequently, for both values of $\mu$, Naturalness arguments favor the regions of parameter space with a heavy Higgs boson, even quite heavier than the LEP bound mass of $114$ GeV.

\begin{figure}
\center
\includegraphics[width=.4\textwidth]{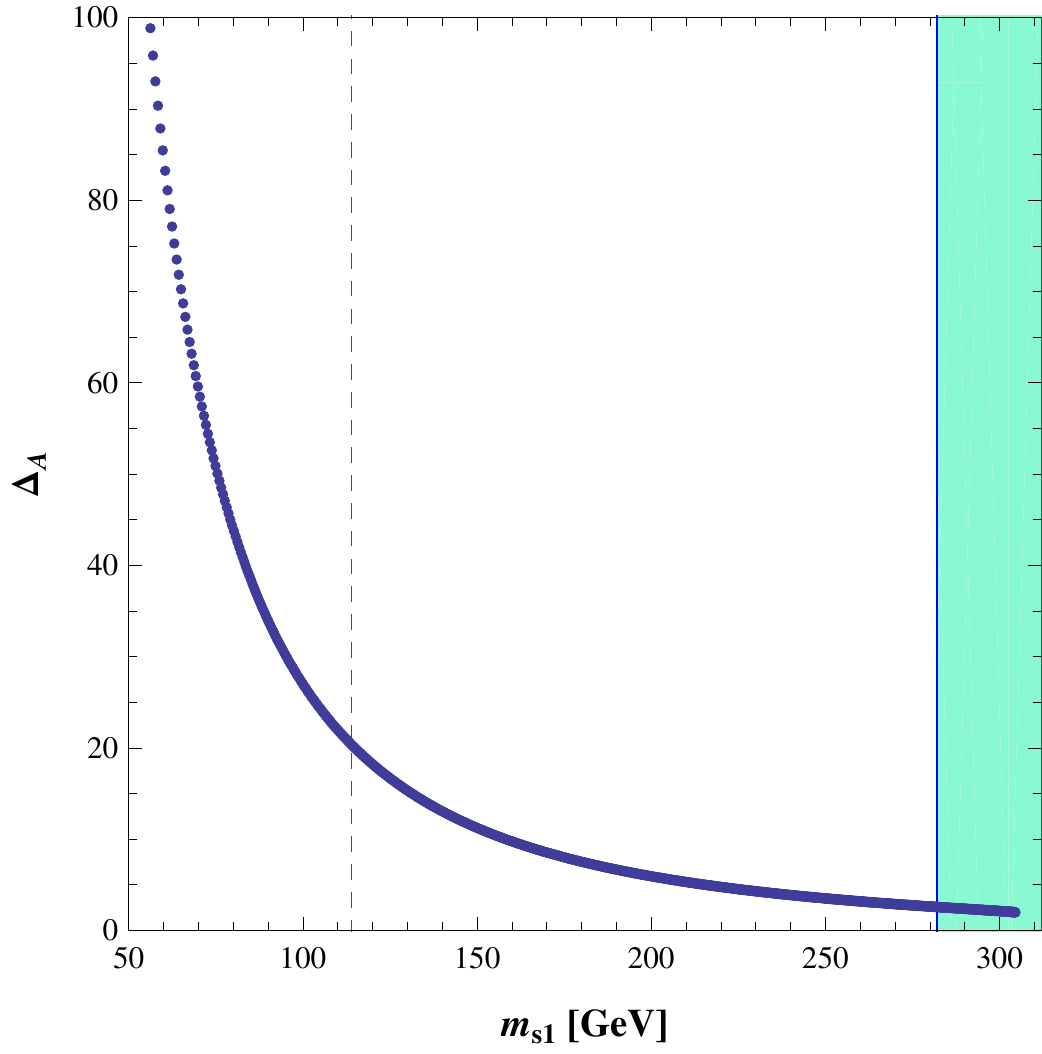}
\includegraphics[width=.4\textwidth]{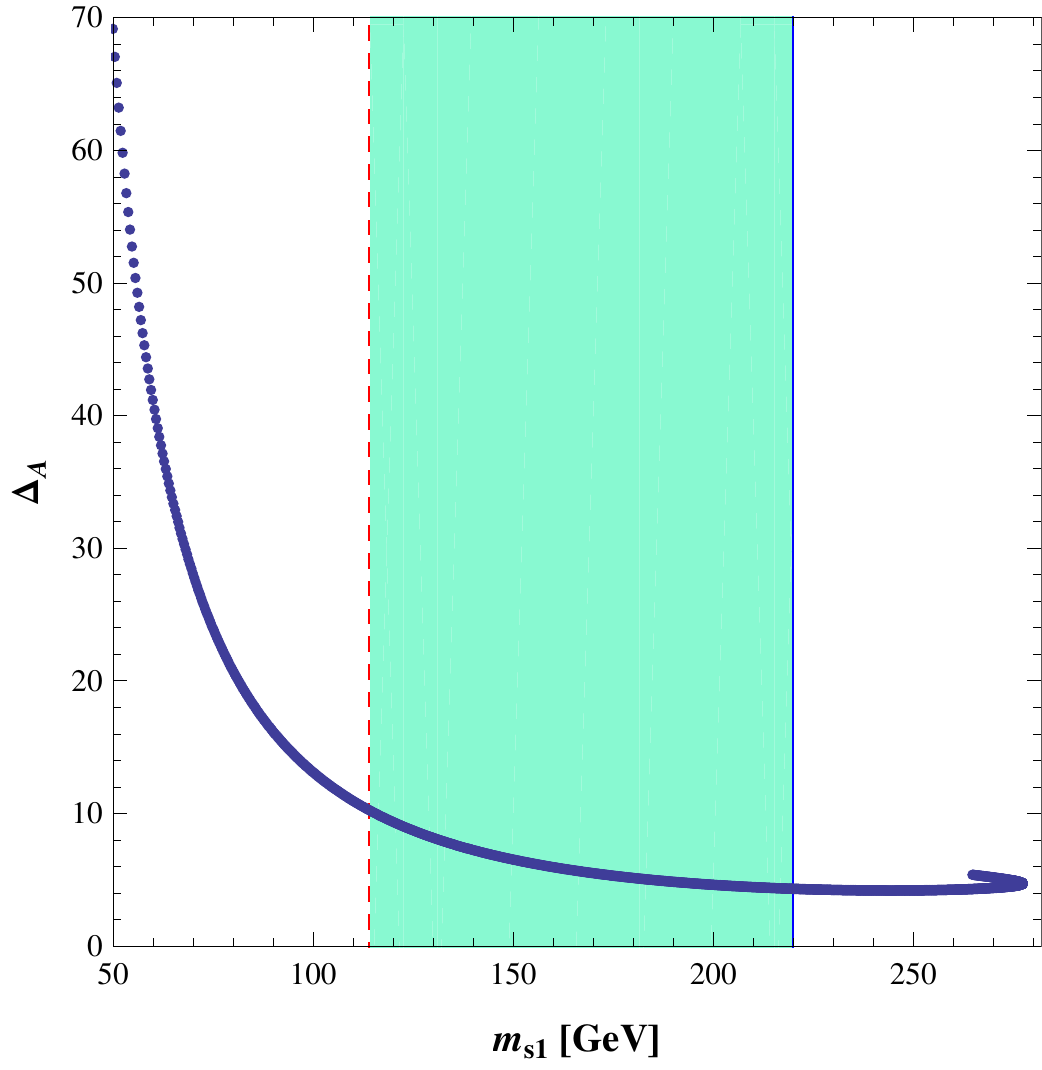}\\

\caption{Logarithmic derivative of the Fermi scale with respect to the dimensional parameter $A$ as a function of the lightest Higgs boson mass $m_{s_1}$, for the representative point in (\ref{point}) and mass of the chargino fixed to $230\,\text{ GeV}$ (on the left), and $400\,\text{ GeV}$ (on the right). The blue area is the region allowed by all the constraints on the scalar potential (see Section \ref{sec:EWSB}) and by the LEP bound on the lightest Higgs boson mass. 
}\label{Naturalness}
\end{figure}

In general we can argue that, with large values of the coupling constant $\lambda$, one can address the fine-tuning problem present in the MSSM Higgs sector and partially present also in the NMSSM with small coupling constants~\cite{FT}.

We could have already guessed this final result, just looking at the minimization conditions (\ref{eq:sin2beta})-(\ref{eq:s}). In fact, considering the limit $\lambda\rightarrow\infty$, the following relations hold

\begin{eqnarray}\label{eq:coslambdalarge}
\cos\beta&\rightarrow& {\rm Constant}\,,\\
\frac{\partial v^2}{\partial A^2}&\rightarrow& \frac{\mu}{2\lambda^2 A}\,,\\
\frac{\partial v^2}{\partial s}&\rightarrow& \frac{A}{\lambda}\,,	\\\label{eq:dsdAlambdalarge}
\frac{\partial s}{\partial A^2}&\rightarrow &\frac{1}{2\lambda A}\,.
\end{eqnarray}

Inserting these limits in the equation for the fine-tuning $\Delta_A$ (\ref{eq:derivatives}), it is obvious that the Naturalness conditions are easily satisfied in the limit of large quartic coupling $\lambda$, independently on the value of the coupling $k$\footnote{A more careful analysis of the large $\lambda$ regime shows that the Naturalness conditions are even easier to satisfy in the particular case of $k$ large, even if the results for the fine-tuning $\Delta_{a_j}$ do not dramatically change, depending on the value of $k$.}.

It is also interesting to understand if the theory continues to be natural, when we allow only the coupling $k$ to be large. Considering the limit $k\rightarrow \infty$, the relations (\ref{eq:coslambdalarge})-(\ref{eq:dsdAlambdalarge}) change in

\begin{eqnarray}
\cos\beta&\rightarrow& {\rm Constant}\,,\\
\frac{\partial v^2}{\partial A}&\rightarrow& {\rm Constant}\,,\\
\frac{\partial v^2}{\partial s}&\rightarrow& k\mu\,,	\\
\frac{\partial s}{\partial A^2}&\rightarrow &\frac{1}{2k A}\,.
\end{eqnarray}

Inserting these limits in (\ref{eq:derivatives}), it follows that the fine-tuning generally does not decrease, increasing the value of the coupling $k$. Consequently, the only way to make the theory natural is to have a large coupling $\lambda$ regime.

\subsection{Naturalness bounds on sparticle masses}\label{sec:boundmasses}

Using Naturalness arguments, we want to set upper-bounds for the masses of sparticles. These bounds are particularly relevant to understand the expected size of the contributions of the sparticles to low-energy processes like the well studied  flavour transitions and LEP precision data.  Moreover with these bounds at hand one can estimate the timescale for the observation of such states at the LHC.

For the case of the stops-sbottoms, in the hypothesis of diagonal squark mass matrices, the basic observation is that the soft mass $ m_{\tilde{Q}}$ which enters in the formula for the masses of the physical stop and sbottom 

\begin{eqnarray}\label{eq:mst}
m_{\tilde{t}_L}&=&\sqrt{m_{\tilde{Q}}^2+m_t^2+m_Z^2\cos2\beta\left(\frac{1}{2}-\frac{2}{3}\sin^2\theta_W \right)}\,,\\\label{eq:msb} m_{\tilde{b}_L}&=&\sqrt{m_{\tilde{Q}}^2+m_b^2+m_Z^2\cos2\beta\left(-\frac{1}{2}+\frac{1}{3}\sin^2\theta_W \right)}\,
\end{eqnarray}

\noindent affects the soft Higgs mass $m_2$ through the one loop renormalization group equation (RGE)~\cite{Martin:1993zk}

\begin{equation}
\frac{d m_{2}^{2}}{dt} 
    =  \frac{3}{8\pi^{2}}
         \lambda_{t}^{2}(m_{\tilde{Q}}^{2}+m_{\tilde{t}_{R}}^{2}) + \cdots\,,
\end{equation}    

\noindent where the ellipsis stands for terms not dependent, in first approximation, on the soft squark masses. One can integrate these equations up to the messenger scale $\Lambda_{\rm mess}$, obtaining, at the leading log,

\begin{equation} \label{runningduetosquarks}
  \delta m_{2}^{2} 
    \simeq -\frac{3}{8\pi^{2}} \lambda_{t}^{2} 
      (m_{\tilde{Q}}^{2}+m_{\tilde{t}_{R}}^{2}) 
      \ln\frac{\Lambda_{\rm mess}}{1~\text{TeV}}\,.
\end{equation} 

To give an estimation of the bound on the masses of stops and sbottoms, one can simply assume the equality of the soft masses $m_{\tilde{Q}}=m_{\tilde{t}_{R}}$. Imposing, then, the Naturalness condition (\ref{eq:finetuning}), with respect to the parameter $m_{\tilde{Q}}$, and using the approximate expression

\begin{equation}
\left|\frac{m_{\tilde{Q}}^2}{v^2}\frac{d\,v^2}{d\,m_{\tilde{Q}}^2}\right|\sim\left|\frac{m_{\tilde{Q}}^2}{v^2}\,\frac{d\,v^2}{d\,m_2^2}\,\frac{d\,m_2^2}{d\,m_{\tilde{Q}}^2}\right|\,,
\end{equation}
one can find analytically the bound

\begin{equation}\label{eq:boundmQ}
m_{\tilde{Q}}\leq v\sqrt{\frac{4\pi^2}{3}}\sin\beta \frac{\sqrt{\Delta}}{ \sqrt{\frac{d v^2}{d m_2^2}}\,\ln\frac{\Lambda_{\rm mess}}{1~\text{TeV}}}\,.
\end{equation}

The result for the bound on the mass of the stop for a messenger scale of $100~\textrm{TeV}$ and an allowed fine-tuning of $10\%$ is shown in the left panel of Figure \ref{boundstop}  for the point of the parameter space  presented in (\ref{point}) and $\mu$ fixed to $400\,\text{GeV}$\footnote{There are no relevant changes in the curve for the bound on the stop mass, for $\mu$ fixed to be equal to the second reference value ($230\,\textrm{GeV}$) discussed in the previous subsection.}. From the Figure it is evident that the upper-bound on the stop mass $m_{\tilde t_L}$ increases when one increases the value of the mass of the lightest Higgs, and, for a Higgs mass satisfying the LEP bound, the mass of the stop can be $\sim 550\text{GeV}$ for an allowed fine-tuning of $10\%$.

\begin{figure}
\center
\includegraphics[width=.4\textwidth]{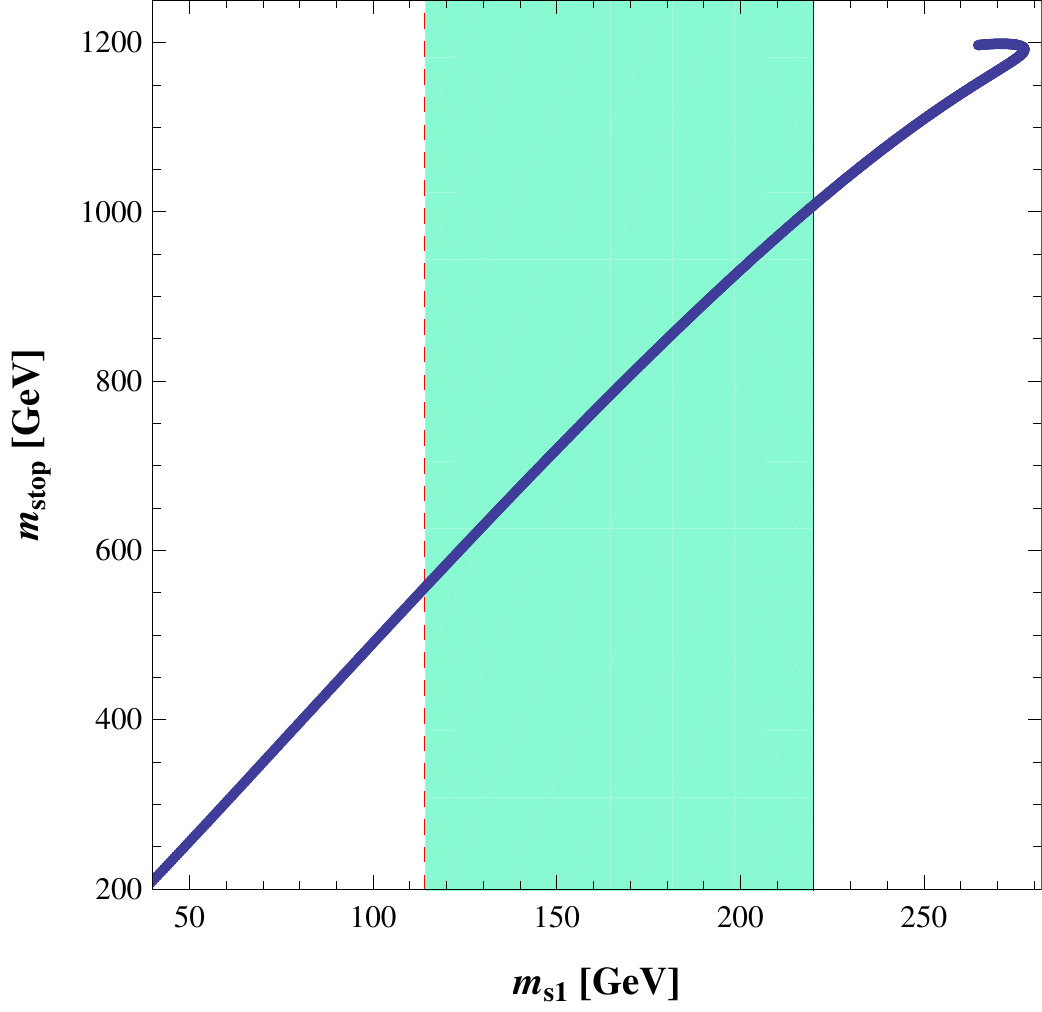}
\includegraphics[width=.4\textwidth]{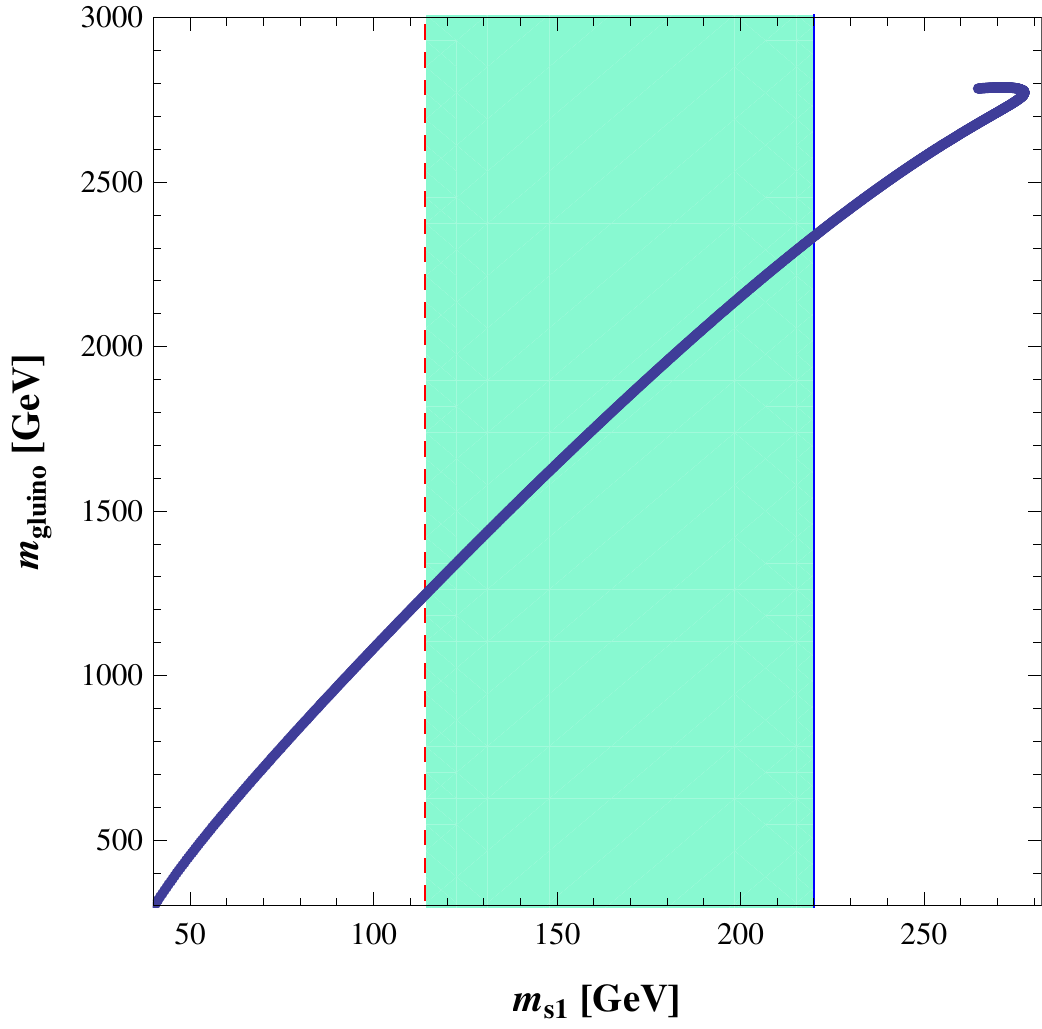}\\

\caption{Upper bound on the mass of the stop (left panel) and of the gluino (right panel) for $\Lambda_{mess}\sim 100 \textrm{ TeV}$ an allowed fine-tuning of $10\%$, computed for the representative point in (\ref{point}) and for the chargino mass $\mu$ equal to $400\,\text{ GeV}$.  The blue area is the region allowed by all the constraints on the scalar potential (see Section \ref{sec:EWSB}) and by the LEP bound on the lightest Higgs boson mass.
\label{boundstop}}
\end{figure}

It is clear from eq. (\ref{runningduetosquarks}) that in the case of the sleptons and the first and second generation of squarks  we would have a bound that is too loose to be useful because of the smaller Yukawa couplings. However, extending  eq. (\ref{runningduetosquarks}) to include two-loops gauge effects the recent analysis of~\cite{Barbieri:2010pd} shows that in our model Naturalness bounds on the sfermion masses are around $10 \textrm{ TeV}$. Such bounds are of no practical impact for our analysis.

Additionally, it is of interest to know the Naturalness bound on the gluino mass $m_{\tilde{g}}$, which, contrary to the stops-sbottoms, contributes to $m_2$  via a two-loop effect enhanced by the large QCD coupling constant $\alpha_s=\frac{g_s^2}{4\pi}$~\cite{Martin:1993zk}. Integrating the soft mass $m_2$ RGE (see Appendix \ref{sec:RGE} for details), one obtains

\begin{equation}\label{eq:deltam2gluino}
\delta m_2^2=\frac{g_s^2}{4\pi^2}\lambda_t^2\, m_{\tilde{g}}^2\ln\frac{\Lambda_{\rm mess}}{1~\text{TeV}} \left(1-\ln\frac{\Lambda_{\rm mess}}{1~\text{TeV}}\right)\,.
\end{equation}

Imposing the Naturalness condition (\ref{eq:finetuning}), with respect to the parameter $m_{\tilde{g}}$, and using the approximate expression

\begin{equation}
\left|\frac{m_{\tilde{g}}^2}{v^2}\frac{d\,v^2}{d\,m_{\tilde{g}}^2}\right|\sim\left|\frac{m_{\tilde{g}}^2}{v^2}\,\frac{d\,v^2}{d\,m_2^2}\,\frac{d\,m_2^2}{d\,m_{\tilde{g}}^2}\right|\,,
\end{equation}
 one can find analytically the bound

\begin{equation}\label{eq:boundmg}
m_{\tilde{g}}\leq 2\pi^2 v\sin\beta\, \frac{\sqrt{\Delta}}{ g_s\sqrt{\frac{d v^2}{d m_2^2}}\,\ln\frac{\Lambda_{\rm mess}}{1~\text{TeV}}\left(1-\ln\frac{\Lambda_{\rm mess}}{1~\text{TeV}}\right)}\,.
\end{equation}

The result for the bound on the mass of the gluino for a messenger scale of 100 TeV	 and an allowed fine-tuning of $10\%$ is shown in the right panel of Figure \ref{boundstop} for the parameter point presented in (\ref{point}) and for $\mu$ fixed to be $400\,\text{GeV}$. From the Figure it is evident that the upper-bound on the gluino mass $m_{\tilde g}$ increases when one considers increasing values of the mass of the lightest Higgs, and, for a Higgs mass satisfying the LEP bound, the mass of the gluino can be $\sim 1.3\text{ TeV}$  for an allowed fine-tuning of $10\%$.

 \bigskip
 
 Our study of Naturalness yielded two interesting results. Firstly, the model is less tuned when the lightest Higgs boson is heavier. Secondly, the bounds on third generation squarks and gluinos are rather loose. In fact, we will see in Section \ref{EWPT} that LEP phenomenology is  rather unaffected by the presence of the superparters of top and bottom. Similarly, the bounds on the masses of the sparticles involved in flavour transitions are quite loose and can naturally yield only negligible contributions with respect to the SM.

These two points are in general quite interesting because they show how the model, in the region where the Higgs is quite massive, is able to settle the tension present in the MSSM for the masses of stops and gluinos. In fact, on the one side in the MSSM Naturalness arguments would require stops-sbottoms and gluinos to be light: if one does not want to allow a $1\%$ or less of fine-tuning, stops, sbottoms and gluinos can not be at the TeV scale in the MSSM; on the other side, $\Delta F=2$ and $\Delta F=1$ flavour transitions would generally push to have heavier sparticles, since, otherwise the new physics effects in the flavour observables ($b \rightarrow s\gamma$ for example) that are well measured and well in agreement with the SM prediction could be too large. On the contrary, in the model at study sparticles masses can be naturally raised at the TeV scale, and consequently the SUSY flavour problem  can be addressed more easily than in the MSSM. Furthermore, the analysis of~\cite{Barbieri:2010pd} confirms this result when first and second generation squarks are included in the discussion.

\section{LEP direct searches}\label{sec:LEP}

Using the formulas for the masses of the Higgs bosons and Higgsinos given in Section \ref{Sec:masses}, we can impose the bounds from LEP direct searches~\cite{pdg,LEPcharged,LEPhiggs,LEPpseudo}

\begin{eqnarray}
m_h&>&114 \text{ GeV,}\label{mh114}\\\label{eq:chargino}
m_{\chi^+}&>&103\, \text{GeV,}\\\label{eq:chargedHiggs}
m_{H^\pm}&>&79\, \text{GeV,}\\
m_{\chi_1}&>&m_Z/2\, \label{mZmezzi},
\end{eqnarray}
and exclude regions of the parameter space accordingly.
For sake of simplicity, instead of taking the limits on the cross-section for neutralino production (summarized in~\cite{review}), we impose the tighter condition to have all the neutralinos with mass above $m_Z/2$. Similarly we are not taking into account the fact that the quoted limits apply to the MSSM and that in the NMSSM they could be less stringent. Thus the limits that we impose are rather conservative and serve the purpose of proving that there is a region parameter space that is certainly allowed by the current data. We are however aware that a larger region of parameter space could be allowed.

The union of the constraints from vacuum stability found in Section \ref{sec:EWSB} and the constraints from LEP is shown in Figure \ref{withLEP} for the  choice of parameters in (\ref{point}) and two variations of $\lambda$ and $\tan\beta$.
The strongest constraint on the parameter space comes from the bound on the neutralino mass given in (\ref{mZmezzi}), which cuts a region in $\mu$ around the value given in (\ref{zeroLSPmass}) where the lightest neutralino is massless.
On the other hand, we see that chargino, charged and neutral Higgs boson searches have basically no impact on the physical parameter space, once that the constraints on the scalar potential of Sections \ref{sec:EWSB} and \ref{Sec:masses} are imposed. Indeed the maximal chargino mass that could be probed at LEP is less than the minimum allowed $\mu$ given in (\ref{mumin}). Analogously, for the neutral and the charged Higgs bosons the typical mass in our model is significantly larger than the bounds and therefore only little restrictions arise from (\ref{mh114}) and (\ref{eq:chargedHiggs}).

\begin{figure}
\center
\includegraphics[width=.8\textwidth]{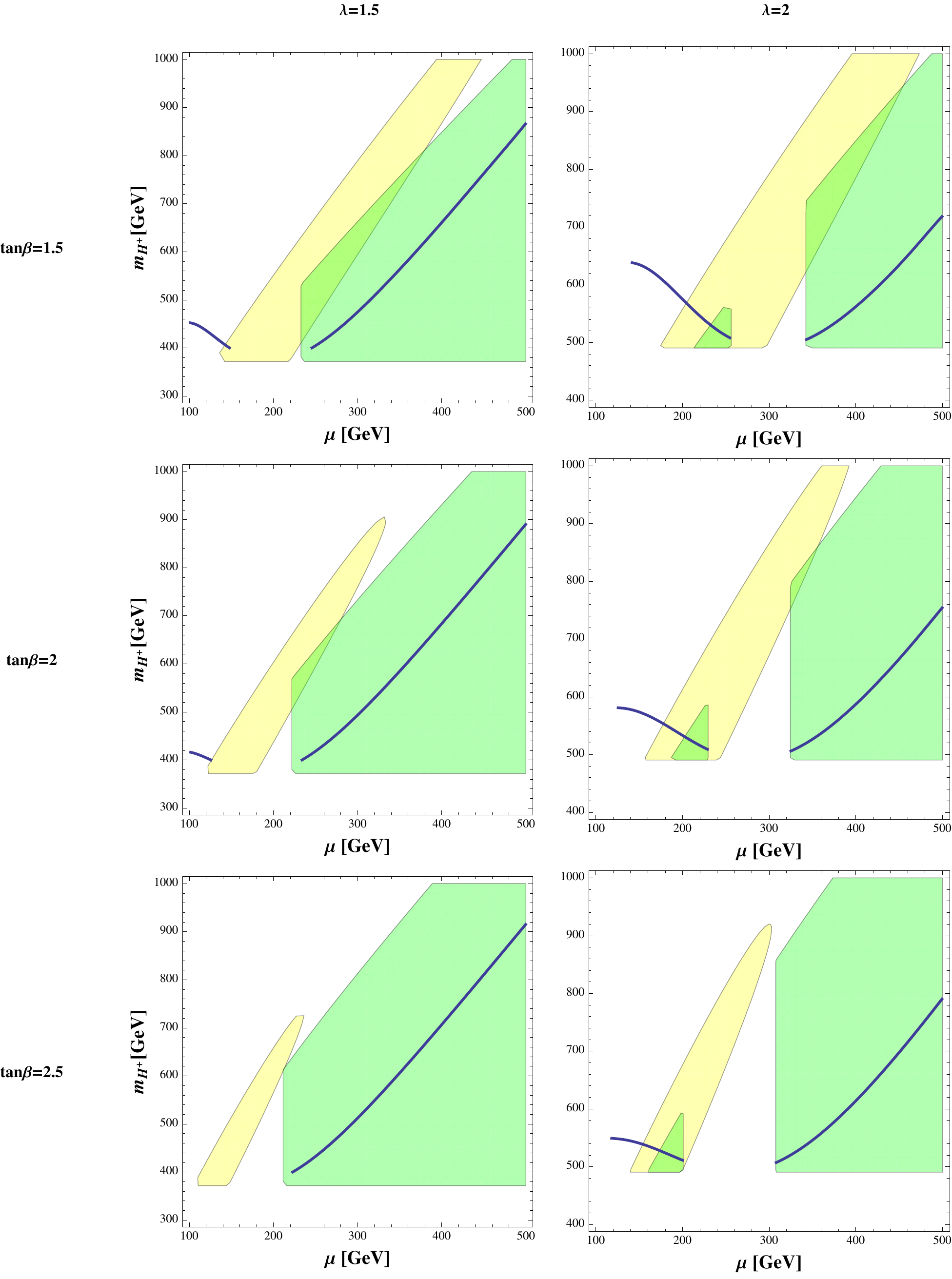}
\caption{\label{withLEP}
Six panels showing the area of the plane $\mu$, $m_{H^+}$ allowed by the constraints of a correct $SU(2)$ symmetry breaking, CP invariance, and LEP direct searches. The left column is for $\lambda=1.5$ and the right column is for $\lambda=2$. In each column is shown, from the top to the bottom, the result for $\tan\beta$ equal to 1.5, 2 and 2.5. In all the cases $k=1.2$ and for $k \sim 1$ the results are qualitatively unchanged.   A solid blue line indicates the points where $ 2m_{\chi_1}=m_{a_1}$. The vertical cut of the green region is due to the LEP bound on $m_{\chi_1}$.}
\end{figure}

\section{Indirect constraints from EWPT} \label{EWPT}
It is well known that the NMSSM in the large $\lambda$ regime can have significant impact on the EWPT~\cite{lsusy}. As such, we compute the new physics contributions to the oblique parameters $T$ and $S$, coming from the new scalars and the new fermions of the model. %
In what follows, we perform the analysis of the EWPT in the $S-T$ plane, with the experimental contours
taken from~\cite{Plot}.

Scalar contributions arise from the enlarged Higgs sector and from the sfermions. All the sleptons, the first and second generation of squarks are not restricted to be particularly light by Naturalness arguments (see Section \ref{FT}) and therefore their contribution can be neglected, assuming they are heavy. Third generation squarks are somehow special, indeed  we have seen explicitly in Section \ref{FT} using Naturalness arguments, that they cannot be taken too heavy. Therefore the stop-sbottom system  can potentially give a sizable contribution to the electroweak precision observables $T$ and $S$ which we compute in Section \ref{Tstsb}. 
The contribution arising from the enlarged Higgs sector with a heavy spectrum will be discussed in Section \ref{SThiggs}.
In the following section, instead, we discuss the contribution from the fermions that, in the limit of heavy gauginos, reduces just to the contribution of the Higgsinos.

\subsection{Higgsinos\label{secSThiggsinos}}

In the basis $N_{1,2,3}$  the interaction Lagrangian of the Higgsinos reads
\bea
\mathcal{L}&=&-\frac{g_2}{2}W^+_\mu \bar{\chi^\_} \left(\gamma^\mu N_1 - \gamma^\mu\gamma^5 N_2\right) +h.c. \\\nonumber
&+&\frac{g_2}{2}W_\mu^3 \left( \bar{\chi^\_}~\gamma^\mu \chi^\_ + N_1 \gamma^\mu\gamma^5 N_2\right)
+\frac{g_1}{2}B_\mu \left( \bar{\chi^\_}~\gamma^\mu \chi^\_ - N_1 \gamma^\mu\gamma^5 N_2\right)\,.
\eea
Therefore the contributions to $T$ and $S$ can be given using the loop functions  $\tilde F$ and $\tilde A$ given in Appendix \ref{Appendix} by the expressions

\begin{eqnarray} 
T&=&\sum_{i=1,2}\sum_{m=1,2,3}V_{mi}^2 \tilde{A}(\mu,m_{\chi_m}) + \\\nonumber
&-&\sum_{m=1,2,3} \sum_{n> m} (V_{1m} V_{2n} + V_{1n}V_{2m})^2 \tilde{A}(m_{\chi_m},-m_{\chi_n})+ \\\nonumber
&-& \frac{1}{2} \sum_{m=1,2,3}(V_{1m}V_{2m}+V_{2m}V_{1m}  )^2\tilde{A}(m_{\chi_m},-m_{\chi_m})\,, \\
S&=& \sum_{m=1,2,3}\sum_{n>m}(V_{1m} V_{2n} + V_{1n}V_{2m})^2 \tilde{F}(m_{\chi_m},-m_{\chi_n})+\\\nonumber
 &+& \frac{1}{2} \sum_{m=1,2,3}(V_{1m} V_{2m} + V_{1m}V_{2m})^2 \tilde{F}(m_{\chi_m},-m_{\chi_m}) - \tilde{F}(\mu,\mu)\,,
\end{eqnarray}
where $V$ is the rotation matrix for the Higgsinos defined in (\ref{Vmatrix}).

In Figure \ref{SThiggsinos} we present the result  as function of $\mu$ and the coupling $k$ for several representative values of $\tan\beta$ and $\lambda$. The Figure shows that the generic value of the contribution to $S$ is sizable but of the same order of the experimental uncertainty; on the other hand the value of $T$ is more problematic and deserves some more discussion.

The result of Figure \ref{SThiggsinos} shows that $T$ prefers small values of both  $\lambda$ and $\tan\beta$, as expected from the dependence of the mass splittings of the Higgsinos on these two parameters. 
Moreover it shows that values of $k$  close to zero or close the perturbative bound of $\sim 1.2$ give the best results, with the latter generically giving a better result than the former. 
This can be understood noting  from (\ref{tracehiggsinos}) that the mass scale of the Higgsinos for fixed $\lambda$ goes like the product $k \mu$, thus, in general, larger values of $k$ and $\mu$ tend to give a smaller contribution to $T$. However, for any finite value of $\mu$,  there is a non-vanishing value of $k$ given by (\ref{zeroLSPmass}) that renders massless the lightest neutralino. In the regions of the $\mu$, $k$ plane where (\ref{zeroLSPmass}) is satisfied, and of course in the vicinity of them, the contribution to $T$ is enhanced by the presence of the light state. From this discussion it is clear that for $k$ large enough the critical value of $\mu$ is pushed to be smaller than the minimal  phenomenologically interesting $\mu\simeq 100 \gev$ and that, away from the line where (\ref{zeroLSPmass}) is satisfied, a larger $k$ gives a smaller contribution to $T$.

This preference of the EWPT for large values of $k$ gives further motivation to consider the regime of the NMSSM with large coupling $k$ (see eq. (\ref{point})), namely the regime where the PQ symmetry is broken by a large coupling and hence all the states in the CP-odd sector are heavy. 

\begin{figure}
\center
\includegraphics[width=.99\textwidth]{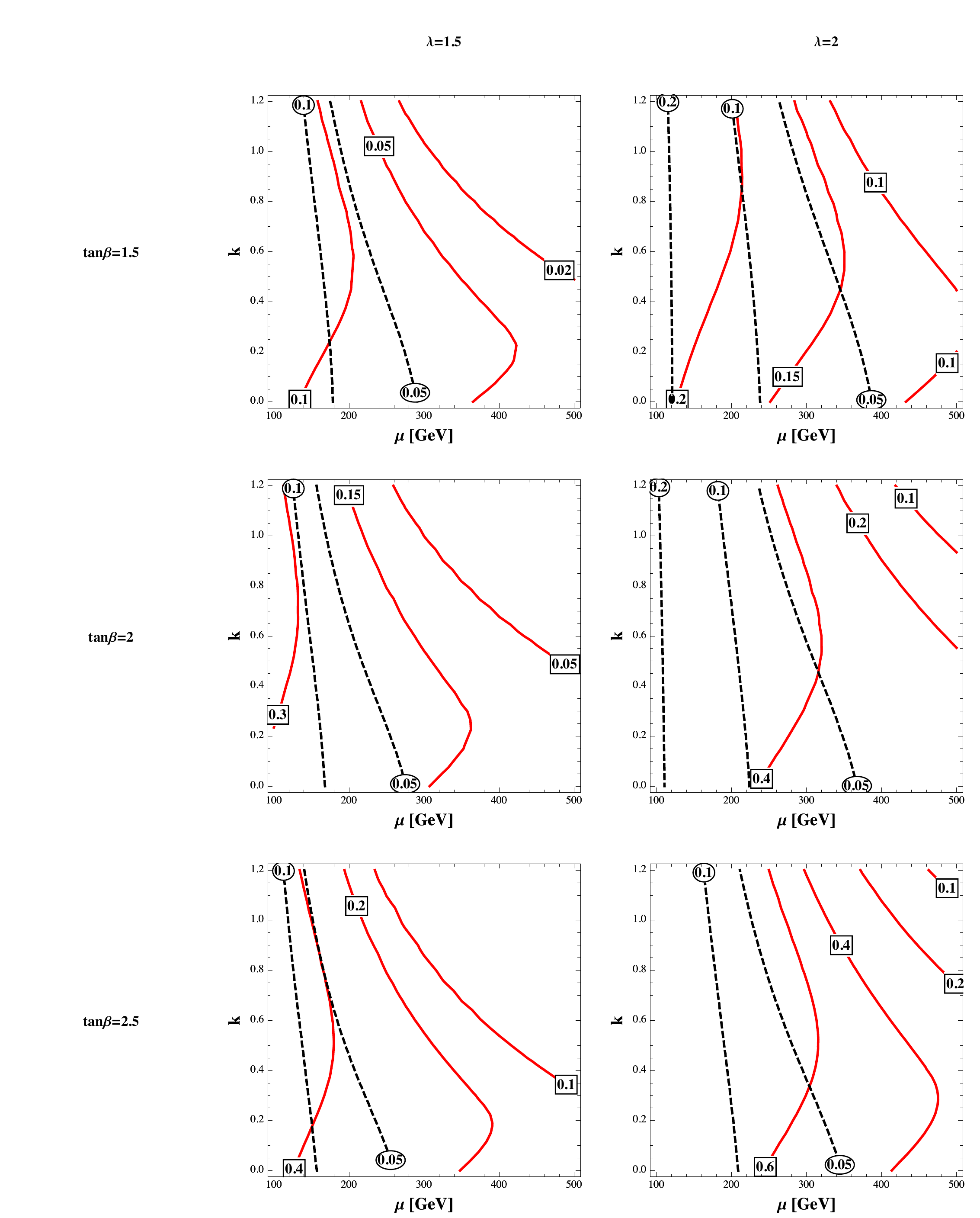}
\caption{
Contribution of the  Higgsinos to  $T$ and $S$  in the parameter space $\mu$, $k$ for different values of $\tan\beta$ and $\lambda$. The left column is for $\lambda=1.5$, the right column is for $\lambda=2$.  From the top to the bottom of each column are shown the results for $\tan\beta$ equal to 1.5, 2 and 2.5.
Solid red lines with squared labels are the contributions to $T$, dashed black lines with round labels are the contributions to $S$. }\label{SThiggsinos}
\end{figure}

\subsection{Stop and sbottom squarks\label{Tstsb}}
We compute the contribution to $S$ and $T$ in the limit of diagonal squark mass matrices so that the interaction eigenstates for squarks coincide with mass eigenstates and the contribution to $S$ and $T$ are just

\begin{eqnarray}
T&=&6 A(m_{\tilde{t}_L} , m_{\tilde{b}_L} )\,,\\
S&=& F(m_{\tilde{b}_L} , m_{\tilde{b}_L} )- F(m_{\tilde{t}_L} , m_{\tilde{t}_L} )\,,
\end{eqnarray}

\noindent where the loop functions $A$ and $F$  are reported in the Appendix \ref{Appendix} and the masses of the third generation squarks are given by (\ref{eq:mst}) and (\ref{eq:msb}) respectively. In order to estimate the minimal effect of the stop-sbottom, we assume that the soft mass $m_{\tilde Q}$ saturates the upper-bound due to fine-tuning constraints given in (\ref{eq:boundmQ}).

Both the contributions to $T$ and $S$ are rather small compared to the experimental uncertainties. In Figure \ref{STstsb} we show the value of $T$ in the plane $\mu$, $m_{H^+}$, for an allowed fine-tuning of $10\%$ and the point of parameter space presented in (\ref{point}).  We find that, for a $10\%$ fine-tuning and for a lightest Higgs boson in accordance with the LEP bound (yellow region in the Figure \ref{STstsb}),  the contribution to $T$ is always smaller than $\sim 0.04$, which is small compared to the contribution coming from Higgsinos. This feature is due to the relatively heavy stops and sbottoms, allowed by Naturalness constraints, and does not change significantly for larger values of $\tan\beta$.

\begin{figure}
\center
\includegraphics[width=.4\textwidth]{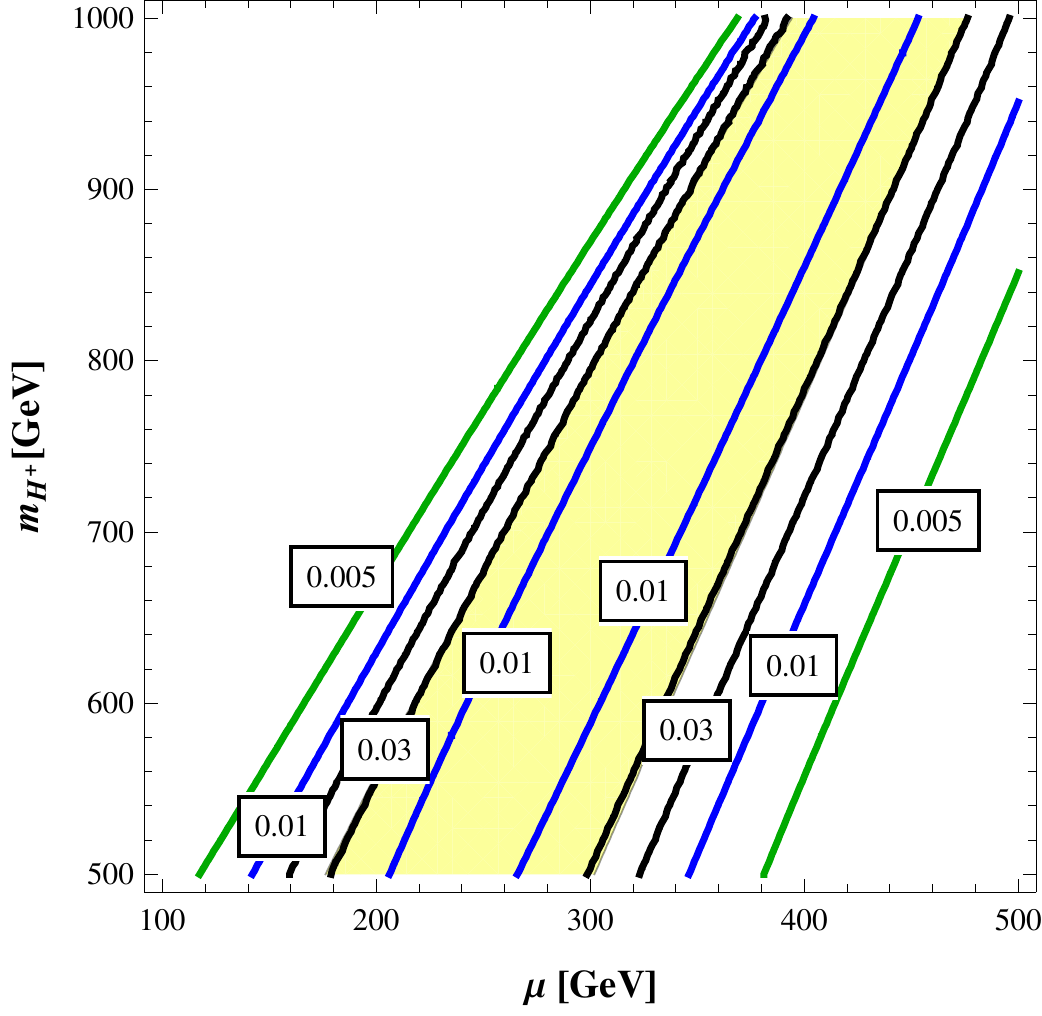}
\caption{Minimal contribution to $T$ coming from the stop-sbottom sector in the plane $\mu,m_{H^+}$ for the reference point of parameter space presented in (\ref{point}), in the approximation of no mixing between $\tilde{t}_L$ and  $\tilde{t}_R$, and with the stop mass saturating the upper-bound (\ref{eq:boundmQ}). The overlaid yellow
area corresponds to $m_{s_1} > 114$ GeV. \label{STstsb}}
\end{figure}

\subsection{CP-odd and CP-even Higgs bosons \label{SThiggs}}

For the Higgs sector we compute the contribution to $T$ and $S$ coming from the three CP-even states and the two CP-odd states.  
 We compute the total value of $S$ and $T$, taking as zero the value of the LEP Electroweak Working Group~\cite{LEPEWWG} minus the contribution that arises in the SM from a Higgs boson with a reference mass just above the LEP bound $m_{h_\textrm{SM}}=115\, \textrm{ GeV}$. This latter contribution is given by 

\bea
T(m_{h_\textrm{SM}})&=&3\left( A(m_{h_\textrm{SM}},m_Z) - A(m_{h_\textrm{SM}},m_W)  \right)\,,\\ S(m_{h_\textrm{SM}})&=&F(m_{h_\textrm{SM}},m_Z)+m_Z^2G(m_{h_\textrm{SM}},m_Z) \label{STSMhiggs}\,,
\eea
with the loop functions $A$, $F$ and $G$ given in the Appendix \ref{Appendix}.

Subsequently, we add the contributions coming from the enlarged Higgs sector of the theory

\begin{eqnarray} 
T&=&\sum_{i=1}^3 U_{i2}^2  ~ T(m_{s_i})+\sum_{i=1}^3 U_{i1}^2 A(m_{H^+},m_{s_i})+ \\ &+&\sum_{j=1}^2 P_{j1}^2 A(m_{H^+},m_{a_i}) 
-\sum_{i=1}^3\sum_{j=1}^2  U_{i1}^2 P_{j1}^2 A(m_{s_i},m_{a_j})\,, \nonumber \\
S&=& \sum_{i=1}^3 U_{i2}^2 S(m_{s_i}) + \sum_{i=1}^3\sum_{j=1}^2 U_{i1}^2 P_{j1}^2 F(m_{a_j},m_{s_i}) - F(m_{H^+},m_{H^+})\,,
\end{eqnarray}

\noindent where $m_{s_i}$ and $m_{a_i}$ are the masses of the scalar and pseudo-scalar Higgs respectively, the rotation  matrices $U$ and $P$ are defined in (\ref{Umatrix}) and (\ref{Pmatrix}) respectively, and the loop functions $F$, $T$, $S$ and $A$ are given in the Appendix \ref{Appendix}.

Once $\tan\beta$, $\lambda$ and $k$ are fixed, the mass spectrum and the mixing matrices still depend on the parameters $\mu,\,m_{H^+}$. Therefore, in place of the customary plot of the position of the model in the $S$-$T$ plane, we show in Figure \ref{STHiggs} the contribution to $T$ and $S$ in the $\mu,m_{H^+}$ plane for representative values of $\tan\beta$ and $\lambda$. The contributions are generically well within the experimental uncertainty and sub-dominant with respect to the contributions from the Higgsinos sector computed before. 

\begin{figure}
\center
\includegraphics[width=.85\textwidth]{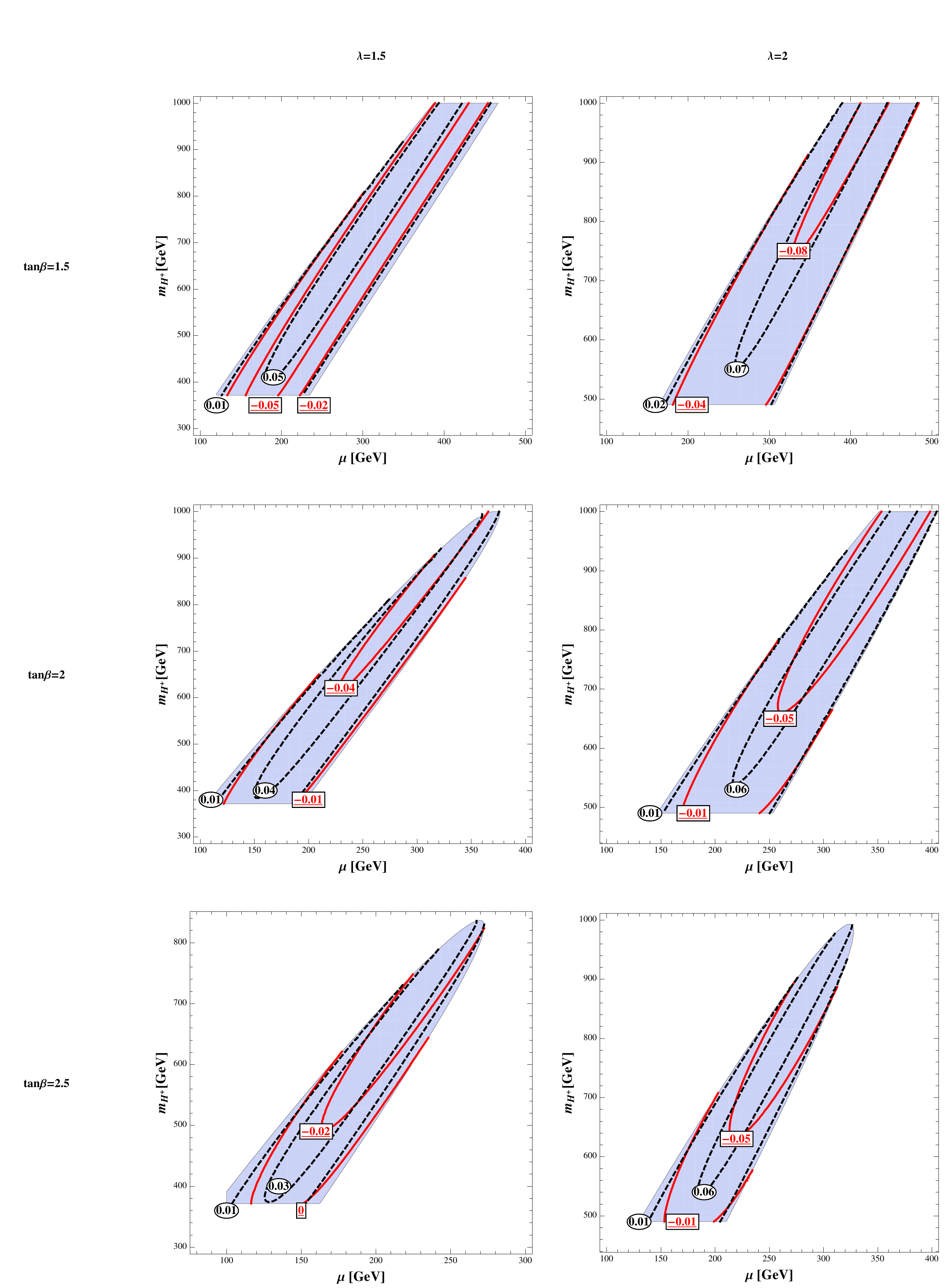}
\caption{Contribution of the scalar sector to $S$ and $T$ in the plane $\mu,m_{H^+}$ for different values of $\tan\beta$ and $\lambda$. The left column is for $\lambda=1.5$, the right column is for $\lambda=2$.  From the top to the bottom of each column are shown the results for $\tan\beta$ equal to 1.5, 2, 2.5. In all the cases $k=1.2$ and for $k \sim 1$ the results are qualitatively unchanged.
Solid red lines with squared labels are the contributions to $T$, dashed black lines with round labels are the contributions to $S$. The blue area denotes the region of the plane where all scalars have positive mass squared.\label{STHiggs}}
\end{figure}

\section{Relic abundance of neutralinos\label{relicDM}}

The relic abundance of a heavy particle that decouples as a non-relativistic specie can be computed solving its Boltzmann equation~\cite{Hut:1977zn,Lee:1977ua,
Vysotsky:1977pe}.
The resulting  relic abundance is given by the approximate formula
\beq
\label{relicOmega} \Omega h^2 \simeq \frac{1.07\times 10^9 \textrm{ GeV}^{-1}}{\sqrt{g_*}M_P (a x_{f.o.}+b x_{f.o.}^2/2)}\,,
\eeq
where $a$ and $b$ are constants related to the annihilation cross-section of the LSP, $g_*=86.25$ for $m_b\ll T_{f.o.}\lesssim m_W$ is the number of SM degrees of freedom at the time of freeze out, $M_P$ the Planck mass and $x_{f.o.}$ the normalized freeze-out point temperature $x_{f.o.}=T_{f.o.}/m_{LSP}$, with $m_{LSP}$ the mass of the LSP, that in our model is supposed to be the lightest Higgsino $\chi_1$. The freeze-out point can be found solving numerically the equation

\beq
\label{xfo} \frac{1}{x_{f.o.}} \simeq \log \frac{ 0.037 g_\chi M_P\, m_{LSP} \left\langle \sigma v \right\rangle \sqrt{x_{f.o.}}}{\sqrt{g_*}},
\eeq
with $g_\chi=2$.

\bigskip

The only relevant quantity from particle physics is the thermally averaged annihilation cross-section $\sigma v$ of the LSP which can be expanded around its non-relativistic limit as 
\beq
\left\langle \sigma v \right\rangle = a + b x\,. \label{sigmav}
\eeq

As far as the cases where $m_{LSP}<m_W$ are concerned, the only available contributions to (\ref{sigmav}) are those coming from the annihilation $\chi\chi \to f \bar{f}$. In general this process can be mediated by a s-channel exchange of a Higgs boson or a $Z$ boson, and by the t-channel exchange of a sfermion. Since in this paper we are interested in the case of heavy sfermions (see Section \ref{sec:boundmasses}), we will only investigate the s-channel contributions. 
The  most important contributions to the s-channel can be understood in terms of symmetry arguments. Indeed the annihilation cross-section, as expanded in (\ref{sigmav}),  corresponds to a partial wave expansion of the annihilation process. Using CP properties of the mediators and the chirality structure of the intervening interactions, one can find that the s-wave annihilation can be mediated only by CP-odd scalars, while the p-wave receives contributions from both a CP-even scalar and the $Z$ boson~\cite{GoldbergSwaveSuppression,DreesNojiri}. The s-wave part of the cross-section is typically suppressed by the smallness of the Yukawa couplings involved, with the only exception of the on-shell production of the CP-odd scalar. The line in the parameter space where this condition is fulfilled is shown in Figure \ref{withLEP} as a solid line.
Away from this line we can just repeat the analysis of the previous investigations of Higgsino DM in the large $\lambda$ regime~\cite{lsusy}. In fact, making the identification $M\rightarrow-2k \mu/\lambda$, the Higgsino sector of our model is equivalent to the one studied in~\cite{lsusy} and therefore the resulting relic abundance is the same in the two cases. 

Above the $WW$ and $ZZ$ threshold new channels open and the LSP can now annihilate via a s-channel exchange of a $Z$ boson or a CP-even Higgs boson, and a t-channel exchange of a chargino. Once again by symmetry arguments the s-channels contribute only to the p-wave, while the t-channel contributes to  the s-wave and the p-wave. Also in this case the resulting relic abundance coincides with that one obtained in~\cite{lsusy}.

In Figure \ref{relic}, the resulting $\Omega h^2$ is given as a function of $\mu$ for different values of $\tan\beta$ for both $\lambda=1.5$ (on the left) and $\lambda=2$ (on the right). The features of the curves of Figure \ref{relic} are mainly due to the dependence of the cross-section on the mass of the LSP. In fact, when $m_Z/2<m_{LSP}<m_{W}$, the only annihilation channel is that one into fermions that is mediated by an off-shell $Z$ boson, which yields a cross-section behaved as $\sigma\sim1/m_{LSP}^2$. This means that the relic abundance increases with $m_{LSP}$ until new channels open. When annihilations into $WW$ and $ZZ$ are available, the relic abundance has to decrease accordingly with the opening of the phase-space for the new modes. Altogether  
the curves have a rise-and-fall shape with a maximum corresponding to $m_{LSP}\simeq m_Z$ and a maximal value being determined by the mixing coefficient that affects the annihilation into fermions.

Generically the annihilation cross-section of the Higgsino DM is too large and the relic abundance is then too low to account for the observed amount of DM, $\Omega h^2\simeq0.11$~\cite{WMAP}. In spite of this general trend there are regions of the parameter space  where $m_{\chi_1}\lesssim m_Z$ and $\tan\beta\lesssim 1.5$ where the LSP has a large singlino component~\cite{lsusy,singlinoDM,singlinoDMnmssm} and the relic abundance is reproduced. 

The necessity of a very large component of singlino in the LSP to reproduce the correct order of magnitude of the relic abundance of the DM is well know~\cite{lsusy,singlinoDM,singlinoDMnmssm}  and is due to the fact that a thermal relic with electroweak couplings should have mass around one TeV rather than few hundreds of GeV to satisfy the requirement of a correct abundance. In this sense the need for a large component of singlino in the LSP should not be regarded as an additional tuning of the model, at least with respect to the usual situation of supersymmetric models in which the dark matter is a weakly interacting  particle with mass close to the weak scale.

In the regions of parameter space that reproduce the correct relic abundance for the DM the model can be tested by direct searches of DM through elastic scattering which we describe in the following section.

\begin{figure}
\begin{center}
\includegraphics[width=.8\textwidth]{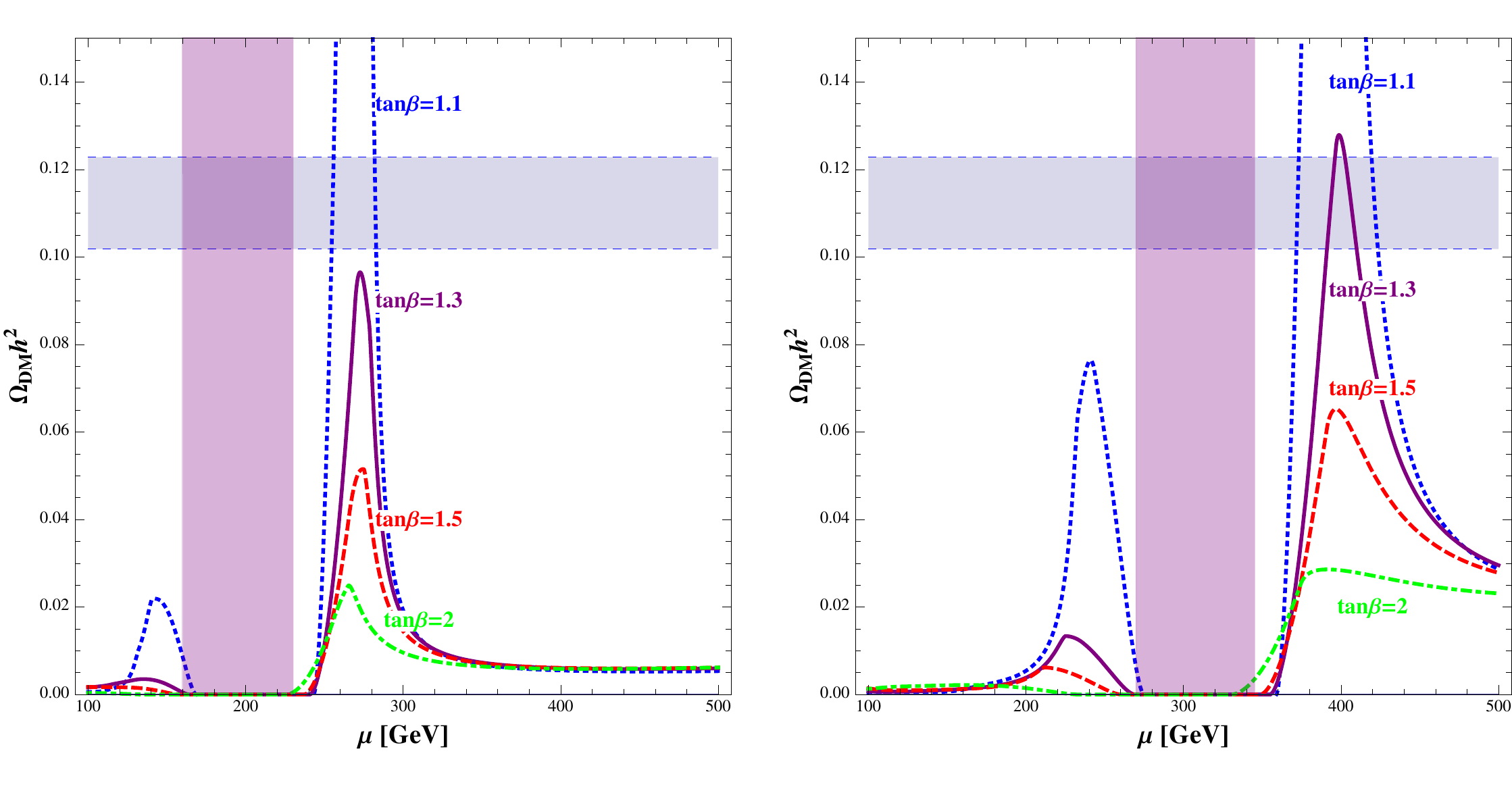}
\end{center}
\caption{ Relic abundance of neutralinos for $\lambda=1.5$ (left) and $\lambda=2$ (right). The horizontal blue band corresponds to the 3$\sigma$ interval from the 7-years WMAP result~\cite{WMAP}. Dotted blue lines are for $\tan\beta=1.1$, purple solid lines for $\tan\beta=1.3$, dashed red lines for $\tan\beta=1.5$ and finally dot-dashed green lines for $\tan\beta=2$. The vertical purple shaded area corresponds to $m_{LSP}<m_Z/2$ in the case $\tan\beta=1.3$. \label{relic}}
\end{figure}

\section{Direct detection of the dark matter\label{directdetection}}

Direct searches of DM particles stored in the halo of our galaxy have been performed and several bounds on the properties of the DM exist~\cite{CDMS,XENON}. The quantity probed by these experiments is the local DM density times the cross-section of an elastic scattering between a DM particle of mass $M$ and the nuclei of the experiment~\cite{BarbierDM,GoodmanWitten}. The local density is typically assumed to be $0.3 \textrm{ GeV/cm}^3$, so that bounds are given directly on the DM-nucleon cross-section. Therefore in the following we will compute the cross-section for the scattering of the LSP of the model on a proton.	

As we have already discussed, in our model the sfermions are heavy and the LSP is a mixture of Higgsinos and singlino. Therefore the only particle that can mediate a DM-nucleon scattering is a Higgs boson interacting via the Yukawa couplings dictated by the superpotential:

$$-\mathcal{L}_{yuk} = { \lambda \over \sqrt{2}} (S_1 \tilde{H}_1 \tilde{H}_2 + h_1 \tilde{S}\tilde{H}_2 + h_2 \tilde{H}_1 \tilde{S}) + {k \over \sqrt{2} } S_1 \tilde{S}^2 + h.c.\,.$$
These interactions contributes to the effective operator 
\beq
\mathcal{O}_{SI}=\frac{1}{\Lambda^2} \bar{\chi}_1\chi_1\bar{N}N\,,
\label{sigmaSIoperator}
\eeq 
which mediates the spin-independent elastic scattering of the lightest neutralino $\chi_1$ on a nucleus $N$ probed in~\cite{CDMS,XENON}.

The elastic cross-section at zero transferred momentum can be written as 
\beq
\sigma_{SI}(\chi_1 p \to \chi_1 p)=\frac{1}{16 \pi (m_p+{m_{LSP})^2}} \left| \mathcal{M} \right|^2\,,
\eeq
where the matrix element is given by
\beq
\mathcal{M}= \sum_m 2 m_{LSP} \frac{1}{m_{s_m}^2} g_{\chi\chi s_m}\left[\frac{2 m_p^2}{v} \left(  \mathcal{U}_{mu} F_u + \mathcal{U}_{md} F_d \right) \right],
\eeq
where 
\beq
  g_{\chi\chi s_m}= {\lambda \over \sqrt{2}} \sum_{(a,b,c)} 2\mathcal{V}_{1a}\mathcal{V}_{1b}\,\mathcal{U}^{-1}_{cm} + \sqrt{2} k \mathcal{V}_{13}^2\mathcal{U}^{-1}_{3m} \label{gLSPhiggs} \, ,
\eeq
and the indices $(a,b,c)$ run over all the ordered permutations of $(1,2,3)$, the matrix $\mathcal{V}$ is the matrix that brings the Higgsinos $\tilde{H}_1,\tilde{H}_2,\tilde{S}$ to the mass eigenstates $\chi_i$ and the matrix $~\mathcal{U}$ brings the scalar CP-even interaction eigenstates  $h_1,h_2,S_1$ to the mass eigenstates $s_m$.
In particular the $~\mathcal{V}$ matrix is related to the matrix $V$ defined in (\ref{Vmatrix}) through $\mathcal{V}= V^t R^{(\pi/4)}$ and $\mathcal{U}$ is related to $U$ of (\ref{Umatrix}) through $\mathcal{U}=U R^{(\beta +\pi/2)}$.

The effect of heavy quarks in the nucleon is taken into account accordingly to~\cite{heavyQnucleon} and incorporated in the values of $F_u$ and $F_d$  which we take from chiral perturbation theory respectively equal to $0.11\pm0.02$ and $0.44 \pm 0.13$~\cite{ChiPTEllis} or, from QCD on the lattice, $0.14\pm0.02$ and $0.23\pm 0.01$~\cite{lattice}. 

The quantity affecting the most the cross-section $\sigma_{SI}$ is the mass of the lightest Higgs boson which is the scale that suppresses the operator of the interaction (\ref{sigmaSIoperator}). In fact, the spin-independent cross section can be estimated  as
\beq \sigma_{SI} \simeq \frac{1}{16 \pi} \frac{m_p^2}{v^2} \frac{m_p^2}{m_{s_1}^4} \simeq 10^{-43}\textrm{cm}^2 \left( \frac{200 \gev}{m_{s_1}} \right)^4\,,
\eeq that is of the order of the sensitivity of current experiments. Therefore  direct searches of WIMPs can significantly restrict the allowed parameter space where $\Omega h^2\lesssim0.1$~\cite{WMAP}. From the previous section, we know that for $1.5<\lambda<2$ such interesting regions are those with $\tan\beta\lesssim 1.5$ and $m_{H^+}$ and $\mu$ in the regions outlined in Figure \ref{withLEP}.

Given the relevance of the mass of the lightest Higgs  for the cross-section, the latter has a significant dependence on the parameters of the scalar sector $\mu$ and $m_{H^+}$. However none of the two is of direct significance for the experiments that, on the contrary, probe $m_{LSP}$. Thus we will show our result trading $\mu$ for $m_{LSP}$ and we will fix some representative values of $m_{H^+}$ taken in the range that is allowed by all the constraints on the scalar potential analysed in Section \ref{sec:EWSB}.
The resulting DM-nucleon cross-section is given in Figure \ref{directDM} as a solid and a dashed thin line, obtained using hadronic matrix elements from~\cite{lattice} and~\cite{ChiPTEllis} respectively and assuming that the LSP accounts for the entire dark matter in the Universe. However the actual rate of DM-nuclei scattering in the model is typically reduced by the scarce amount of thermal relic neutralinos, therefore in Figure \ref{directDM} we give also  thick lines which correspond to the prediction of the model taking this reduction into account.

From the thin lines in Figure \ref{directDM}, we see that the predicted cross section is typically above the lower-bounds from experiments. This shows that in this model is difficult to reproduce the relic abundance of the dark matter without violating the experimental limits on WIMP scattering.

On the other hand when one takes into account the actual abundance of LSP computed in Section \ref{relicDM} the limits from direct detection experiments are not so stringent. The  region of low $m_{H^+}$ is well below the limits because of a low nucleon-DM cross-section and a low relic abundance, while for larger $m_{H^+}$ the regions corresponding to $m_{\chi_1}>60 \gev$ are typically excluded. In both cases the region that is compatible with direct WIMP searches indicates a preference for small of $\mu \sim (200-300) \gev$.

It is noteworthy to mention that for $m_{\chi_1}\sim (50-60) \gev$ the prediction is compatible with the cross section favored by the fit~\cite{CDMSfit} to  the recent claims~\cite{CDMS} of observation of a DM signal in direct detection experiments. Typically the regions compatible with this intriguing hint of dark matter detection  have relatively large $\mu$ and $m_{H^+}$, however, the  statistical significance of these claims is still relatively low and thus in the following we will concentrate on the region with lower $m_{H^+}$ that is safely compatible with the direct searches of DM.

\begin{figure}
\begin{center}
\includegraphics[width=.32\textwidth]{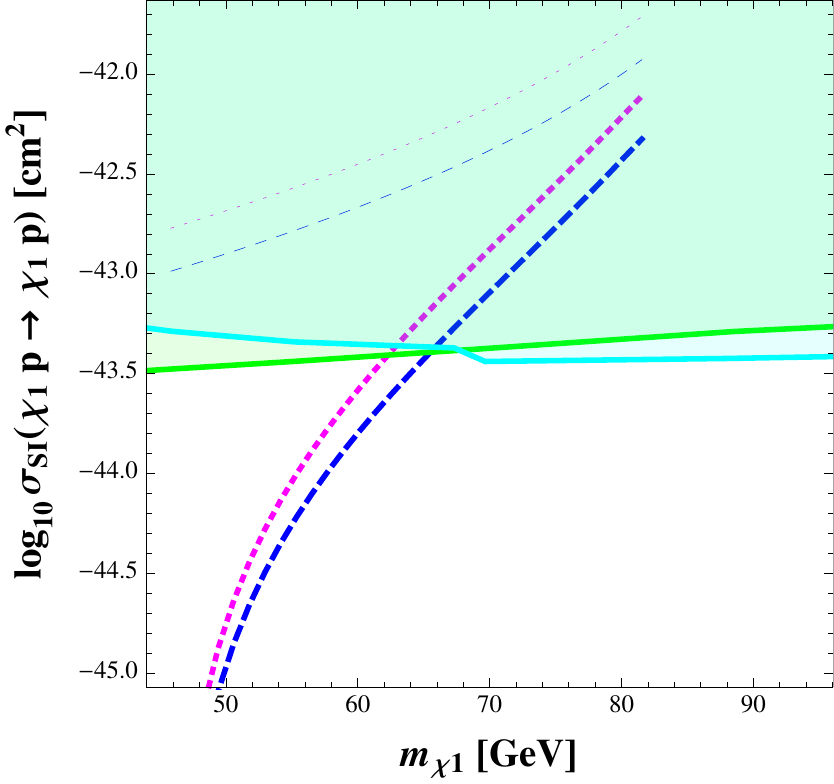}
\includegraphics[width=.32\textwidth]{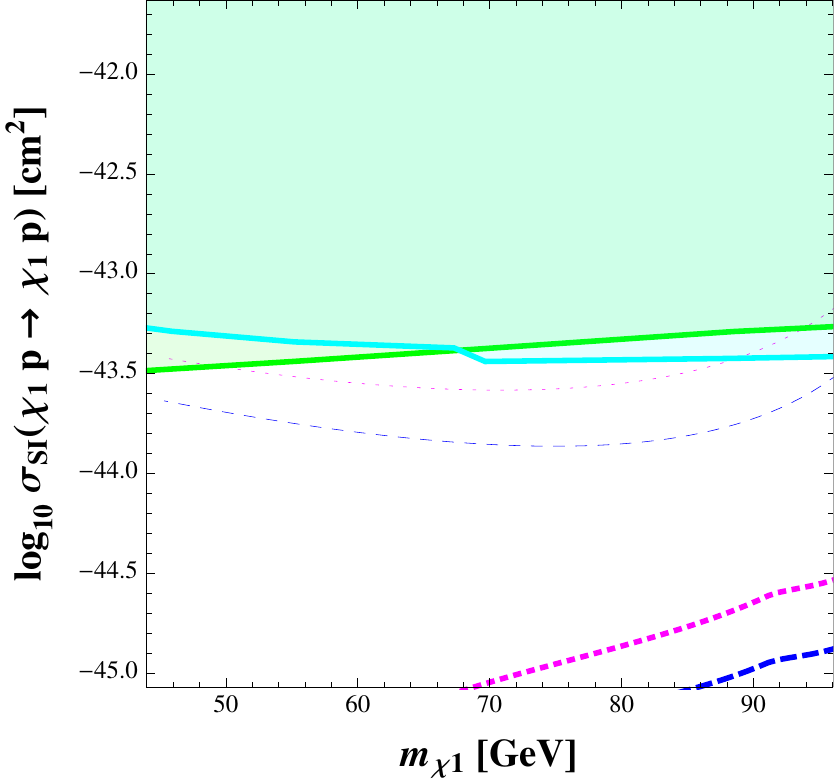}
\includegraphics[width=.32\textwidth]{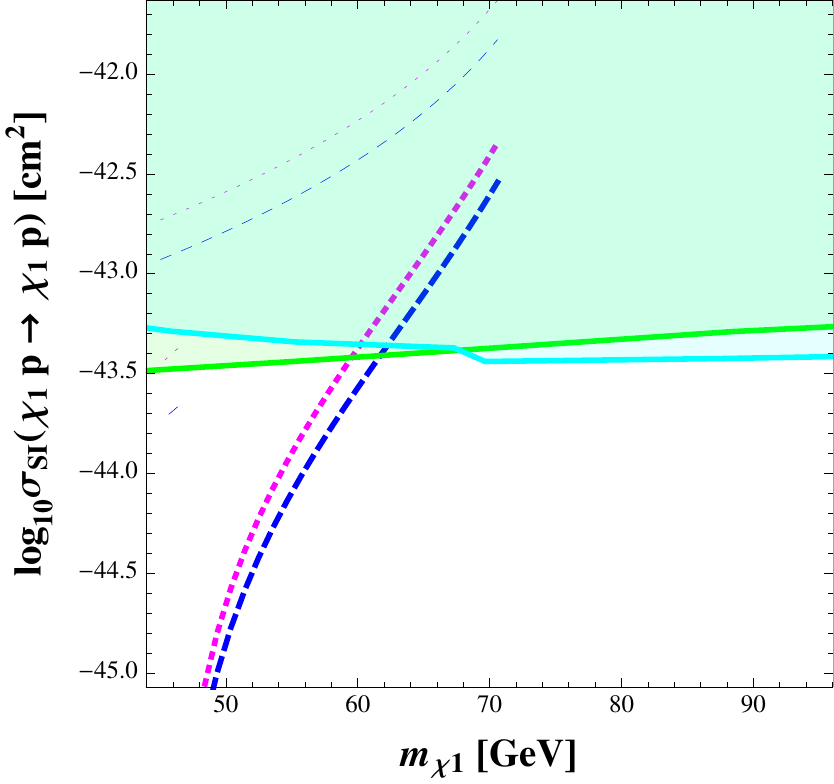}
\end{center}

\caption{Three panels showing $\log_{10}(\sigma_{SI})$ for the prediction of the spin-independent DM-proton scattering cross-section fixing $\tan\beta=1.5$. On the left for $\lambda=1.5$ and $m_{H^+}=500 \gev$, in the middle $\lambda=2$ and $m_{H^+}=550 \gev$, on the right $\lambda=2$ and $m_{H^+}=700 \gev$. The dashed and solid lines correspond to the prediction obtained taking the values of the hadronic matrix element from~\cite{ChiPTEllis} and from~\cite{lattice} respectively. Thick lines show the  prediction of the model once the actual relic abundance is taken into account. The shaded green and cyan areas are those excluded by Xenon~\cite{XENON} and CDMS~\cite{CDMS} respectively. \label{directDM} }

\end{figure}

\section{LHC phenomenology} \label{LHC}
\begin{figure}
\begin{center}
\includegraphics[width=.5\textwidth]{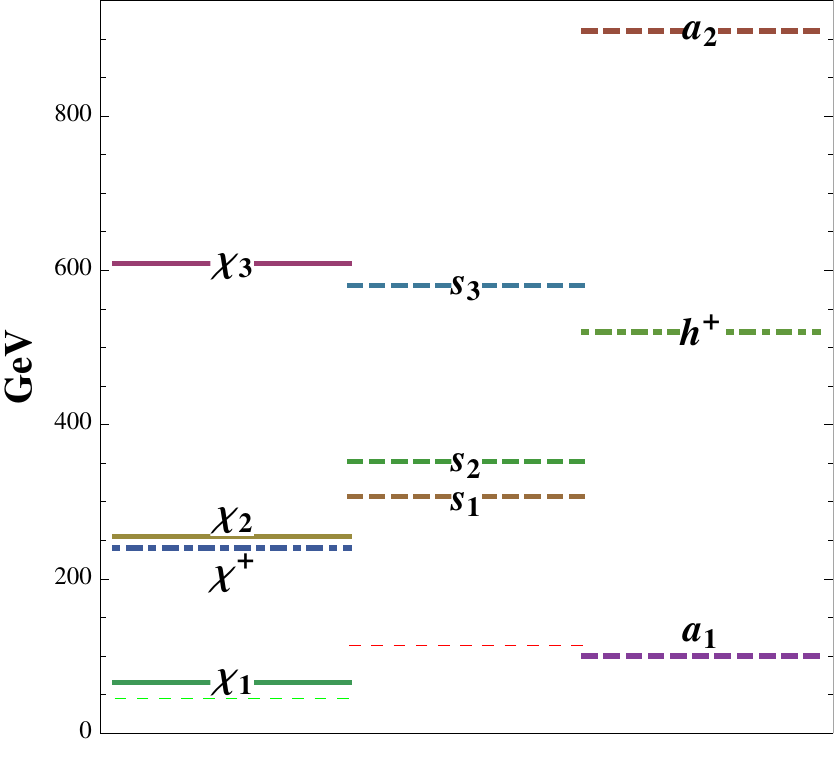}
\end{center}
\caption{The spectrum of the model for fixed parameters as in eqs. (\ref{point}) and (\ref{point2}). Solid lines correspond to neutral fermions, dashed lines to CP-even and CP-odd scalars, dot-dashed lines to charged particles. The two thin dashed lines correspond to $m_Z/2$ (in green) and $114 \gev$ (in red).	\label{spectrumlhc} }
\end{figure}

After the discussion of the restriction of the parameter space from past experiments, we discuss now how the LHC can probe our model.
The spectrum of the model has already been studied in general in Section \ref{Sec:masses}. Here we specialize the point selected in (\ref{point}) fixing the remaining parameters to
\beq \label{point2}
\mu=240 \gev\,,~~~m_{H^+}=520 \gev\,.
\eeq
The resulting spectrum is given in Figure \ref{spectrumlhc} and shows several interesting features. We notice that there is a light neutralino as LSP, $\chi_1$, a second lightest neutralino, $\chi_2$, rather close to the chargino, $\chi^+$, of mass $\mu$, and a significantly heavier $\chi_3$. In the scalar sector we find a relatively heavy lightest CP-even Higgs boson, $s_1$, that
 is rather close to its maximal mass (\ref{boundNMSSM}) and to the second CP-even state, $s_2$. Both these scalar states have a mass of order $\lambda v$ and from the relation (\ref{murange}) we deduce that their closeness to the fermionic states $\chi^+$ and $\chi_2$ is in fact generic in the entire region of parameter space allowed by all the constraints discussed previously. The pseudo-scalars always have a heavy state close to the TeV and a light state of roughly few hundreds of GeV. This large separation  between the pseudo-scalars is the result of a significant level-repulsion effect arising from the large mixing between the two states. Because of this effect, it is rather typical to have a spectrum with $2 m_{a_1}<m_{s_1}$. 

The precise values of the masses in the point defined by eqs. (\ref{point}) and (\ref{point2}) are 

\begin{eqnarray}
m_{s_{1}}=307 \gev \,, & m_{s_{2}}= 352 \gev\,, & m_{s_{3}}= 580 \gev\,, \\
m_{\chi_{1}}= 66 \gev\,, & m_{\chi_{2}}= 255 \gev\,, & m_{\chi_{3}}= 609 \gev\,, \\
m_{a_{1}}= 99 \gev\,, & m_{a_{2}}= 910 \gev\,.
\end{eqnarray}

From the spectrum it is evident that the regime of the NMSSM that we are considering is radically different from the case with perturbative couplings. Indeed in the spectrum we have Higgs and Higgsino states with masses of several hundreds of GeV, with the lightest state typically heavier than the $Z$ boson.  As such, a detailed analysis of the couplings and branching fractions is needed. This is performed in the next sections.

\begin{figure}
\begin{center}
\includegraphics[bb=170bp 0bp 648bp 700bp,clip,width=0.85\linewidth]{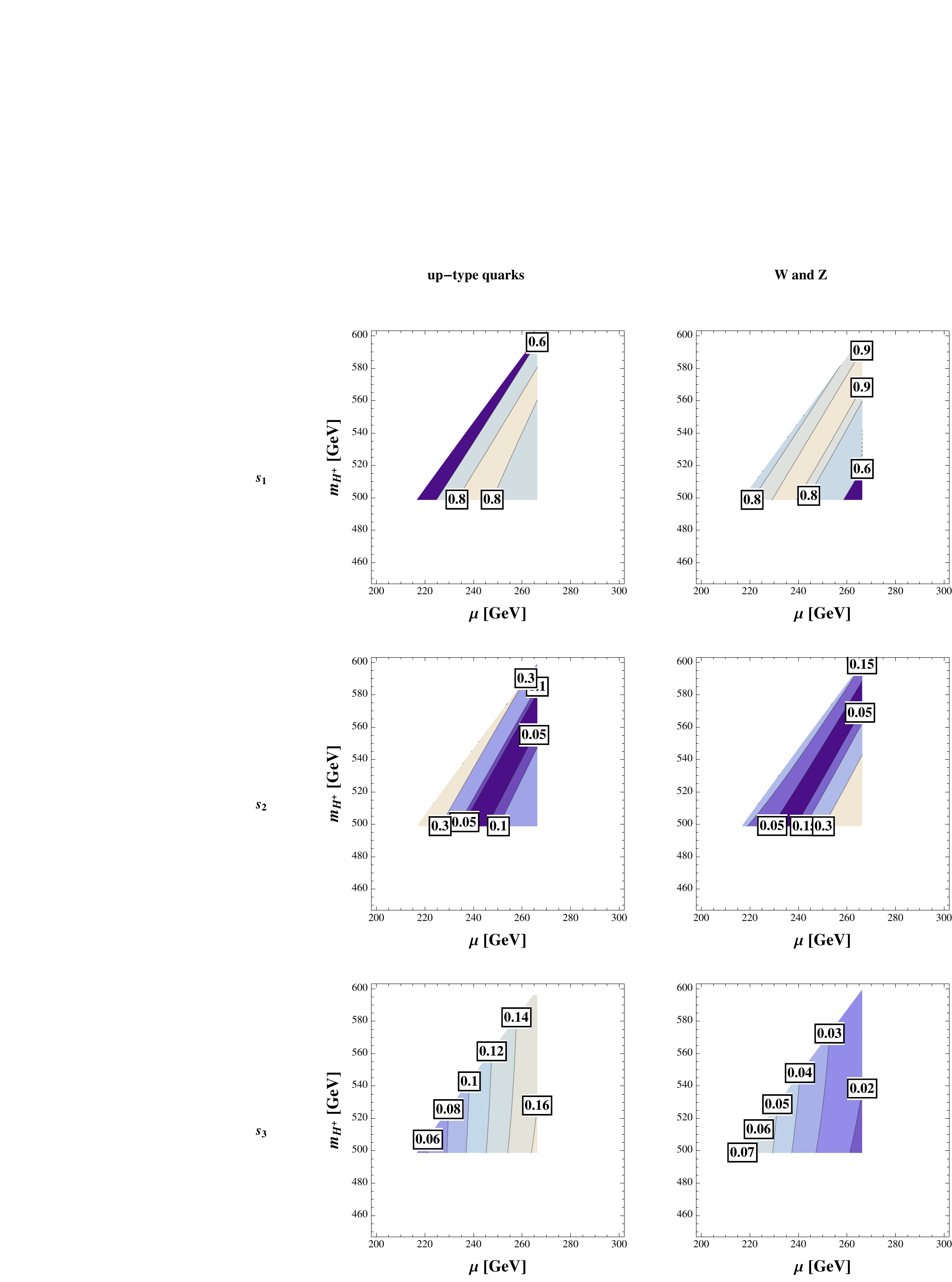}\end{center}
\caption{ (Left column) Reduced couplings squared of the up-type quarks with the CP-even scalars as defined in (\ref{xiup}). (Right column) Reduced couplings squared of the SM vectors with the CP-even scalars as defined in (\ref{xiV}). \label{reducedtt} }
\end{figure}

\subsection{Production of the new states}

We want to focus on the production of the scalar particles as they  provide a direct handle on $\lambda$, the characteristic coupling of the model. Furthermore the production of scalars is well studied in the SM and can be easily understood in terms of  ratios between a coupling of a scalar of our model and the corresponding coupling of the SM Higgs boson of equal mass. In such a way one inherits all the studies available in the literature about (differential and inclusive) QCD effects in Higgs production.
The ratios directly affecting the  cross-section of the partonic production process  \beq gg \to s_i,a_j\eeq
are those of the couplings of up-type quarks to CP-even and CP-odd scalars denoted respectively by  $\xi_{tts_i}$ and $\xi_{tta_j}$ and given by
 \begin{eqnarray}\label{xiup}
 \xi_{tts_i}&=&\left(\sin\beta U_{i2}^*-\cos\beta U_{i1}^*\right)^2\,,\\
  \xi_{tta_j}&=&\cos\beta^2P_{j1}^2\,,
   \end{eqnarray}
where the matrices $U$ and $P$ are defined in (\ref{Umatrix}) and (\ref{Pmatrix}) respectively.

  The reduced couplings squared $\xi_{tts_i}$ for the the CP-even Higgs bosons are shown in Figure \ref{reducedtt} in the relevant region of the plane $\mu$, $m_{H^+}$ after that the constraints on the scalar potential, the constraints from direct searches at LEP, and from direct dark matter searches are imposed. 
  
   The state $s_1$ is in general significantly coupled to the up-type quarks and it has, over the majority of the parameter space, a cross-section of at least 50$\%$ of the one of the SM Higgs boson with the same mass. The state $s_2$, on the contrary, has at most  30$\%$ of the cross-section of the SM Higgs boson of equal mass. The converse of this slight decoupling of $s_2$ is the relatively large coupling of $s_3$, which can have a cross-section up to $15\%$ of the one of the SM Higgs boson of the same mass.
This quantitative analysis  shows that the scalar CP-even sector consists altogether of three  states all significantly coupled to up-type quarks. In particular we find remarkable to have such a large coupling for the heaviest state, $s_3$. In general such mixing scenario seems favourable for the discovery of states beyond the lightest one.

For the pseudo-scalars the value of the couplings to the up-type quarks are mainly determined by the fact that the two mass eigenstates are almost maximal admixtures of singlet and doublet interaction eigenstates. As such, the two states nearly equally share a coupling equivalent to a fraction $\cos\beta$ of the Yukawa of the SM Higgs boson.  In particular, for the case of the point of parameter space (\ref{point}) and (\ref{point2}), we obtain $\xi_{tta_1}\simeq 0.16$ and $\xi_{tta_2}\simeq0.14$.
  
\bigskip

For the partonic production processes  \beq qq\to VV qq \to qq s_i \textrm{ and } q\bar{q}\to VV q\bar{q} \to q\bar{q} s_i\,, \label{vbf}\eeq we define the reduced couplings squared
\beq
 \xi_{VVs_i}=U_{i2}^2\,.\label{xiV}
\eeq
In this case, as we observe in Figure \ref{reducedtt}, the heavy states $s_2$ and $s_3$ are more decoupled, thus their discovery in the vector boson fusion processes (\ref{vbf}) is considerably more difficult than for a SM Higgs boson of the same mass.

\bigskip

Finally, the fermionic states $\chi_1,\chi_2,\chi_3,\chi^+$ are produced through their gauge couplings, as in the MSSM. Similarly, the production of $H^+$ will occur through its gauge and Yukawa interactions, as it happens in the MSSM.


\subsection{Decays}
The ordering of the masses shown in Figure \ref{spectrumlhc} and dimensional considerations on the decay width allows for a rough determination of the relevant decay channels of each state. We discuss some of them beginning from the lowest lying states.

The state $\chi_1$ is the lightest particle with negative R-parity and therefore it is stable. As well known its production results in large amount of missing transverse momentum.

The next lightest state is typically $a_1$ that, due to kinematics, can decay only in SM fermion pairs. Therefore its main decay modes are \beq a_1\to b \bar{b}\,,\eeq \beq a_1\to \tau \bar{\tau} \,.\eeq

The lightest CP-even state, $s_1$, has the usual decays into SM vector bosons 
\beq s_1\to ZZ\,,\quad s_1\to WW\,, \label{stoVV}\eeq
and the non-SM decays 
\beq s_1\to a_1 a_1\,,  \label{chainwithpseudo}\eeq
\beq s_1\to a_1 Z\,,\eeq
\beq s_{1} \to \chi_{1} \chi_{1}\,, \label{s1lspdecay}\eeq
\beq s_{1} \to \chi_{1} \chi_{2} \,.\label{s1chi1chi2}\eeq

However, due to the fact that $m_{s_1}\simeq \mu \simeq m_{\chi_2}$, the decay $s_1\to \chi_2 \chi_1$ is not available over the majority of the interesting parameter space.

The state $s_{1}$ has a total width  of several tens of GeV, as show in Figure \ref{gammaS1}. The contribution from the non SM-like decays is sizable and therefore the decays into $W^{-}W^{+}$, which would be dominant for a SM-like state, is always subdominat, as shown by the branching fraction into $W^{-}W^{+}$ shown in Figure \ref{BRs1WW}.

\begin{figure}
\begin{center}
\includegraphics[width=0.5\linewidth]{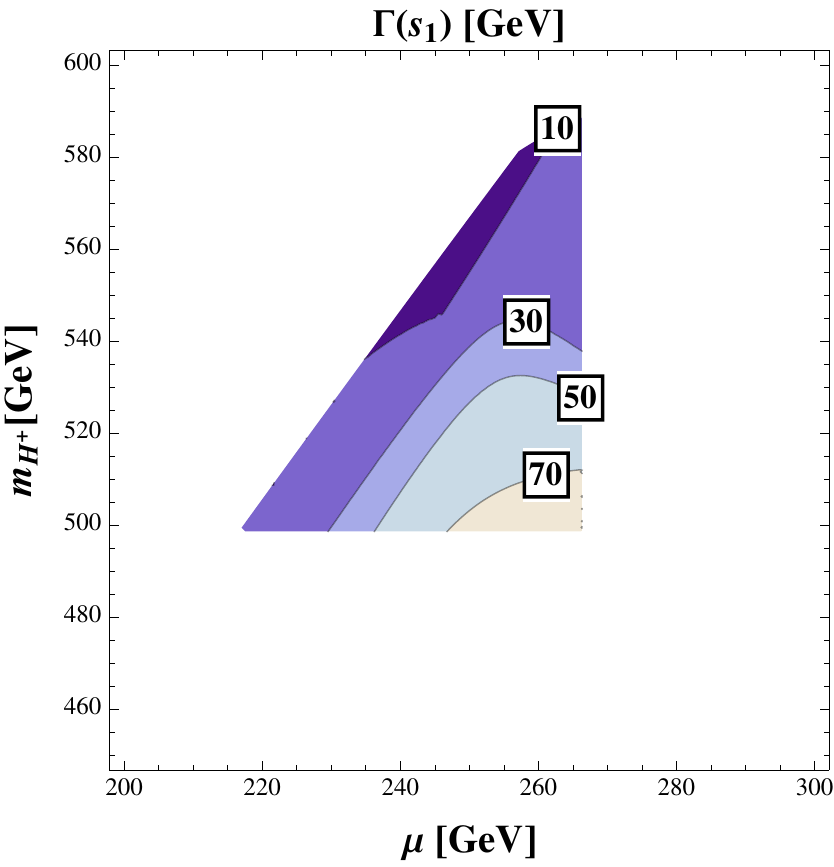}\includegraphics[width=0.5\linewidth]{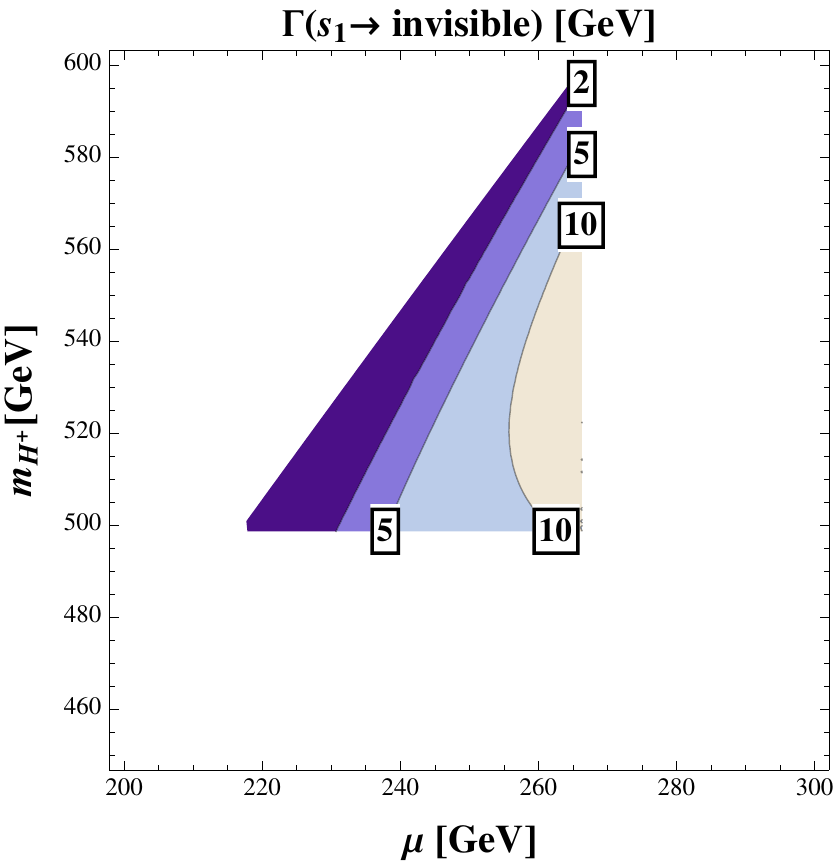}
\end{center}
\caption{ 
\label{gammaS1}\label{s1invisiblewidth} The total width (left) and the invisible width (right) of  the lightest CP-even Higgs boson in the region of parameters space compatible with direct constraints from LEP, dark matter searches and the stability of the potential. }
\end{figure}

The reduced rate of the resonant production of $W^{-}W^{+}$ through Higgs states that results from the lessening of branching fraction and the Higgs production cross section is the first evidence of the non-SM nature of the Higgs boson that we expect to show up at the LHC. The reduced production rate of vectors motivates the search of the Higgs boson in other final states. As show in Figure \ref{BRs1a1a1}, when kinematically accessible, the decay into pseudo-scalars $s_{1} \to a_{1} a_{1}$ is dominant. The process 
\beq
gg\to h \to a_{1}a_{1} \to \tau\bar{\tau}b\bar{b}\,,\label{higgsinpseudo}
\eeq
is potentially interesting to clarify the reason of the reduced production rate of resonant $W^{+}W^{-}$ pairs. However we observe that the decay into pseudo-scalar is usually dominant, especially where the SM-like decay into vectors is more suppressed. As such, we find that the Higgs boson could be observed earlier in the non-SM decay $s_{1}\to a_{1} a_{1}$. A search for the Higgs boson in the channel eq. (\ref{higgsinpseudo}) might well be in the reach of the first few $fb^{-1}$ of luminosity of LHC at 7 TeV center of mass energy.

\begin{figure}
\begin{center}
\includegraphics[width=0.5\linewidth]{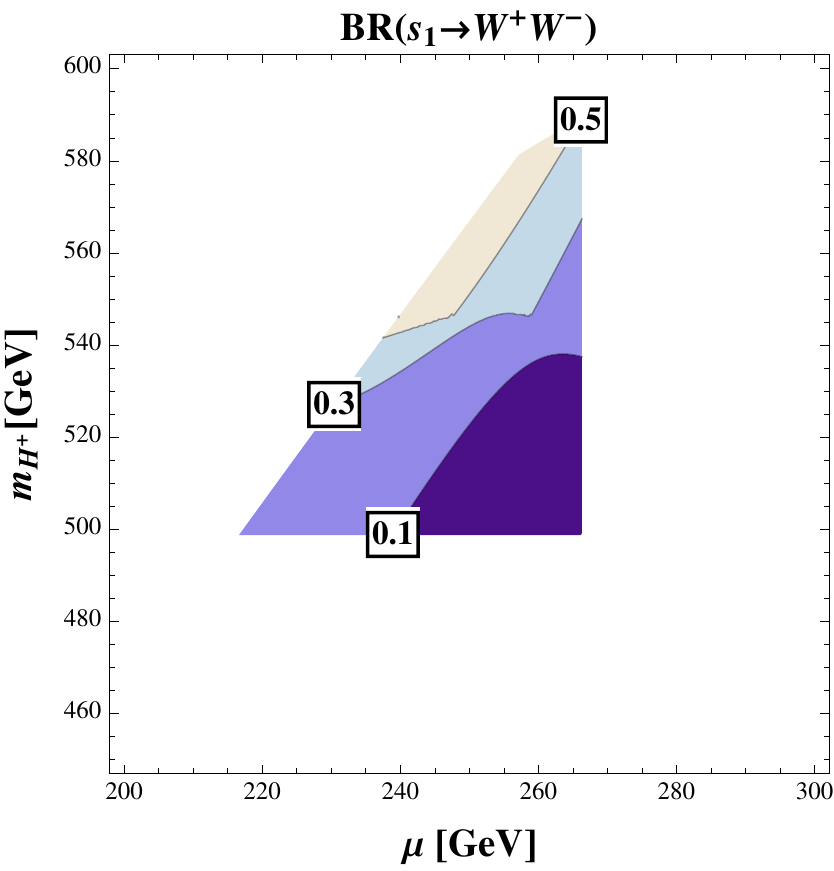}\includegraphics[width=0.5\linewidth]{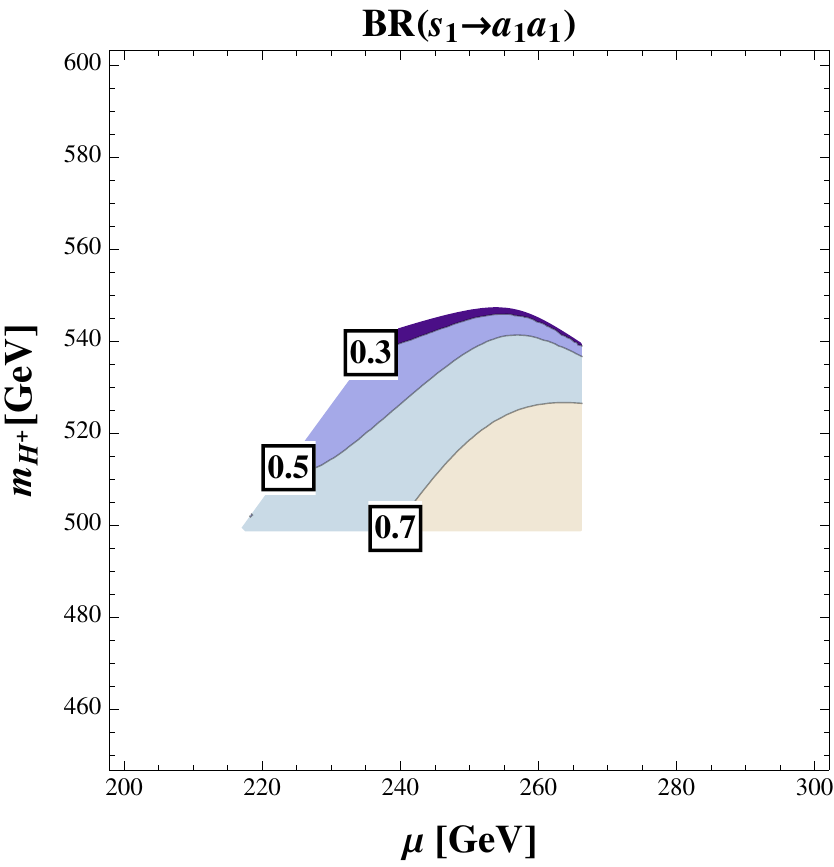}

\end{center}
\caption{ 
\label{BRs1WW} The branching fraction of  the lightest CP-even Higgs boson into $W^-W^+$ (left) and into $a_{1}a_{1}$ (right) in the region of parameters space compatible with direct constraints from LEP, dark matter searches and the stability of the potential. \label{BRs1a1a1}}
\end{figure}

Among the non-SM decay modes of the state $s_{1}$ the decay into two LSPs is particularly interesting because it results in an invisible decay of the Higgs boson of up to $10$ GeV,
 as shown in Figure \ref{s1invisiblewidth}. 
 The presence of a sizable invisible width is a sign of the large coupling $\lambda$ and therefore represents an indication of $\lambda$SUSY. The invisible width can be measured reconstructing the line-shape of the Higgs resonance using decay modes in clean final states as $s_{1}\to ZZ\to 4\ell$ and comparing the total width of the resonance with the widths of the observable channels. The measurement becomes easier when the decay into pseudo-scalars is absent as the branching fraction of the decay into $ZZ$ gets larger. As such the measurement of the invisible width helps to cover the portion of parameters space where the decay into pseudo-scalars is not available.

\bigskip

The states $s_2$ and $s_3$ are typically not heavy enough to decay into pairs of $s_1$, thus the results of previous studies contained in~\cite{lsusylhc} are not applicable to this case. We have seen in Figure \ref{reducedtt} that   $s_2$ and $s_3$ are relatively decoupled from SM vectors, therefore the relevant modes  are \beq s_{2,3} \to a_1 Z \textrm{ and } s_{2,3} \to a_1a_1\,,\label{s23a11}\eeq  and those into fermions. For the latter case we notice that the modes 

\beq s_{2,3}\to\chi_1\chi_2 \textrm{ and } s_{2,3} \to  t \bar{t} \label{s23tofermions}\eeq
are always kinematically allowed. However, because of the mass dependence of the partial widths and of the largeness of $\lambda$ and $k$, we expect  the decays involving $a_1$ of (\ref{s23a11}) to be dominant. 

\bigskip
The branching fractions and the total widths of the CP-even scalars computed with parameters fixed as in (\ref{point}) and (\ref{point2}) are given in Table \ref{br}. 
\bigskip

\begin{table}
\begin{center}

\begin{tabular}{|c|c|c|c|c|c|c|c|c|}
\hline 
 & $ZZ$ & $WW$ & $t\bar{{t}}$ & $\chi_{1}\chi_{1}$ &$\chi_{1}\chi_{2} $ & $a_{1}Z$ & $a_{1}a_{1}$ & $\Gamma$ {[}GeV{]} \tabularnewline
\hline
\hline 
$s_{1}$ & 0.088 & 0.196 & 0 & 0.090 & 0 & 0.059 & 0.568 & 30.3  \tabularnewline
\hline 
$s_{2}$ & 0.004 & 0.008 & 0.002 & 0.179 & 0.027 & 0.001 &0.782 & 33.6\tabularnewline
\hline 
$s_{3}$ & 0.023 & 0.047 & 0.039 & 0.461 & 0.013 & 0.165 & 0.255 & 48.2\tabularnewline
\hline
\end{tabular}
\end{center}
\caption{The branching fractions and the total widths of the CP-even scalars computed with parameters fixed as in eqs. (\ref{point}) and (\ref{point2}).\label{br}}

\end{table}

Due to the heaviness of the charged Higgs boson $H^+$, the only two-body decay mode available for the chargino is 
$$\chi^+ \to \chi_1 W^+\,.$$ 
Similarly, the fact that $m_{s_1}\gtrsim m_{\chi_2}$ implies that  the only relevant modes for $\chi_2$ are the two-body decays 
$$\chi_2 \to \chi_1 Z \textrm{ and } \chi_2 \to \chi_1 a_1\,.$$

The situation for $a_2$ is more involved as it has all the following decay channels: $a_2 \to s_j a_1$, $a_2 \to Z s_k$, $a_2 \to f_{SM} \bar{f}_{SM}$ , $a_2 \to \chi_i \chi_j$. Among the available decay modes, more than one involves large couplings and/or large final state multiplicities. Hence, a detailed computation of the partial width is needed to determine which channel actually dominates.
 
\section{Conclusions} \label{conclusions}
 In this work, pursuing as much as possible an analytic approach, we closely examined the low energy phenomenology of a supersymmetric model with a strong
 coupling phase in the Higgs sector at some scale $\sim (10-100) \textrm{ TeV}$. The presence of a strong coupling phase allows large couplings in the effective superpotential at the weak scale and helps to increase the tree-level mass of the Higgs boson, thus relaxing the tension between direct LEP searches and the requirement of a little level of fine-tuning of the model.  

The model under study is defined by the superpotential  (\ref{superpotential}), which is a restriction without dimensionful parameters of the more general form (\ref{generalNMSSM}). Models of the more general form have been shown to be realized as low energy effective superpotentials arising from UV complete models that are compatible with the unification of gauge couplings~\cite{nomura}. In this work we have shown that in the model defined by (\ref{superpotential}) one can address the so called $\mu$ problem generating  dynamically the $\mu$-term for the mass of the Higgsinos with a value of order $m_Z$ in the same way as it is realized in the perturbative NMSSM. This, however, is non-trivial because the model under study  is in a completely different r\'egime characterized by large couplings $\lambda$ and $k$.

\bigskip

Our quantitative study of the generation of  $\mu$ shows that in this model the size of the $\mu$ term is connected to the mass of the lightest Higgs boson such that
\beq\label{eq:boundlightestHmu}
\frac{1}{\sqrt{2}}m_{s_1} \lesssim \mu \lesssim \frac{3}{2}m_{s_1}\,.
\eeq

The existence of such relation is welcome and not surprising. In fact, the generation of $\mu$ is a EWSB effect and the characteristic scale of the scalar sector responsible for the EWSB is $\lambda v$, that is essentially controlling the Higgs mass. Hence the model is in a r\'egime where the $\mu$ term is generated always in connection with the mass of a heavy lightest Higgs and thus it is natural to have a $\mu$ term not too small compared to LEP direct searches though still close to the weak scale.  
We also studied the fine-tuning of the model in particular in those regions where $\mu$ is phenomenologically acceptable, by mean of the sensitivity of the VEV of the Higgs doublets to variations of the fundamental parameters and we found that the model requires the least fine-tuning when the mass of the lightest Higgs is large.  Altogether we find that making the requirement of a heavy lightest Higgs one also gets a phenomenologically viable solution of the $\mu$ problem with low fine-tuning.

\bigskip

In the second part of the paper, we checked the model against a great deal of experimental constraints and we found that the mass of the lightest Higgs plays an essential r\^ole. Some regions of the parameter space are already excluded by current data, but most of the more ``natural'' parameter space have still to be explored. The experimental situation can be summarized as follows.

As far as the spectrum is concerned, the possibility of having a large quartic term $\lambda$ in the tree-level potential results in the fact that the Higgs bound from LEP can be easily satisfied, in particular for low $\tan\beta$. The relation between the lightest Higgs and the chargino masses eq. (\ref{eq:boundlightestHmu}) explains  why the mass of the chargino  is generically above direct searches. The lightest pseudo-scalar and the lightest neutralino still have significant dependence on the choice of $k$. A small $k$ would result in a tiny breaking of the PQ symmetry and thus in a   ``split'' spectrum with a very light state in the CP-odd sector and all the rest of the scalars much heavier. On the other hand, a large coupling $k$ would give a more coherent picture with a spectrum made entirely of heavy states. We pursued this latter possibility obtaining  pseudo-scalars that easily fulfill LEP bounds.  On the contrary, in the large $k$ regime LEP limits on the lightest neutralino impose more severe bounds to the parameter space. 

Great care has been devoted to the study of the EWPTs, with the result that the larger contributions to $S$ and $T$ come from the Higgsino/Higgs sector. The model  can be in good agreement with precision data and has a preference for large values of $k$ and small values of $\tan\beta$ that give rise to a heavy lightest CP-odd and a heavy lightest CP-even Higgs, respectively, and altogether support once more the idea of a spectrum entirely made of heavy scalars.

We also studied the thermal production of the lightest neutralino and we found that generically the abundance of relic neutralinos is too low to explain the whole dark matter in the Universe. Notable exceptions are the regions of parameter space where $m_{\chi_1}\lesssim m_Z$ and $\tan\beta\lesssim 1.5$, that can give relic abundance within the interval given by WMAP. These regions can be further probed considering direct searches of WIMPs. It turns out that it is generically not possible to account for the dark matter in the Universe without violating WIMP searches limits. On the contrary when the relic abundance of neutralinos is less that the WMAP observation one can be in agreement with WIMP searches when the mass of the lightest Higgs is above $200 \gev$. This corresponds to cut away almost all the parameter space where $\lambda\lesssim 1.5$ and give further support to the idea of a heavy lightest Higgs boson and a large $\lambda$.

The experimental bounds  are summarized in  Figure \ref{tutto20} where the limits from direct searches are shown in the plane $\mu\,,\,m_{H^+}$. The Figure also shows the contribution of the Higgsinos to $S$ and $T$, which can be taken as representative of the overall contribution of our model to the EWPTs. 
Altogether we find that  for $\lambda\simeq 1.5$ the limits from WIMP searches  and LEP rule out most of the parameter space. For larger values of $\lambda$, WIMP searches allow a larger fraction of the parameter space, but limits from LEP  still give significant bounds; in particular the limit on $\chi_1$ reduces the acceptable range of $\mu$ and the EWPTs require $\tan\beta\lesssim1.5$.

We did not discuss the issues connected to flavour physics, however we expect them to not pose any problem. For example one can estimate the contribution to  $b\to s \gamma$ and realize that is rather generic to have small contributions beyond those of the SM. As consequence of our assumption of diagonal squark mass matrices, the only new contributions to $b\to s\gamma$ would be the loop of chargino-stop  and the loop of charged Higgs and top.
These two diagrams must cancel in the supersymmetric limit~\cite{Ferrara:1974wb} but even barring the possibility of a cancellation we expect the arising contributions to be small. Indeed, the particles involved in the loops are in general relatively heavy (see the Naturalness limits on those masses in Section \ref{FT}), and, in addition, the value of $\tan\beta$ suitable for our model is quite small.
Indeed one can estimate this contribution taking the limit of large charged Higgs mass such that only the chargino loop is left. Given the small values of $\tan\beta$ that we consider the resulting contribution is always small compared to the SM loop of $W$ and top.
A recent  analysis~\cite{bsgamma} shows that for the values of $\mu\sim 200 \gev$ of interest for us is enough to take the lightest stop mass above 400 GeV to be in good agreement with the SM prediction even in the large $\tan\beta$ case. As discussed in Section \ref{FT} such values of the mass of the stop are perfectly natural in our model.

\bigskip

Finally, we want to point out the future possibilities to probe experimentally our model. Firstly,
future experiments on WIMP direct detection have  the potential to probe largely the parameter space of our model. Furthermore new experiments will clarify the nature of current low statistical significance claims of the observation of WIMP scattering out of a handful of events. In view of further clarification from forthcoming experiments we limit ourselves to report that our model has regions of the parameter space where the mass of the LSP and its cross section with the proton do agree with the rates and recoil energies of these claims.

Secondly, the other main candidate to probe the model is the LHC. Naturalness arguments allowed us to estimate that scalar top partners and gluinos might well be in the 1-2 TeV range and are therefore in the mid-term reach of LHC. There are no constraints which forbid gluinos and stops as light as the reach at the 2010-2011 run of LHC, but there is no preference for such lightness. This is the counterpart of raising the Higgs mass at the tree-level.

The scalar sector has light states that might well be in the reach of early LHC runs. The lightest CP-even state is substantially different from a SM Higgs boson. In a large fraction of the parameters space it decays into $WW$ and $ZZ$ only subdominatly, while it decays dominantly in a pair of pseudo-scalars $a_{1}a_{1}$, when kinematically possible, or to a pair of LSPs. It seems optimistic to reach a discovery in the standard channels into $WW$ and $ZZ$ within the scheduled 2010-2011 run. Indeed the reduced production rate of resonant $W^{-}W^{+}$ pairs is the first signal of the non-standard nature of the Higgs boson that we expect to show up in standard searches. 
Rescaling the cross-section from Ref.~\cite{haa14} for the LHC at 7 TeV center of mass energy, we estimate that at least a hint of the process
\beq pp\to s_1 \to a_1 a_1 \,,\eeq
that results in final states of the type $b\bar{b}\tau\bar{\tau}$, $4\tau$,$4b$ \cite{haa7} might be visible with few 1/fb of integrated luminosity at the 2010-2011 run.  

Furthermore, a sizable invisible decay width of the Higgs boson is a further signal of the presence of non-SM decay modes, as for instance the decay into two LSPs.

It is also rather generic to have a significant production of the light CP-odd with main decay mode into $b\bar{b}$ pairs. However, this signal seems difficult to observe  because of the large QCD background of $b\bar{b}$. As such, it seems more promising either to search for the light CP-odd boson in decay chains like those of the lightest CP-even Higgs,  or to look for cleaner but suppressed decay modes like the one into $\tau\bar{\tau}$.

In the long-term run one can search for the remaining states of the CP-even sector, that can be all produced significantly at the LHC via gluon fusion and decay in detectable final states. Also the fermionic states of the neutralino and chargino sector are in the reach of a longer run of LHC. Their discovery and the determination of their mass is a crucial step to test the model, as the mass parameter $\mu$ can be determined directly from their spectrum. These states can be searched in final states with on-shell or off-shell gauge bosons and Higgs bosons and missing transverse energy.  The long-term run of LHC should allow for the search of other heavy sates like the charged Higgs boson and the heavy CP-odd.

\begin{figure}
\begin{center}
\subfigure[]{\includegraphics[width=.8\textwidth]{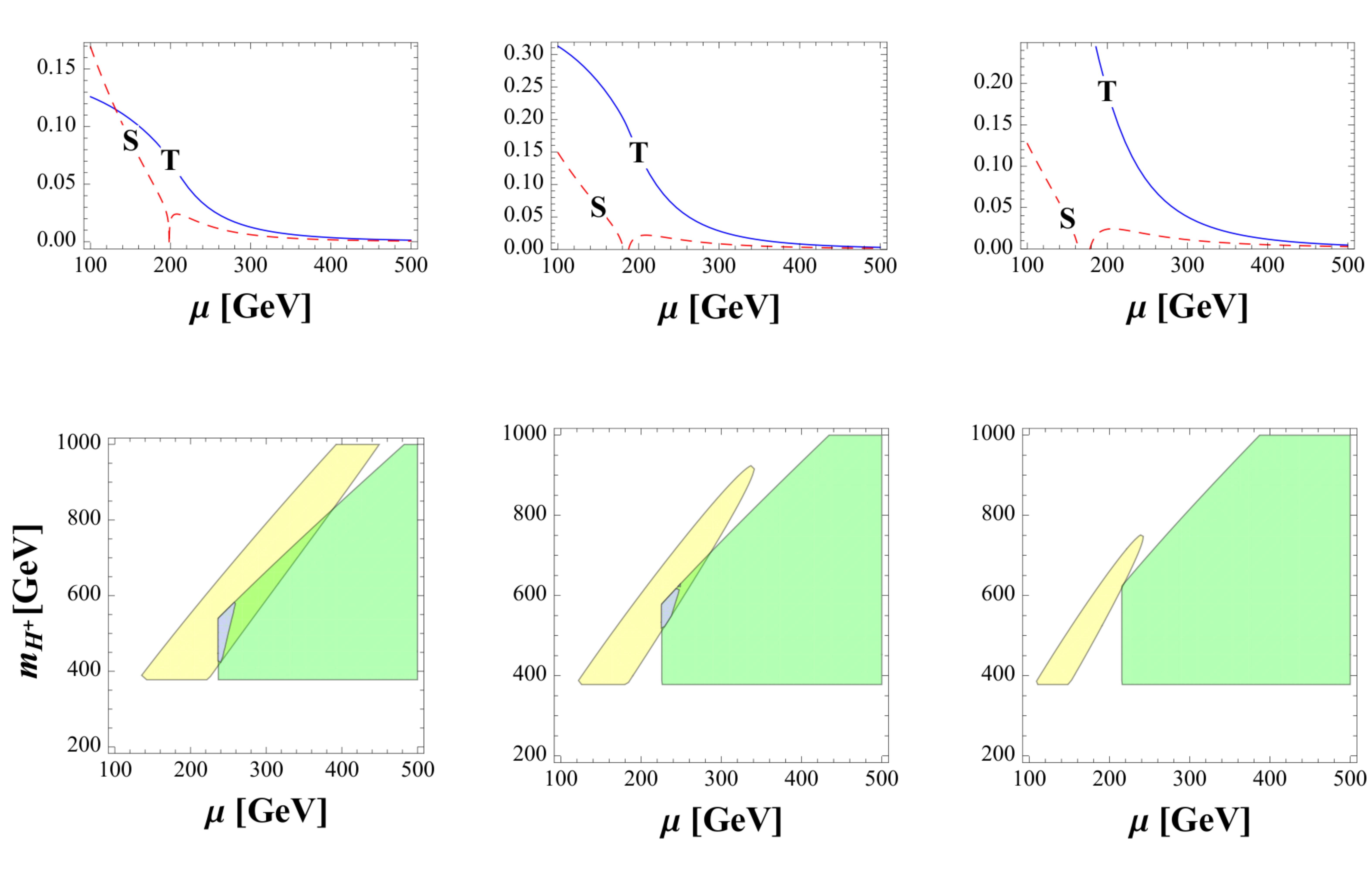}}
\\
\subfigure[]{\includegraphics[width=.8\textwidth]{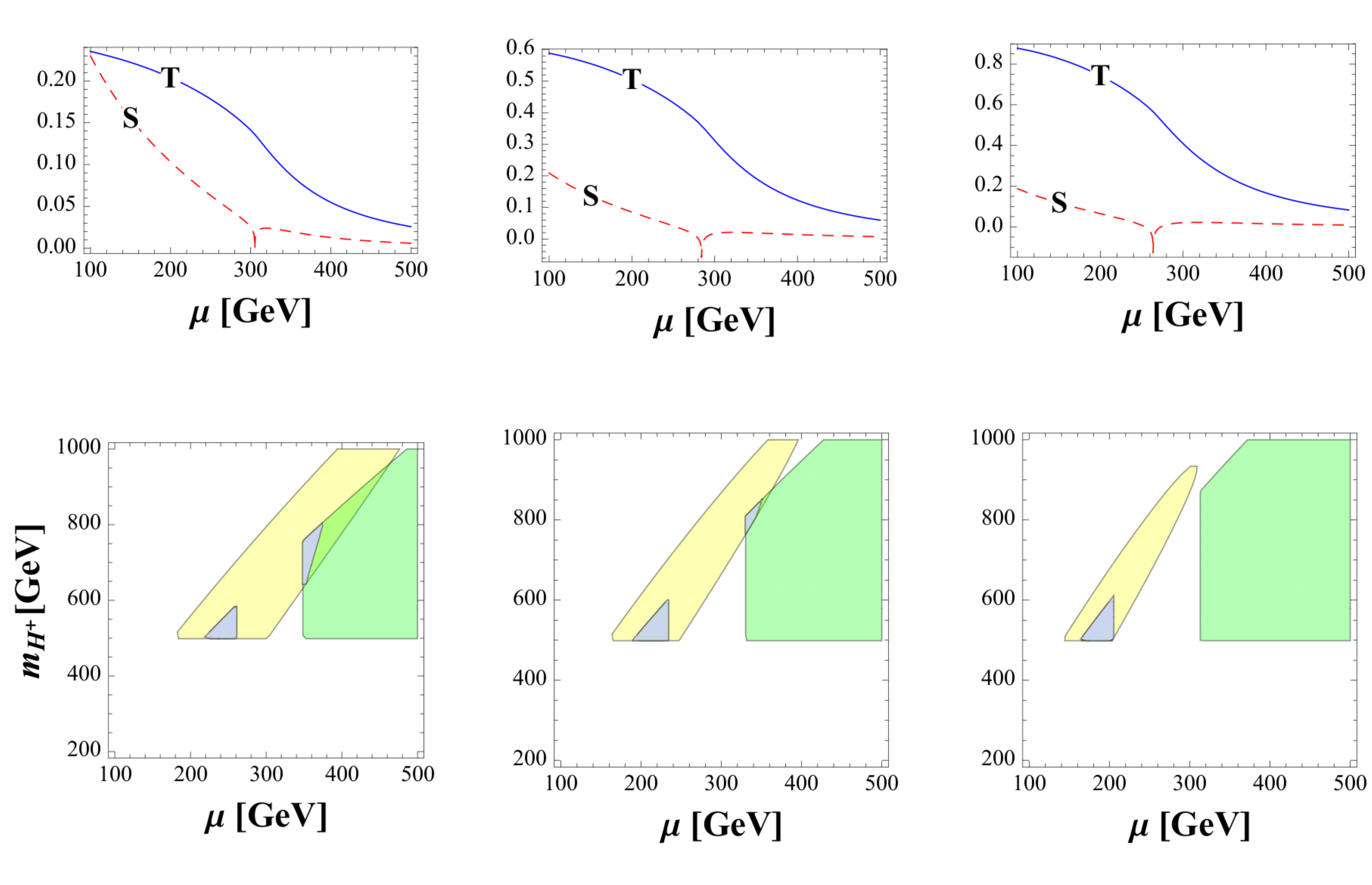}}
\end{center}
\caption{
Summary of the experimental constraints for $\lambda=1.5$ (a)  and $\lambda=2$ (b). In both the panels (a) and (b) the lower row shows the constraints on the plane $\mu$, $m_{H^+}$ for fixed $k=1.2$ and $\tan\beta$ equal to 1.5, 2 and 2.5 from left to right.  The green area corresponds to all the LEP constraints and all the stability constraints except that for the CP-even scalars, which is represented separately by the yellow area. The blue area corresponds to a DM-nucleon interaction rate below current bounds. 
In the upper row we show the contributions to  $S$ (dashed line) and $T$ (solid line) coming from the Higgsino sector as a function of $\mu$ for the same choices of parameters.
 }\label{tutto20}
\end{figure}

\subsubsection*{Acknowledgements}
We thank Riccardo Barbieri for inspiring this work and for reading the manuscript. We  thank Riccardo Rattazzi for discussions and for reading the manuscript. We also thank Vyacheslav S. Rychkov for reading the manuscript and for suggestions.  We thank W. Altmannshofer,  D. Pappadopulo and A.Wulzer for useful discussions.

SG thanks the Galileo Galilei Institute for Theoretical Physics for the hospitality during the completion of this work and acknowledges support by the European Community's Marie Curie Research Training Network under contract MRTN-CT-2006-035505 [``HEP-TOOLS'']. RF thanks the Galileo Galilei Institute for Theoretical Physics for hospitality during the workshop "Searching for New Physics at the LHC". The work of RF is supported by the Swiss National Science Foundation under contract No. 200021-116372
\begin{appendix}

\section{One loop contributions to $S$ and $T$}\label{Appendix}
We collect here the one loop functions used through the paper. 

For a boson loop with internal masses
$m_1$ and $m_2$ and coupling the the gauge boson $W_\mu$ given by $iW_\mu\phi_1^*\partial_\mu\phi_2$, we have

\begin{eqnarray}
2 \alpha_{\rm em} v^{2} A(m_{1},m_{2})&\equiv&\frac{1}{16\pi^{2}} \biggl[ \frac{m_{1}^{2}+m_{2}^{2}}{2}
    - \frac{m_{1}^{2}m_{2}^{2}}{m_{1}^{2}-m_{2}^{2}}
      \ln\frac{m_{1}^{2}}{m_{2}^{2}} \biggr]\,,\\
\frac{1}{4\pi} F(m_{1},m_{2})&\equiv&\frac{1}{96\pi^{2}} \biggl[ -\ln\frac{\Lambda^{4}}{m_{1}^{2}m_{2}^{2}}
    + \frac{4m_{1}^{2}m_{2}^{2}}{(m_{1}^{2}-m_{2}^{2})^{2}}+\nonumber\\
  & +& \frac{m_{1}^{6}+m_{2}^{6}-3m_{1}^{2}m_{2}^{2}(m_{1}^{2}+m_{2}^{2})}
      {(m_{1}^{2}-m_{2}^{2})^{3}} \ln\frac{m_{1}^{2}}{m_{2}^{2}} \biggr]\,.
\end{eqnarray}

Additionally, also the diagram in which the gauge boson $W_\mu$ and the Higgs boson $\phi$ propagate in the 
loop contributes to the parameter $S$, giving as loop function

\begin{equation}
  G(m_{1},m_{2}) \equiv \frac{1}{2\pi}
    \left[ \frac{2m_{1}^{2}m_{2}^{2}}{(m_{1}^{2}-m_{2}^{2})^{3}}
      \ln\frac{m_{1}^{2}}{m_{2}^{2}}
    - \frac{m_{1}^{2}+m_{2}^{2}}{(m_{1}^{2}-m_{2}^{2})^{2}} \right].
\end{equation}

For a fermion loop with internal masses $m_1$ and $m_2$ and a vector coupling $W_\mu \bar\psi_i\gamma^\mu\psi_2$, we have

\begin{eqnarray}
  2 \alpha_{\rm em} v^{2} \tilde{A}(m_{1},m_{2}) &\equiv& \frac{1}{16\pi^{2}} \biggl[ (m_{1}-m_{2})^{2}\ln\frac{\Lambda^{4}}
      {m_{1}^{2}m_{2}^{2}} - 2m_{1}m_{2}+
\nonumber\\
  & +& \frac{2m_{1}m_{2}(m_{1}^{2}+m_{2}^{2})-m_{1}^{4}-m_{2}^{4}}
      {m_{1}^{2}-m_{2}^{2}} \ln\frac{m_{1}^{2}}{m_{2}^{2}} \biggr]\,,
\\
 \frac{1}{4\pi}\tilde{F}(m_{1},m_{2}) &\equiv&
  \frac{1}{24\pi^{2}} \biggl[ -\ln\frac{\Lambda^{4}}{m_{1}^{2}m_{2}^{2}}
    - \frac{m_{1}m_{2}(3m_{1}^{2}-4m_{1}m_{2}+3m_{2}^{2})}
      {(m_{1}^{2}-m_{2}^{2})^{2}}+
\nonumber\\
  & +& \frac{m_{1}^{6}+m_{2}^{6}-3m_{1}^{2}m_{2}^{2}(m_{1}^{2}+m_{2}^{2})
      +6m_{1}^{3}m_{2}^{3}}{(m_{1}^{2}-m_{2}^{2})^{3}} 
      \ln\frac{m_{1}^{2}}{m_{2}^{2}} \biggr] \,.
\end{eqnarray}

Differently, for an axial coupling, the result can be obtained by letting $m_{1} \rightarrow 
-m_{1}$.  These expressions are valid for both Dirac and Majorana fermions, 
with an extra factor of 2 in the case of identical Majorana fermions.

\section{Renormalization group running at two loops}\label{sec:RGE}

At the two loop level, the mass of the gluino $m_{\tilde{g}}$ contributes to the renormalization group running of the Higgs soft mass $m_2$. In this Appendix we collect the main equations and assumptions used in Section \ref{sec:boundmasses} to get eq. (\ref{eq:deltam2gluino}) for the running of $m_2$~\cite{Martin:1993zk}.

At two loops, the main dependence of $m_2$ on the gluino mass is given by 

\begin{equation}\label{eq:RGEm2}
\frac{dm_2^2}{dt}\sim\frac{3}{8\pi^2}\left(\lambda_t^2(m_{\tilde{Q}}^{2}+m_{\tilde{t}_{R}}^{2})+h_t^2\right)+\frac{1}{8\pi^4}g_s^2m_{\tilde{g}} \left(2\lambda_t^2 m_{\tilde{g}}-\lambda_t^\dagger h_t-\lambda_t h_t^\dagger\right)\,.
\end{equation}

The second term is a purely two loop contribution directly dependent on the gluino mass, instead, in the first (one loop) term, we have dependence on the squark masses of third generation and on the third generation trilinear term $h_t$. At one loop, they are related to the mass of the gluino through the equations

\begin{eqnarray}\label{eq:runmqmt}
\frac{d m_{\tilde{Q}}^{2}}{dt}\sim\frac{d m_{\tilde{t}_{R}}^{2}}{dt}&\sim& -\frac{2}{3\pi^2}\, g_s^2\, m_{\tilde{g}}^2\,,\\
\frac{d h_t}{dt}&\sim& \frac{2}{3\pi^2}\,g_s^2 \,\lambda_t \,m_{\tilde{g}}\,,
\end{eqnarray}

\noindent that can be solved, using the leading log approximation. Assuming then that at the high scale $h_t$ is small, if compared to its running, then it is easy to prove that numerically the most relevant contributions in (\ref{eq:RGEm2}) are those not involving the trilinear term $h_t$.

Replacing then, inside the RGE for $m_2$ the squark masses, obtained integrating the differential equation (\ref{eq:runmqmt}), one can finally prove eq. (\ref{eq:deltam2gluino}) for the running of $m_2$ due to the gluino mass.

\end{appendix}

\section*{Addendum}
After that this paper was accepted for publication, Ref.~\cite{Kanehata:2011ei}  appeared and showed the possible appearance of absolute minima with a structure of the VEVs different than what considered in this paper. Generically the extremal points discussed in Ref.~\cite{Kanehata:2011ei} can be deeper than the electroweak breaking minimum discussed in our text  in significantly large regions of the parameter space of the NMSSM, however in our regime the potential evaluated at our extremal point is safely deeper than in most of the extremal points considered in~\cite{Kanehata:2011ei}.

The only exceptions are the extremal points with vanishing VEV of the Higgs doublets and non vanishing VEV for the singlet that can become deeper than our extremal point in sizable regions of the parameter space. Around these minima the electroweak symmetry is not broken and therefore the points of parameter space where these minima are the absolute minimum should be discarded ~\footnote{
One could still keep these points if the $SU(2)$ breaking vacuum is metastable with a life-time longer than the age of the Universe.}.

Imposing on our parameter space that the minima discussed in Ref.~\cite{Kanehata:2011ei} are not deeper than the one that  correctly breaks the $SU(2)$ symmetry,
we observe that for $k\simeq 1$ or larger this constraint only mildly affects the parameter space, cutting 
off only regions where the Higgs boson mass is not far from the LEP limit.  
According 
to the results of Section \ref{FT} and Section \ref{directdetection} the regions excluded by this constraint are 
the most fine-tuned and are typically in tension with the bounds from 
direct detection of the dark matter.
\bigskip

After that this paper was accepted for publication a new analysis of our same 
superpotential, but with $k<0$, has been presented in Ref.~\cite{bertuzzo}. 
They find that the constraint found in~\cite{Kanehata:2011ei} is much less restrictive 
for $k<0$ than for $k>0$, in particular there is acceptable parameter space for both small and 
large $k$.  The conclusions of Ref.~\cite{bertuzzo} on the expected 
phenomenology of a scale invariant $\lambda$SUSY for $k<0$ substantially agree 
with our results obtained for $k>0$.


\begin{thebibliography}{9}



\bibitem{LEPhiggs}
  R.~Barate {\it et al.}  [LEP Working Group for Higgs boson searches and
                  ALEPH Collaboration and  and],
  Phys.\ Lett.\  B {\bf 565}, 61 (2003)
  \linkart[hep-ex/0306033].


\bibitem{EFTBMSSM}
 A.~Brignole, J.~A.~Casas, J.~R.~Espinosa and I.~Navarro,
  Nucl.\ Phys.\  B {\bf 666} (2003) 105
  [arXiv:hep-ph/0301121].
    M.~Dine, N.~Seiberg and S.~Thomas,
  Phys.\ Rev.\  D {\bf 76}, 095004 (2007)
  [arXiv:0707.0005 [hep-ph]].
  M.~Carena, K.~Kong, E.~Ponton and J.~Zurita,
  Phys.\ Rev.\  D {\bf 81} (2010) 015001
  [arXiv:0909.5434 [hep-ph]].
  I.~Antoniadis, E.~Dudas, D.~M.~Ghilencea and P.~Tziveloglou,
  Nucl.\ Phys.\  B {\bf 831}, 133 (2010)
  [arXiv:0910.1100 [hep-ph]].
  I.~Antoniadis, E.~Dudas, D.~M.~Ghilencea and P.~Tziveloglou,
  Nucl.\ Phys.\  B {\bf 808}, 155 (2009)
  [arXiv:0806.3778 [hep-ph]].
  I.~Antoniadis, E.~Dudas, D.~M.~Ghilencea and P.~Tziveloglou,
  AIP Conf.\ Proc.\  {\bf 1078}, 175 (2009)
  [arXiv:0809.4598 [hep-ph]].
 I.~Antoniadis, E.~Dudas and D.~M.~Ghilencea,
  JHEP {\bf 0803}, 045 (2008)
  [arXiv:0708.0383 [hep-th]].

\bibitem{Tobe:2002zj}
  K.~Tobe and J.~D.~Wells,
  Phys.\ Rev.\  D {\bf 66}, 013010 (2002)
  \linkart[hep-ph/0204196].



\bibitem{Espinosa:1998re}
  J.~R.~Espinosa and M.~Quiros,
  Phys.\ Rev.\ Lett.\  {\bf 81}, 516 (1998)
  \linkart[hep-ph/9804235].

\bibitem{newvectors}
  S.~P.~Martin,
  Phys.\ Rev.\  D {\bf 81}, 035004 (2010)
  [\linkart[0910.2732] [hep-ph]].

  K.~S.~Babu, I.~Gogoladze and C.~Kolda,
  \linkart[hep-ph/0410085].

\bibitem{dterms}
  A.~Maloney, A.~Pierce and J.~G.~Wacker,
  JHEP {\bf 0606}, 034 (2006)
  [\linkart[hep-ph/0409127]].






\bibitem{SU2nondecDterms}
  P.~Batra, A.~Delgado, D.~E.~Kaplan and T.~M.~P.~Tait,
  JHEP {\bf 0402}, 043 (2004)
  [arXiv:hep-ph/0309149].

\bibitem{Barbieri:2010pd}
  R.~Barbieri, E.~Bertuzzo, M.~Farina, P.~Lodone and D.~Pappadopulo,
  arXiv:1004.2256 [hep-ph].

\bibitem{Lodone:2010kt}
  P.~Lodone,
  arXiv:1004.1271 [hep-ph].

\bibitem{fathiggs}
  R.~Harnik, G.~D.~Kribs, D.~T.~Larson and H.~Murayama,
  Phys.\ Rev.\  D {\bf 70}, 015002 (2004)
  \linkart[hep-ph/0311349];
  S.~Chang, C.~Kilic and R.~Mahbubani,
  Phys.\ Rev.\  D {\bf 71}, 015003 (2005)
  \linkart[hep-ph/0405267];
    A.~Delgado and T.~M.~P.~Tait,
  JHEP {\bf 0507}, 023 (2005)
  \linkart[hep-ph/0504224].


\bibitem{lUV}
  A.~Birkedal, Z.~Chacko and Y.~Nomura,
  Phys.\ Rev.\  D {\bf 71}, 015006 (2005)
  \linkart[hep-ph/0408329].


\bibitem{BarbieriPQ}
  R.~Barbieri, L.~J.~Hall, A.~Y.~Papaioannou, D.~Pappadopulo and V.~S.~Rychkov,
  JHEP {\bf 0803}, 005 (2008)
  \linkart[0712.2903] [hep-ph].

\bibitem{lambdamax}
  M.~Masip, R.~Munoz-Tapia and A.~Pomarol,
  Phys.\ Rev.\  D {\bf 57}, R5340 (1998)
  \linkart[hep-ph/9801437].

\bibitem{lsusy}
  R.~Barbieri, L.~J.~Hall, Y.~Nomura and V.~S.~Rychkov,
  Phys.\ Rev.\  D {\bf 75}, 035007 (2007)
  \linkart[hep-ph/0607332].


\bibitem{lsusylhc}
  L.~Cavicchia, R.~Franceschini and V.~S.~Rychkov,
  Phys.\ Rev.\  D {\bf 77}, 055006 (2008)
  \linkart[0710.5750] [hep-ph].
  
\bibitem{DGP}
  G.~R.~Dvali, G.~F.~Giudice, A.~Pomarol,
  Nucl.\ Phys.\  {\bf B478}, 31-45 (1996).
  [hep-ph/9603238].

\bibitem{musol} 
  M.~Dine and A.~E.~Nelson, ``\emph{Dynamical supersymmetry breaking at low-energies}'',
  Phys.\ Rev.\  D {\bf 48}, 1277 (1993), \href{http://arXiv.org/pdf/hep-ph/9303230}{\it arXiv:hep-ph/9303230}.
    M.~Dine, A.~E.~Nelson, Y.~Nir and Y.~Shirman, ``\emph{New tools for low-energy dynamical supersymmetry breaking}'',
  Phys.\ Rev.\  D {\bf 53}, 2658 (1996), \href{http://arXiv.org/pdf/hep-ph/9507378}{\it arXiv:hep-ph/9507378}.
    A.~de Gouvea, A.~Friedland and H.~Murayama, ``\emph{Next-to-minimal supersymmetric standard model with the gauge mediation  of supersymmetry breaking}'',
  Phys.\ Rev.\  D {\bf 57}, 5676 (1998), \href{http://arXiv.org/pdf/hep-ph/9711264}{\it arXiv:hep-ph/9711264}.
    Z.~Chacko and E.~Ponton, ``\emph{Yukawa deflected gauge mediation}'',
  Phys.\ Rev.\  D {\bf 66}, 095004 (2002), \href{http://arXiv.org/pdf/hep-ph/0112190}{\it arXiv:hep-ph/0112190}.
%
   T.~S.~Roy and M.~Schmaltz, ``\emph{A hidden solution to the $\mu/B_\mu$ problem in gauge mediation}'',
  Phys.\ Rev.\  D {\bf 77}, 095008 (2008), \href{http://arxiv.org/abs/arXiv:0708.3593}{\it arXiv:0708.3593}.
  H.~Murayama, Y.~Nomura and D.~Poland, ``\emph{More Visible Effects of the Hidden Sector}'',
  Phys.\ Rev.\  D {\bf 77}, 015005 (2008), \href{http://arxiv.org/abs/arXiv:0709.0775}{\it arXiv:0709.0775}.
  G.~F.~Giudice, H.~D.~Kim and R.~Rattazzi, ``\emph{Natural mu and $B_\mu$ in gauge mediation}'',
  Phys.\ Lett.\  B {\bf 660}, 545 (2008), \href{http://arxiv.org/abs/arXiv:0711.4448}{\it arXiv:0711.4448}.

\bibitem{csakietal}
C.~Csaki, A.~Falkowski, Y.~Nomura and T.~Volansky, ``\emph{New Approach to the $\mu-B_\mu$ Problem of Gauge-Mediated Supersymmetry Breaking}'',
  Phys.\ Rev.\ Lett.\  {\bf 102} (2009) 111801, \href{http://arxiv.org/abs/arXiv:0809.4492}{\it arXiv:0809.4492}.
  
  
\bibitem{GiuRa}
  G.~F.~Giudice, R.~Rattazzi,
  Phys.\ Rept.\  {\bf 322}, 419-499 (1999).
  [hep-ph/9801271].
  
  \bibitem{GiuMa}
  G.~F.~Giudice, A.~Masiero,
  Phys.\ Lett.\  {\bf B206}, 480-484 (1988).
  
  \bibitem{DGS}
       A.~Delgado, G.~F.~Giudice and P.~Slavich, ``\emph{Dynamical $\mu$ Term in Gauge Mediation}'',
  Phys.\ Lett.\  B {\bf 653}, 424 (2007), \href{http://arxiv.org/abs/arXiv:0706.3873}{\it arXiv:0706.3873}.
  
  


  
  


\bibitem{CY}
  J.~Cao and J.~M.~Yang,
  Phys.\ Rev.\  D {\bf 78}, 115001 (2008)
  \linkart[0810.0989 [hep-ph].


\bibitem{review}
  U.~Ellwanger, C.~Hugonie and A.~M.~Teixeira,
  \linkart[0910.1785].

\bibitem{romao}
  J.~C.~Romao,
  Phys.\ Lett.\  B {\bf 173}, 309 (1986).


\bibitem{RGE}
  J.~P.~Derendinger and C.~A.~Savoy,
  Nucl.\ Phys.\  B {\bf 237}, 307 (1984);
  N.~K.~Falck,
  Z.\ Phys.\  C {\bf 30}, 247 (1986);
    R.~B.~Nevzorov and M.~A.~Trusov,
  Phys.\ Atom.\ Nucl.\  {\bf 64}, 1513 (2001)
  [Yad.\ Fiz.\  {\bf 64}, 1589 (2001)]
  \linkart[hep-ph/0112301].

\bibitem{largeS}
  U.~Ellwanger, M.~Rausch de Traubenberg and C.~A.~Savoy,
  Phys.\ Lett.\  B {\bf 315}, 331 (1993)
  \linkart[hep-ph/9307322];
  U.~Ellwanger, M.~Rausch de Traubenberg and C.~A.~Savoy,
  Nucl.\ Phys.\  B {\bf 492}, 21 (1997)
  \linkart[hep-ph/9611251].

\bibitem{Barbieri:1987fn}
  R.~Barbieri and G.~F.~Giudice,
  Nucl.\ Phys.\  B {\bf 306} (1988) 63.


\bibitem{FT}
  M.~Bastero-Gil, C.~Hugonie, S.~F.~King, D.~P.~Roy and S.~Vempati,
  Phys.\ Lett.\  B {\bf 489}, 359 (2000)
  \linkart[hep-ph/0006198];
  P.~C.~Schuster and N.~Toro,
  \linkart[hep-ph/0512189].
  
    \bibitem{Martin:1993zk}
  S.~P.~Martin and M.~T.~Vaughn,
  Phys.\ Rev.\  D {\bf 50} (1994) 2282
  [Erratum-ibid.\  D {\bf 78} (2008) 039903]
  \linkart[hep-ph/9311340].
  
  
  
\bibitem{pdg}
  C.~Amsler {\it et al.}  [Particle Data Group],
  Phys.\ Lett.\  B {\bf 667}, 1 (2008).
  
  
  \bibitem{LEPcharged}
    [LEP Higgs Working Group for Higgs boson searches and ALEPH Collaboration
                  an],
  \linkart[hep-ex/0107031].
  
  
\bibitem{LEPpseudo}
  S.~Schael {\it et al.}  [ALEPH Collaboration and DELPHI Collaboration and
                  L3 Collaboration and ],
  Eur.\ Phys.\ J.\  C {\bf 47}, 547 (2006)
  \linkart[hep-ex/0602042].
  
  
  \bibitem{Plot}
\begin{verbatim}
http://lepewwg.web.cern.ch/LEPEWWG/plots/summer2005/s05_stu_contours.eps
\end{verbatim}

\bibitem{LEPEWWG}
The LEP Electroweak Working Group; http://lepewwg.web.cern.ch/LEPEWWG/

\bibitem{Hut:1977zn}
  P.~Hut,
  Phys.\ Lett.\  B {\bf 69}, 85 (1977).
\bibitem{Lee:1977ua}
  B.~W.~Lee and S.~Weinberg,
  Phys.\ Rev.\ Lett.\  {\bf 39}, 165 (1977).
\bibitem{Vysotsky:1977pe}
  M.~I.~Vysotsky, A.~D.~Dolgov and Y.~B.~Zeldovich,
  JETP Lett.\  {\bf 26}, 188 (1977)
  [Pisma Zh.\ Eksp.\ Teor.\ Fiz.\  {\bf 26}, 200 (1977)].
  
  \bibitem{GoldbergSwaveSuppression}
  H.~Goldberg,
  Phys.\ Rev.\ Lett.\  {\bf 50}, 1419 (1983)
  [Erratum-ibid.\  {\bf 103}, 099905 (2009)].
  
\bibitem{DreesNojiri}
  M.~Drees and M.~M.~Nojiri,
  Phys.\ Rev.\  D {\bf 47}, 376 (1993)
  \linkart[hep-ph/9207234].
  
  \bibitem{WMAP}
  N.~Jarosik {\it et al.},
  \linkart[1001.4744] [astro-ph.CO].
  
  


\bibitem{singlinoDM}
  B.~R.~Greene and P.~J.~Miron,
  Phys.\ Lett.\  B {\bf 168}, 226 (1986);
 S.~A.~Abel, S.~Sarkar and I.~B.~Whittingham,
  Nucl.\ Phys.\  B {\bf 392}, 83 (1993)
  \linkart[hep-ph/9209292];
  R.~Flores, K.~A.~Olive and D.~Thomas,
  Phys.\ Lett.\  B {\bf 245}, 509 (1990);
  K.~A.~Olive and D.~Thomas,
  Nucl.\ Phys.\  B {\bf 355}, 192 (1991);
 
  \bibitem{singlinoDMnmssm}
  G.~Belanger, F.~Boudjema, C.~Hugonie, A.~Pukhov and A.~Semenov,
  JCAP {\bf 0509}, 001 (2005)
  \linkart[hep-ph/0505142];
  V.~Barger, P.~Langacker and H.~S.~Lee,
  Phys.\ Lett.\  B {\bf 630}, 85 (2005)
  \linkart[hep-ph/0508027].



\bibitem{CDMS}
  Z.~Ahmed {\it et al.}  [The CDMS-II Collaboration and CDMS-II
                  Collaboration],
  \linkart[0912.3592] [astro-ph.CO].
  
\bibitem{XENON}
  J.~Angle {\it et al.}  [XENON Collaboration],
  Phys.\ Rev.\ Lett.\  {\bf 100}, 021303 (2008)
  \linkart[0706.0039] [astro-ph].



\bibitem{BarbierDM}
  R.~Barbieri, M.~Frigeni and G.~F.~Giudice,
  Nucl.\ Phys.\  B {\bf 313}, 725 (1989).
\bibitem{GoodmanWitten}
  M.~W.~Goodman and E.~Witten,
  Phys.\ Rev.\  D {\bf 31}, 3059 (1985).


\bibitem{heavyQnucleon}
  M.~A.~Shifman, A.~I.~Vainshtein and V.~I.~Zakharov,
  Phys.\ Lett.\  B {\bf 78}, 443 (1978).


\bibitem{ChiPTEllis}
  J.~R.~Ellis, K.~A.~Olive and C.~Savage,
  Phys.\ Rev.\  D {\bf 77}, 065026 (2008)
  \linkart[0801.3656] [hep-ph].

\bibitem{lattice}
  J.~Giedt, A.~W.~Thomas and R.~D.~Young,
  \linkart[0907.4177] [hep-ph].
  
  
  
  \bibitem{CDMSfit}
  M.~Farina, D.~Pappadopulo and A.~Strumia,
  \linkart[0912.5038] [hep-ph];

  J.~Kopp, T.~Schwetz and J.~Zupan,
  JCAP {\bf 1002}, 014 (2010)
  \linkart[0912.4264] [hep-ph].
  
  
  
  
\bibitem{nomura}
  Z.~Chacko, Y.~Nomura and D.~Tucker-Smith,
  Nucl.\ Phys.\  B {\bf 725}, 207 (2005)
  \linkart[hep-ph/0504095].



\bibitem{Ferrara:1974wb}
  S.~Ferrara and E.~Remiddi,
  Phys.\ Lett.\  B {\bf 53}, 347 (1974).


\bibitem{bsgamma}
  R.~Barbieri and D.~Pappadopulo,
  JHEP {\bf 0910}, 061 (2009)
  \linkart[0906.4546] [hep-ph].

\bibitem{haa14}
D.~Dominici, G.~Dewhirst, A.~Nikitenko, S.~Gennai, and L.~Fano,
      Search for radion decays into Higgs boson pairs in the 
                      $\gamma \gamma b b,\tau^+ tau^- b b$ and $bbbb$ final state, 
{\bf CMS-NOTE-2005-007}

\bibitem{haa7}

R.~Franceschini and R.~Torre, {\it in preparation}

  \bibitem{Kanehata:2011ei}
  Y.~Kanehata, T.~Kobayashi, Y.~Konishi, O.~Seto, T.~Shimomura,
  ``Constraints from Unrealistic Vacua in the Next-to-Minimal Supersymmetric Standard Model,''
    [arXiv:1103.5109 [hep-ph]].
 \bibitem{bertuzzo}
  E.~Bertuzzo, M.~Farina,
  ``Higgs boson signals in lambda-SUSY with a Scale Invariant Superpotential,''
   [arXiv:1105.5389 [hep-ph]].



\end{thebibliography}
\end{document}